\documentclass[fleqn,usenatbib]{mnras}

\usepackage{newtxtext,newtxmath}
\usepackage{booktabs}
\usepackage{array}
\usepackage{csvsimple}
\usepackage{rotating}
\usepackage{longtable}
\usepackage{pdflscape}
\usepackage{float}
\usepackage{caption}
\usepackage{xspace} 
\usepackage[export]{adjustbox}
\usepackage{afterpage}
\usepackage{longtable}
\usepackage{threeparttable}
\usepackage{subcaption}

\usepackage[T1]{fontenc}

\DeclareRobustCommand{\VAN}[3]{#2}
\let\VANthebibliography\thebibliography
\def\thebibliography{\DeclareRobustCommand{\VAN}[3]{##3}\VANthebibliography}

\usepackage{graphicx}	\usepackage{amsmath}

\newcommand{\Fermi}{\textit{Fermi}\xspace}

\title[Blazar Spectroscopy]{Spectroscopic Analysis of \textit{Fermi}-detected Blazars using SDSS-V}

\author[M. I. Nlowie et al.] {Mohammed Iddrisu Nlowie,$^{1}$\thanks{E-mail: m.i.nlowie@sms.ed.ac.uk}
James Aird,$^{1}$ Eli Kasai,$^{2}$ Amy L. Rankine,$^{1}$ 
Paloma Guetzoyan,$^{1}$
\newauthor
Scott Anderson,$^{3}$ Catarina Aydar,$^{4}$ William N. Brandt,$^{5}$ Joel R. Brownstein,$^{6}$ Xinyu Dai,$^{7}$ \newauthor
Lorena Hernández-García,$^{8,9}$ Paola Rodríguez Hidalgo,$^{10}$ Pranavi Hiremath,$^{1}$Anton Koekemoer,$^{11}$ \newauthor
Andrea Merloni,$^{4}$ Sean Morrison,$^{12}$ Mara Salvato,$^{4}$ Donald P. Schneider,$^{5}$ Axel Schwope,$^{13}$ \newauthor
Benny Trakhtenbrot$^{14}$ \\ \\
$^{1}$University of Edinburgh, Institute for Astronomy, Royal Observatory,
Blackford Hill, Edinburgh EH9 3HJ, UK\\
$^{2}$Department of Physics, Chemistry \& Material Science, University of Namibia, Private Bag 13301, Windhoek, Namibia \\
$^{3}$Department of Astronomy, University of Washington, Box 351580, Seattle, WA 98195, USA\\
$^{4}$Max-Planck-Institut f\"ur extraterrestrische Physik (MPE), 
Gie{\ss}enbachstra{\ss}e 1, 85748 Garching bei M\"unchen, Germany\\
$^{5}$Department of Astronomy and Astrophysics, 525 Davey Lab, The Pennsylvania State University, University Park, PA 16802, USA\\
$^{6}$ Department of Physics and Astronomy, University of Utah, 270 S. 1400 E. \#E2108, Salt Lake City, UT 84112, USA \\
$^{7}$Homer L. Dodge Department of Physics and Astronomy, University of Oklahoma, Norman, OK 73019, USA\\
$^{8}$Instituto de Estudios Astrof\'isicos, Facultad de Ingenier\'ia y Ciencias, Universidad Diego Portales, Av. Ej\'ercito Libertador 441, Santiago, Chile\\
$^{9}$Centro Interdisciplinario de Data Science, Facultad de Ingenier\'ia y Ciencias, Universidad Diego Portales, Av. Ej\'ercito Libertador 441, Santiago, Chile\\ 
$^{10}$Department of Astronomy, University of Washington, Box 351580, Seattle, WA 98195, USA\\
$^{11}$Space Telescope Science Institute, 3700 San Martin Drive, Baltimore, MD 21218, USA\\
$^{12}$ Department of Astronomy, University of Illinois at Urbana-Champaign, Urbana, IL 61801, USA\\
$^{13}$ Leibniz-Institut fur Astrophysik Potsdam (AIP), An der Sternwarte 16, D-14482 Potsdam, Germany \\
$^{14}$School of Physics and Astronomy, Tel Aviv University, Tel Aviv 69978, Israel\\
}

\date{Accepted XXX. Received YYY; in original form ZZZ}

\pubyear{\the\year{}}

\begin{document}
\label{firstpage}
\pagerange{\pageref{firstpage}--\pageref{lastpage}}
\maketitle

\begin{abstract} 
The automated spectroscopic pipeline of the Sloan Digital Sky Survey (SDSS) systematically assigns Galactic star, galaxy, or typical quasar classifications to jet-dominated blazars, owing to the absence of a non-thermal jet continuum component in its template library. In this study, we present a new, physically motivated, multi-component spectral fitting pipeline that we apply to 746 optical counterparts of Fermi/4FGL-DR4 $\gamma$-ray sources in the SDSS-V Data Release 20 spectroscopic database, yielding 707 well-fitted blazar candidates dominated by Power-law+Galaxy (59.4\%), Power-law+Lines (22.6\%), and Power-law+QSO (15.0\%) model families. Independent WISE infrared (IR) photometry confirms that 96.6\% of the 88 sources originally misclassified as Galactic stars by SDSS fall within the canonical blazar region, with 90.9\% reclassified as BL Lacertae object (BL Lac) candidates, demonstrating the success of our new pipeline.
The new classification scheme naturally recovers the known cosmological separation between BL Lac and Flat Spectrum Radio Quasar (FSRQ) candidates and a separation of approximately one order of magnitude in median $\gamma$-ray luminosity.
We compare our redshift estimates to a validation sample of 111 sources in the Third \textit{Fermi}-LAT Catalogue of High-Energy Sources (3FHL), finding a 10.4\% reduction in catastrophic failures ($\eta = 0.387$ vs $0.432$) over the SDSS pipeline. The equivalent width analysis validates the traditional $|{\rm EW}| = 5$\,\AA\ classification boundary at the population level (BL Lac: $4.47 \pm 0.15$\,\AA; FSRQ: $22.86 \pm 0.86$\,\AA), although 49.5\% of individual BL Lac candidates exceed this threshold and 51.4\% show simultaneous emission and absorption features. This hybrid population challenges the traditional binary blazar classification and points towards a more physically continuous description of blazar properties.

\end{abstract}

\begin{keywords}
galaxies: active -- BL Lacertae objects: general -- techniques: spectroscopic -- quasars: general
\end{keywords}

\section{Introduction}
Blazars are a special subclass of radio-loud Active Galactic Nuclei (AGNs) \citep{kellermann1989vla} that contain powerful beamed outflows in the form of relativistic jets originating close to the nucleus \citep{blandford2019relativistic}, with their jets closely aligned along our line of sight. They show the most extreme observational features of all AGNs, dominated by non-thermal emission processes, which span the entire electromagnetic spectrum from radio to TeV $\gamma$-rays and exhibit strong and variable polarization in the radio and optical bands, with flux and polarization variability observed on timescales ranging from minutes to years \citep{wagner1995intraday, urry1995unified, raiteri2025variability}. These variations in time are partly influenced by the jet emission undergoing a strong Doppler amplification, a relativistic boosting of the flux due to the motion of the emitting source towards the observer \citep{weaver2022kinematics}. The spectral energy distribution (SED) of blazars is characterized by the broad double-peaked structure: a low-energy component peaking between IR and X-ray frequencies, which is attributed to the synchrotron emission processes from the relativistic electrons in the jet, and a high-energy component peaking in the GeV-TeV $\gamma$-ray range, generally interpreted as inverse Compton scattering of either synchrotron photons (synchrotron self-Compton, SSC) or an external photon field \citep[external Compton, EC,][]{maraschi1992jet, sikora1994comptonization}. 

 Traditionally, blazars are divided into two main subclasses based on their optical spectroscopic properties: (1) Flat-spectrum radio quasars (FSRQ) and (2) BL Lacertae objects (BL Lacs). FSRQs are characterized by the presence of strong, broad optical emission lines with rest equivalent width (EW) $> 5$~\AA\ and quasar-like optical spectra, indicative of radiatively efficient accretion and a significant broad-line region (BLR); whereas BL Lacs in contrast, show weak or absent emission lines (EW $< 5$~\AA), featureless optical spectra often dominated by non-thermal jet continuum, and signs of radiatively inefficient accretion \citep{stickel1991complete, stocke1991einstein}. Blazars are said to be hosted in elliptical galaxies, with BL Lac host galaxy properties enabling absorption-line redshift determination when the non-thermal continuum is sufficiently weak \citep{oke1974distance}. The difference in optical spectroscopic properties reflects a physical distinction between the two blazar populations: FSRQs are powered by a radiatively efficient accretion disk that fuels the BLR, providing abundant seed photons for external Compton scattering, while BL Lacs represent lower accretion states where the jet dominates the observed emission, and the synchrotron self-Compton process plays the primary role in high-energy emissions \citep{fossati1998unifying,ghisellini2011transition}.
 
The identification of blazars has evolved through multiple complementary approaches. Gamma-ray detections from the \textit{Fermi}-Large Area Telescope (LAT) onboard NASA's \textit{Fermi} Gamma-ray mission have proven particularly effective as blazars dominate the extragalactic $\gamma$-ray sky \citep{atwood2009large, urry1995unified, ackermann2015third}. IR signatures revealed by the Wide-field Infrared Survey Explorer (\textit{WISE}) have also emerged as a powerful diagnostic tool, as blazars occupy distinct regions in the \textit{WISE} colour-colour space due to non-thermal jet emission, enabling efficient candidate identification \citep{massaro2011identification, d2012infrared}. As high radio emitters, radio surveys identify blazars through flat spectra or inverted spectra characteristic of Doppler-boosted synchrotron emission \citep{healey2008cgrabs}.

Despite the various characteristic emission properties of blazars at different wavelengths, optical spectroscopy remains the definitive method for classification and redshift estimation. However, optical spectroscopic classification of blazars remains challenging when the non-thermal jet continuum overwhelms the thermal emission from the host galaxy and accretion disk \citep{vermeulen1995bl}. Automated spectroscopic pipelines such as that used by the Sloan Digital Sky Survey \citep[SDSS;][Morrison et al., in prep]{bolton2012spectral}     are optimized for generic quasar and galaxy templates and hence fail to account for the dominant jet continuum characteristic of blazars. This leads to a systematic misclassification -- particularly of  BL Lac-like objects as stars.

In this work, we address these limitations by applying multi-component spectral fitting to a sample of  746 \textit{Fermi}-detected sources using SDSS optical spectra (see Section \ref{sec:samp_defini} for definition of our sample). Our approach (described in Section \ref{sec: 3}) explicitly models the non-thermal jet continuum alongside galaxy and quasar templates, enabling robust classification even when thermal features are suppressed. We demonstrate that this physically informed method significantly improves upon the SDSS automated classifications and redshift measurements, recovering the canonical BL Lac/FSRQ dichotomy and revealing systematic misclassification patterns in the SDSS catalogue (Section~\ref{sec:results}). We validate our classifications through independent diagnostics, including WISE IR colour properties across the two main blazar populations (Section \ref{sec:discuss}). We end with our overall conclusions and a discussion of future directions (Section~\ref{sec:conclusions}).

\section{Telescopes and Sample Definition}
\label{sec:samp_defini}
In this section, we provide details of our $\gamma$-ray--selected sample (Section~\ref{sec:fermi}) and the spectroscopic observations provided by SDSS-V (Section~\ref{sec:sdssv}). In  Section \ref{corr}, we present the cross-matching between the SDSS-V spectroscopic catalogue and \Fermi/4FGL-DR4 $\gamma$-ray catalogue, which yields our working sample. We then present supporting multi-wavelength imaging properties from the Wide-field Infrared Survey Explorer \citep[\textit{WISE}][]{wright2010wide} and DESI Legacy Imaging Survey \citep{dey2019overview} that confirm the extragalactic nature of the vast majority of these sources (Section \ref{support_wave}). 

\subsection{The \Fermi 4FGL-DR4 $\gamma$-ray  catalogue}
\label{sec:fermi}

The Large Area Telescope (LAT), which is the main instrument onboard the \textit{Fermi Gamma-ray Space Telescope (Fermi)} mission, is an imaging, wide field-of-view (FoV), high-energy $\gamma$-ray telescope, which has been making observations of the entire $\gamma$-ray sky each day since science operations began in August 2008 \citep{atwood2009large}. Over the years, the LAT survey has been releasing successive incremental versions of catalogues based on a comprehensive analysis of the LAT data. For all the catalogues, including the 4th iteration of the \Fermi Gamma-ray LAT (hereafter, 4FGL) catalogue, new analysis techniques and event reconstructions were employed, most notably \textit{Pass 8}, an overhaul of the LAT data processing pipeline \citep{Atwood_2013arXiv1303.3514A}. \textit{Pass 8} basically redefined how $\gamma$-ray events are reconstructed, which results in an enhanced effective area at low energies ($<{\rm 100}$ MeV). Compared with earlier data releases, these improvements make a significant difference in sensitivity and source classification accuracy.  

Based on the first 14 years of research data, the latest incremental data release (\textit{Release 4}) of the 4FGL  catalogue (4FGL-DR4) employs the same analysis techniques as the 4FGL-DR3 (used for 12 years of scientific data) to detect sources between 50 MeV and 1 TeV \citep{abdollahi2022incremental}. The catalogue contains 7194 $\gamma$-ray sources detected above a test-statistic threshold of $TS > 25$, corresponding to approximately a $4\sigma$ detection threshold. DR4 adds 546 new sources and 4 new extended sources relative to DR3, while 14 sources were deleted and 10 were relocalized. Among the new sources, 237 have been associated with counterparts at other wavelengths \citep{ballet2023fermi}. 

The spectral modelling of the \Fermi sources primarily uses three functional forms depending on the observed spectral properties: Log-Parabola (LP) for sources showing significant spectral curvature  (primarily AGNs), Power-law (PL) for sources without significant curvature, particularly faint sources, and a Power-law with exponential cutoff (PLEC) for pulsars with nebulae. These spectral models are used in the estimation of the $\gamma$-ray fluxes for those sources.  
In DR4, Gaussian priors on curvature parameters for faint sources (${\rm TS} < 100$)  are used to prevent detection of curvature driven by statistical fluctuations rather than genuine spectral features, resulting in more robust estimates of spectral peak energies \citep{ballet2023fermi}. 

 \subsubsection{Source Association}
 Out of the 7194 sources detected in 4FGL-DR4, 2425  objects remain unassociated with known $\gamma$-ray emitting sources  \citep{neumann2025classification}, whereas 4769 sources have plausible counterparts at other wavelengths identified through statistical methods. The source association of the object is implemented using two complementary approaches: the Bayesian association and Likelihood Ratio (LR) method, applied as described for earlier data releases \citep{ballet2020fermi, ballet2023fermi, abdollahi2020fermi}. The Bayesian association method uses the positional coincidence, local source density, and a prior on the flux distribution of counterparts to estimate the probability that a radio, optical, or X-ray source is the correct counterpart. 
 The likelihood-ratio method is instead used near the Galactic plane ($ |b| < 10^{o} $) where crowded fields tend to produce frequent false matches due to the overlap of sources \citep[see the original 4FGL catalogue,][for a detailed description of the methods]{abdollahi2020fermi}. The photon index distribution for the blazar sources also shows an evolutionary trend across data releases. Blazar candidates from older data releases (4FGL; DR1, DR2, DR3) exhibit photon indices peaked at $\Gamma \approx 2.2$, whereas the newer DR4 blazar candidates peaked at $\Gamma \approx 2.5$, suggesting detection of sources with spectra dominated by lower energy photons, which is likely due to variability bias in long-term energy exposure \citep{ajello2022fourth}. This trend likely reflects the detection of sources with softer $\gamma$-ray spectra in newer data releases, where a higher proportion of lower-energy photons relative to higher-energy ones corresponds to lower $\gamma$-ray luminosities for a given observed flux. These sources are therefore either intrinsically less luminous or more distant objects that only become detectable with the longer exposures and improved sensitivity of DR4, and whose optical counterparts are correspondingly fainter and more challenging to characterise spectroscopically.

\subsubsection{Fermi blazar classification}
The main classification scheme within the \Fermi-LAT catalogues relies on a multi-wavelength approach that combines radio observations, IR photometry, $\gamma$-ray detections, and optical spectroscopy (when available). 

Blazars dominate the 4769 \Fermi-associated sources and constitute the largest class of extragalactic $\gamma$-ray emitters. Of these, $ 1490$ are identified as BL Lac type objects and $820$ are identified as FSRQs based on pre-existing optical spectroscopy. A further $1623$ constitute "Blazar Candidates of Uncertain Type" (BCU) as they lack an optical classification. BCUs are sources that show  a  two-humped SED typical of blazars but lack an optical spectrum for distinct classification \citep{kang2019evaluating}. The rest are classified as non-blazar AGN and stellar sources \citep{ballet2023fermi}. While ongoing optical spectroscopic follow-up programmes have steadily reclassified BCUs into confirmed blazar subclasses, the high fraction of BCUs among newly added sources ($72 \%$ of sources added from DR 2 to DR4; \citealt{ballet2023fermi}) has kept the overall BCU fraction roughly constant, constituting $\sim34 \%$ of 4FGL - DR4 sources with plausible counterparts -- comparable to the $35\%$ fraction in the original DR1 catalogue.

With optical spectroscopy as the primary classification tool for blazars, the SDSS spectroscopy has been widely used to classify and characterise \Fermi-associated sources across multiple data releases \citep{shaw2012spectroscopy, massaro2014optical, alvarez2016optical}. These works relied on SDSS spectra as an archival resource for characterization and confirmation of the blazar nature of a significant number of sources. However, these studies treated the SDSS spectra as a passive archive without addressing the systematic limitation of the automated SDSS pipeline itself. The pipeline (see Section~\ref{sec:sdssv}) lacks an explicit non-thermal jet continuum in its template library and provides no distinct blazar classification, leading to the systematic misclassification of some sources (mostly jet-dominated) as stars, passive galaxies, or quasars. This motivates the development of a jet-aware multi-component spectral fitting approach to correctly characterise, classify, and estimate redshifts for these sources within the SDSS-V DR20 spectroscopic database.

\subsection{SDSS-V DR20 Spectroscopic catalogue and Pipeline}
\label{sec:sdssv}
The Sloan Digital Sky Survey, now in its fifth generation (SDSS-V; 2020 -- 2027), provides large-scale optical spectroscopy for multi-object studies spanning stars, galaxies, quasars, and AGNs \citep{almeida2023eighteenth, adamane2025nineteenth,  kollmeier2026sloan} using two 2.5 metre telescopes: the Sloan Foundation Telescope at Apache Point Observatory in New Mexico \citep{gunn20062} and the Ir\'en\'ee du Pont Telescope at Las Campanas Observatory in Chile \citep{bowen1973optical}. This dual-site configuration enables comprehensive coverage of both northern and southern skies. In this work, we use the  $({\rm spAll-lite-v6\_2\_1})$ Daily catalogue, which contains roughly 12 million spectra obtained using the BOSS spectrograph \citep{smee2013multi} during the SDSS-V programme up until MJD 60708 (2nd February 2025), which will form the forthcoming SDSS Data Release 20 (DR20; SDSS-V Collaboration et al, in prep).
The BOSS spectra span wavelengths of $3600 - 10400$ \AA\ at a resolution of ${\rm R} \approx 2000$.  
 Each object has been processed through version 6.2.1 of the automated SDSS spectroscopic pipeline \citep[${\rm idlspec2d}$;][Morrison et al., in prep]{bolton2012spectral} to assign a spectroscopic classification (CLASS: STAR, QSO, or GALAXY), estimated redshift ($z$), uncertainty $(z_{\rm ERR})$, and a redshift quality indicator (given by the ZWARNING flag). 
The Daily catalogue is based on co-adds of multiple exposures of the same source obtained on a given night, while observations obtained on different nights are retained as distinct entries in the catalogue. 

\subsubsection{SDSS Spectroscopic Pipeline Classification}
The SDSS spectroscopic pipeline (${\rm idlspec2d}$) uses a $\chi^2$ minimization approach to classify spectra and also estimate redshifts by comparing observed spectra against libraries of template spectra spanning stars, galaxies, and quasars \citep[Morrison et al, in prep]{bolton2012spectral}. For each observed spectrum, the pipeline evaluates multiple template classes across a grid of trial redshifts. The templates include: (1) galaxy eigenspectra derived from principal component analysis of 480 high-quality SDSS galaxy spectra, which capture dominant modes of spectral variation; (2) quasar eigenspectra derived from 849 SDSS-IV Reverberation Mapping (RM) quasars;  
and (3) stellar templates combining observational libraries (Indo-US, ELODIE) with theoretical atmosphere models for different stellar types \citep[Morrison et al, in prep]{bolton2012spectral, prugniel2001database, valdes2004indo}.

At each trial redshift, the pipeline performs error-weighted least square fitting to determine how well each template matches the observed spectrum. The quality of fit is accounted for by the reduced $\chi^2$. The combination of redshift and template class, which produces the lowest reduced $\chi^2$ across all template families, is adopted as the pipeline classification (CLASS) and redshift ($z$). A low confidence warning flag (\textsc{ZWARNING}) is assigned when the best fit is not statistically distinguishable from an alternative solution, indicating potential classification ambiguity \citep{bolton2012spectral}.

\subsubsection{Systematic Misclassification of Blazars}

A critical limitation of the SDSS spectroscopic pipeline is its potential to misclassify non-thermal jet-dominated blazars, particularly BL Lacs with featureless optical spectra. The pipeline classifies spectra using template fitting across stellar, galaxy, and QSO libraries \citep[Morrison et al, in prep]{bolton2012spectral}, but lacks blazar-informed templates. BL Lac objects mostly exhibit relatively featureless optical continua dominated by non-thermal synchrotron emission from the relativistic jet, described by power-law spectral energy distributions \citep{urry1995unified}. These power-law continua closely resemble the featureless spectra of DC (ie., featureless) white dwarfs ---stellar remnants whose optical emission is a smooth, featureless blackbody with no detectable absorption or emission lines. For that reason, the SDSS automated pipeline cannot reliably distinguish whether a featureless optical continuum originates from a BL Lac object or a Galactic stellar source, leading to the systematic misclassification of genuine BL Lac candidates as \texttt{CLASS = STAR}. This confusion between BL Lacs and DC white dwarfs in purely optical classification schemes has been widely noted \citep[e.g.][]{plotkin2010optically}. Along with the multi-wavelength validation presented in Section~\ref{support_wave}, which also demonstrates that \textit{Fermi}-LAT blazars are frequently misclassified by the SDSS automated pipeline, these issues motivate the multi-component fitting approach developed in this work (see Section~\ref{sec: 3}).

\subsection{Cross-matching between the SDSS-V and 4FGL-DR4 catalogues}
\label{corr}
We perform a crossmatch between the 4FGL-DR4 associated sources and the SDSS-V $({\rm spAll-lite-v6\_2\_1})$ spectroscopic catalogue. 
While \textit{Fermi}-LAT's arcminute-scale positional uncertainties \citep[typically  $\sigma_{\rm pos}\sim 30''$ for the $95\%$ confidence error ellipse;][]{ballet2023fermi} present a challenge for direct source identification and matching, the 4FGL-DR4 catalogue provides refined multi-wavelength counterpart positions through the extensive Bayesian and likelihood association procedures \citep{abdollahi2020fermi}.
These procedures identify counterparts from high-precision surveys, including radio catalogues \citep[NVSS at 1.4\,GHz with 
$\sim 1''$ positional accuracy for bright sources;][]{condon1998nrao}, IR surveys \citep[SUMSS at 843\,MHz with $1$--$2''$ accuracy;][]{mauch2003sumss}, all-sky infrared 
photometry \citep{wright2010wide}, and spectroscopically confirmed blazar catalogues \citep[Roma-BZCAT;][]{massaro2009roma, massaro2015refining}. The catalogues report these counterpart positions in the \texttt{RA\_counterpart} and \texttt{DEC\_counterpart} columns, which represent the survey catalogue positions instead of the $\gamma$-ray-derived localizations. These associated positions typically have arcsecond-level uncertainties, enabling precise cross-matching with optical spectroscopic surveys. We use a 2 arcsec matching radius for cross-matching the multi-wavelength counterpart positions for the 4769 objects to SDSS spectroscopic observational positions. 

\begin{figure}
    \centering
    \includegraphics[width=0.99\linewidth]{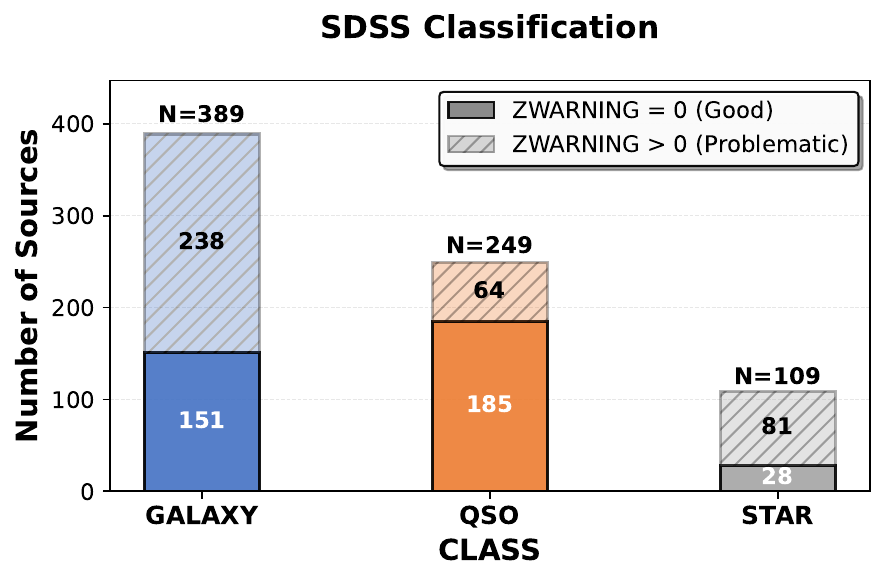}
    \caption{Distribution of 746 unique optical spectroscopic matches for Fermi-detected sources among the 3 main classification schemes within the SDSS pipeline and divide according to their redshift quality (ZWARNING) flags, as indicated.}
    \label{fig:s_cl}
\end{figure}
A total of 1400 optical spectra were obtained, which corresponds to 746 unique sources. The discrepancy between total spectra and unique sources arises from multiple spectroscopic observations of the same objects (313 sources with repeat observations). For sources with multiple spectra, we retain the observation with higher signal-to-noise (S/N). Figure~\ref{fig:s_cl} shows the distribution of the SDSS pipeline classification for our sample. The GALAXY class dominates with 389 sources ($52 \%$), followed by 249 QSOs ($33\%$) and 109 ($15\%$) sources classified as STAR ($15 \%$). The substantial fraction of STAR classifications is striking, considering the fact that our sample is selected from $\gamma$-ray sources associated with the 4FGL-DR4 catalogue, which is expected to be dominated by an extragalactic population consisting of blazars and AGN rather than Galactic stellar sources. 
The SDSS pipeline assigns quality flags to each spectroscopic classification through the ZWARNING parameter, where ${\rm ZWARNING} = 0$ indicates a reliable redshift measurement and classification, while ZWARNING > 0 flags potential issues with the spectroscopic data, fit quality, or redshift determination. $364~ (49\%)$ have ${\rm ZWARNING} = 0$, indicating reliable classifications, while $383~(51\%$) have ${\rm ZWARNING} > 0$, flagging potential classification or redshift uncertainties. The high fraction of problematic classifications motivates this work to distinctively classify and correctly estimate the redshifts, particularly for the jet-dominated spectra. 
Table~\ref{tab:number_of_sources} summarises 
the full sample construction from the initial 
4FGL-DR4 catalogue through to the final 
working samples used for spectroscopic, 
IR, and optical-infrared colour analyses.

\subsection{Supporting multi-wavelength imaging}
\label{support_wave} 

\subsubsection{Wide-field Infrared Survey Explorer}
\begin{figure*}
    \centering
    \begin{subfigure}[t]{0.48\linewidth}
        \centering
        \includegraphics[width=0.99\linewidth]{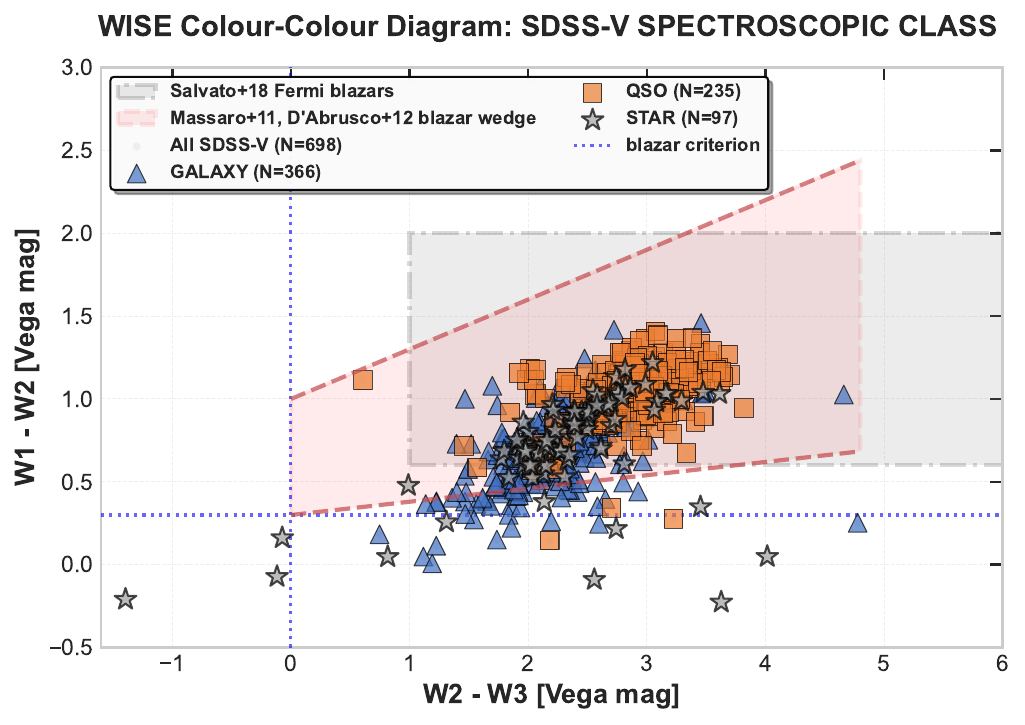}
        \phantomcaption
         \label{fig:sdss_wise}
    \end{subfigure}
    \hfill
    \begin{subfigure}[t]{0.48\linewidth}
        \centering
        \includegraphics[width=0.99\linewidth]{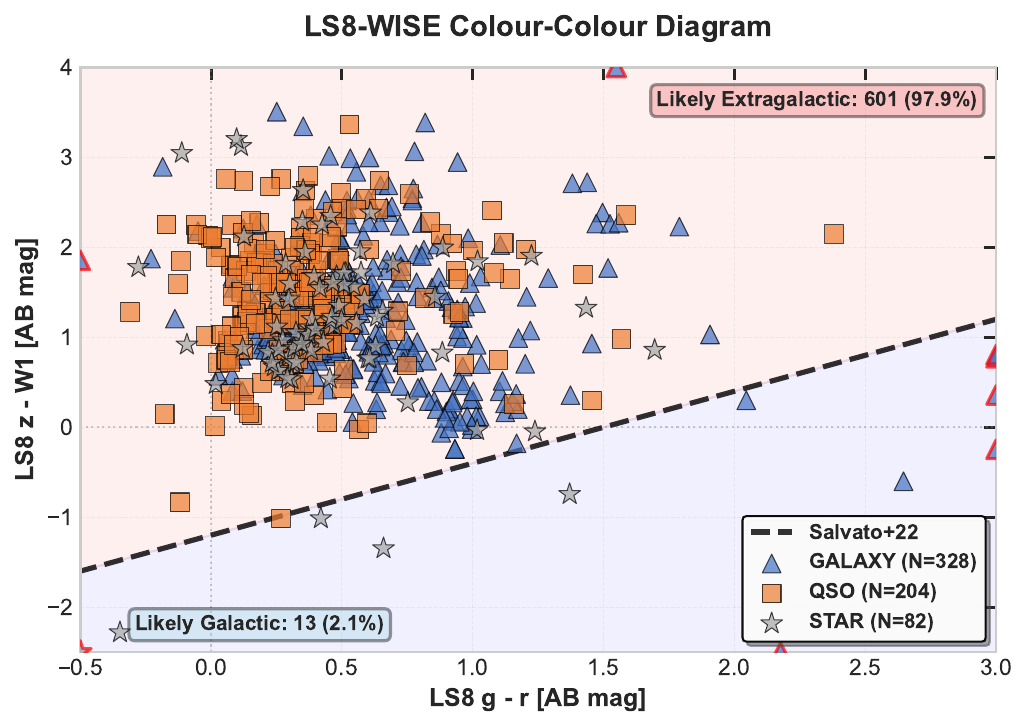}
        \phantomcaption
         \label{fig:ls8_wise}
    \end{subfigure}
    \caption{Multi-wavelength colour diagnostics for the Fermi-detected SDSS-V spectroscopic sample. \textit{Left panel:} \textit{WISE}  colour-colour diagram for the 698 sources with valid \textit{WISE} magnitudes,  colour-coded by SDSS pipeline classification: GALAXY (blue triangles, $N = 366$), QSO (orange squares, $N = 235$), and STAR (grey stars, $N = 97$). The coloured regions show established blazar selection criteria from 
    \citet{massaro2011identification} and 
    \citet{d2012infrared} (red dashed wedge) and  \citet{salvato2018finding} (grey box). Over $80\%$ of sources fall within the canonical blazar region, confirming their IR signatures are consistent with non-thermal blazar emission. \textit{Right panel:} Scatter plot of 614 sources in ($g - r$) versus ($z - W1$) colour space, colour-coded by SDSS pipeline classification. The dashed line \citep{salvato2022erosita} separates 
    likely Galactic sources from likely 
    extragalactic sources. $93.9\%$ (77/82) of 
    SDSS STAR-classified sources lie above the 
    extragalactic line, confirming that these 
    objects exhibit photometric signatures 
    indicating that an extragalactic origin is more likely than a Galactic origin. Sources lying outside plot boundaries are clipped to the axis limits and indicated by red-edged markers.}
    \label{fig:wise_ls8}
\end{figure*}

\textit{WISE} took observations of the sky in four mid-IR bands: $W1 ~(3.4 ~\mu m$), sensitive to stellar light and dust-obscured AGN due to reduced extinction; $W2 ~(4.6 \mu m$), effective for detection of dust-obscured AGN; $W3~(12 ~\mu m$), which captures thermal emission from warm dusty star-forming regions or AGN torus emission; and $W4~(22 ~\mu m$), not heavily used due to lower resolution and sensitivity \citep{cluver2014galaxy}. \textit{WISE} IR colours provide independent confirmation of SDSS misclassifications, as blazars occupy a distinct region of \textit{WISE} colour space --- the blazar strip --- clearly separated from the Galactic stellar locus \citep{massaro2011identification, d2012infrared}. Sources misclassified as stars or galaxies by the SDSS automated pipeline can therefore be identified as blazar candidates through their IR colours alone, without relying on optical spectral features. Out of the 746 sources, 698 (GALAXY = 366, QSO = 235, STAR = 97) have valid $W1$, $W2$, and $W3$  magnitudes. In Figure~\ref{fig:sdss_wise}, we investigate the \textit{WISE} colour-colour IR signatures of our sample using the \textit{WISE} magnitudes reported in the SDSS-V catalog against canonical demarcated blazar regions. We compare our sources to the WISE blazar region  by \cite{massaro2011identification} and \cite{d2012infrared}, approximating the strip boundaries from their published colour-colour diagrams as $W1-W2 > 0.3 + 0.05 \times (W2-W3)$ and $W1-W2 < 1.0 + 0.3\times (W2-W3)$ for $0 < W2-W3 < 4.8$. We also annotate a region identified by \cite{salvato2018finding} as containing X-ray detected \textit{Fermi} blazars. Over 80\% of all sources fall squarely within the canonical blazar colour space, consistent with other extragalactic sources and far from the Galactic stellar locus \citep{stern2012mid}.

\subsubsection{DESI Legacy Survey colour-colour Diagnostics} 
To further validate our suspected misclassified SDSS objects, we examine the extragalactic and Galactic signatures of the objects within the \textit{Legacy Survey}  magnitude space. The Legacy Survey Data Release 8 (LS DR8; \cite{dey2019overview}) provides deep optical imaging in the $g, r$ and $z$ bands with typical $5\sigma$ depth of $g = 24.0$, $r = 23.4$, $z = 22.5$, reaching approximately 1-2 magnitudes deeper than earlier SDSS photometry. This deeper photometric coverage 
was used as one of the parent catalogues for SDSS-V targets \citep{almeida2023eighteenth} 
and provides independent constraints on source classification through multi-wavelength colour-colour diagnostics.

668 out of our 746 SDSS-V spectroscopic counterparts of \textit{Fermi} sources have photometry available from LS8. 
Of these 668 objects, 614 have measured magnitudes in the $g, r$ and $z$ bands as well as in the \textit{WISE} $W1$ photometry incorporated witin the LS8 catalogues. We construct a colour-colour diagram plot of (${\rm g-r}$) versus (${\rm z-W1}$), where the optical colours (${\rm g - r}$) trace stellar population properties, and the optical-IR colour (${\rm z-W1}$) is sensitive to dust content, separating stellar sources with low dust emission from extragalactic sources with AGN-heated dust emission. The colour combinations have been extensively used to separate likely extragalactic from Galactic sources based on their spectral energy distribution \citep[e.g.,][]{ruiz2018xmmpzcat, salvato2022erosita}. We adopt the relation $(z - W1) = 0.8 \times (g - r) - 1.2$) by \citet{salvato2022erosita} to establish the demarcation between the likely Galactic and extragalactic sources. 

The demarcation (see Figure~\ref{fig:ls8_wise}) confirms that our Fermi-selected sources with SDSS-V spectroscopy, which we expect to be associated with distant blazars, indeed fall in the "extragalactic" region of this diagram.  $99.2 \%~ (203/204) $ of SDSS QSO-classified sources populate the extragalactic region, validating both the demarcation criterion and the accuracy of the QSO classification. Notably, $97.9 \%~(321/328)$ of SDSS GALAXY-classified sources fall above the extragalactic line, consistent with their association with $\gamma$-ray emission likely driven by AGN activity. For the SDSS STAR-classified objects, $93.9 \%~(77/82)$ consistently occupy the extragalactic region, demonstrating that these objects exhibit photometric signatures typical of extragalactic objects. 
While the multi-wavelength properties enable efficient candidate selection, optical spectroscopy remains essential for definitive classification and redshift determination. However, the standard automated pipelines fail when the jet continuum dominates the optical spectrum \citep{goldoni2021optical}. A parametric multi-component fitting based on the expected characteristics of blazar spectra---combining a non-thermal power-law continuum with host galaxy and AGN templates---provides more reliable classification and redshift estimates \citep{pita2014spectroscopy, goldoni2021optical}. This motivates a new spectral classification pipeline that we have developed to automatically fit the SDSS-V optical spectra of Fermi-detected sources with suitable templates that capture both the featureless jet emission and any signature of either an underlying host galaxy or QSO emission (both from the accretion disk and BLR).

\begin{table*}
\centering
\caption{Summary of Fermi sources and their optical counterparts.}
\label{tab:number_of_sources}
\begin{tabular}{l c c l}
\hline\hline
\textbf{Category} & \textbf{Objects} & \textbf{Catalogue} & \textbf{Notes} \\
\hline
Total Fermi sources & 7194 & 4FGL-DR4 & All-sky gamma-ray catalogue \\
Fermi sources with counterparts &4769 & -- & As identified in 4FGL-DR4 catalogue \\
No. of SDSS spectra of Fermi sources & 1400 & SDSS-V & After crossmatch within 2\arcsec\\
Unique SDSS counterparts to Fermi sources & 746 & SDSS-V & Using unique SDSS-ID \\
No. with valid optical photometry & 614 & LegacySurvey8 & Closest matches within 2\arcsec\\
No. with valid Infrared photometry & 698 & \textit{WISE}  & As identified in our SDSS working sample \\
\hline
\end{tabular} 
\end{table*}

\section{Multi-component Fitting} 
\label{sec: 3} 
In this section, we provide a detailed description of the multi-component spectral fitting pipeline developed in this work. The pipeline explicitly models the non-thermal jet continuum in combination with galaxy, QSO, and emission line-only templates that captures physically-informed contributions from the host galaxy, accretion disk or BLR to recover robust classifications and redshift measurements.

\subsection{Template Libraries and preparation}
\label{sec:templates}
We employ three complementary template libraries for our optical spectral fitting that we have found to be effective in describing the optical spectroscopic properties of our Fermi-SDSS cross-matched sample. 

Firstly, to capture host galaxy light,  we adopt three empirical elliptical galaxy templates from the Spitzer Wide-area InfraRed Extragalactic (SWIRE) library:  Ell2, Ell5, and Ell13, which represent stellar populations with ages of 2, 5, and 13 Gyr, respectively \citep{polletta2007spectral, de2025origin}. These templates span the range of stellar ages expected for massive elliptical galaxies and include stellar absorption features such as the  Ca II H$\&$K break (4000\AA) and Mg b absorption ($\sim$5175 \AA), which are often the most detectable spectral features in jet-dominated BL Lac spectra. While originally developed for multi-wavelength photometric SED modelling, the SWIRE elliptical templates span a wavelength range of 1000~ \AA\ to $1000~ \mu$m, providing full coverage of the SDSS optical window (3600 -- 10400~ \AA) and are sufficient to reproduce the broadband continuum shape and characteristic features of evolving elliptical galaxies \citep{polletta2007spectral}.  

We generate eight synthetic QSO templates and three emission line-only templates as the second and third sets of templates using the QSOGEN code \citep{temple2021modelling}. QSOGEN is a parametric quasar SED model that generates synthetic spectra that account for continuum emission, empirically derived emission line contributionshost galaxy light, dust reddening, and intergalactic medium (IGM) absorption. QSOGEN is calibrated on the SDSS DR16 quasar catalogue and reproduces optical-to-infrared colours across wide ranges of redshifts ($0 < z < 5$). The redshift parameter in QSOGEN controls 
the relative contribution of the host galaxy, as lower-redshift SDSS quasars exhibit a small-but-significant host galaxy contribution at the red wavelengths, while higher-redshift quasars are dominated by the AGN continuum \citep{vanden2001composite}.  
We construct six base templates corresponding to two redshifts ($z = 0.2\,{\rm and} \,1.5$) with varying emission line properties: default emission lines, high equivalent width (high-EW), and high blueshift. The $z = 0.2$ templates represent stronger host galaxy contributions, characteristic of lower-redshift quasars, while the $z = 1.5$ templates represent quasars with minimal host contamination; we note that all templates are considered across our full redshift range when used to fit observed spectra as described in Section~\ref{sec:chi} below. We add two reddened templates at intermediate host contribution (i.e. at $z = 0.5$) with default emission line properties but dust extinction of ${\rm E(B-V) = 0.1 ~and ~0.2}$, accounting for quasars with redder optical continua. This results in eight QSO templates in total. 

In addition to the full QSO templates, we include three emission-line-only templates corresponding to median-line, high-EW-line, and high-blueshift components in QSOGEN. These 3 templates isolate emission line contributions in the absence of continuum, enabling flexible modelling of sources where lines are present but the QSO continuum is weak or absent. This gives eleven (11) QSO-derived templates in total.  

These QSO and emission-line-only templates capture the characteristic FSRQ spectral properties, including strong emission lines such as broad features from the BLR -- Mg II ($\lambda~2798$), C IV ($\lambda~1549$) and H$\alpha$($\lambda~6563$)-- as well as narrow forbidden lines from the narrow-line region (NLR) such as [O III] ($\lambda~5007$), and [O II] ($\lambda~3727$). The reddened templates are particularly important for blazars, as dust extinction can significantly alter the observed continuum shape and line strengths \citep{gaskell2004nuclear, gaskell2007agn}.

We prepare the templates to cover the wavelength range of $500 - 15~000 $~\AA\ ensuring full coverage of the SDSS BOSS spectrograph window ($3600 - 10400\, \AA$) across the entire redshift grid with ($z = 0.01 - 5$), with the blue limit mapping to a rest-fram wavelength of $\sim 630$~\AA\,. All spectral fitting is performed directly on the native log-spaced BOSS wavelength grid of each individual spectrum. Each set of templates is resampled onto this native observed-frame grid using  \textsc{SpectRes} \citep{carnall2017spectresfastspectralresampling}, a  flux-conserving resampling algorithm that preserves integrated fluxes when regridding, avoiding correlated residuals at spectral features that would systematically bias $\chi^{2}$ statistics. This native approach preserves the natural pixel-to-pixel noise properties of the BOSS spectrograph and avoids the flux redistribution artefacts that would arise from resampling the observed data onto an external grid.  

\subsection{Spectral Model Families} 
\label{sec:models}

The fitting procedure evaluates six distinct model families---where a family is defined as a set of models sharing the same physical components but varying in template choice--- each representing a physical scenario for the optical emission properties of blazars.  All six models are evaluated across a fixed logarithmic redshift grid spanning $z = 0.01$ to $5.0$ with 150 grid points (Section \ref{sec:chi^2}).  \\ 

\noindent\textit{Single-component models:}
\begin{enumerate}
    \item \textbf{Galaxy} --- SWIRE elliptical galaxy templates (Ell2, Ell5, Ell13), each scaled by amplitude $s_{\rm gal}$, representing passively evolving host galaxy continua (Section~\ref{sec:templates}).
    \item \textbf{QSO} --- eight synthetic QSO templates scaled by $s_{\rm QSO}$, capturing quasar continuum and emission line contributions 
    (Section~\ref{sec:templates}).
    \item \textbf{Powerlaw (PL)} --- a flexible dual-form power-law representing non-thermal synchrotron jet emission with spectral curvature.
\end{enumerate}

\noindent\textit{Multi-component models:}
\begin{enumerate}
    \setcounter{enumi}{3}
    \item \textbf{PL + Galaxy} --- non-thermal jet continuum combined with a host galaxy template, characterising BL Lac candidates where jet emission 
    dilutes stellar absorption features.
    \item \textbf{PL + QSO} --- non-thermal jet continuum combined with a QSO 
    template, capturing FSRQs where the jet coexists with accretion 
    disc and BLR emission.
    \item \textbf{PL + Lines} --- non-thermal jet continuum combined with 
    emission-line-only templates, enabling the identification of sources where the accretion disc continuum is absent or weak but emission lines remain detectable and masquerading BL Lacs. 
\end{enumerate}

In the single-component PL model and the three multi-component models, we adopt a 
a flexible dual-form power-law, which represents non-thermal synchrotron jet emission with additional spectral curvature:  

\begin{equation}
F_{\rm PL}(\lambda) = \begin{cases}
    \displaystyle\frac{A}{0.5x^{-\alpha} + 0.5x^{-(\alpha + \delta)}} & \text{if } \delta \ge 0 \\[12pt]
    A \left(0.5x^{\alpha} + 0.5x^{\alpha + \delta}\right) & \text{if } \delta < 0
\end{cases}
\end{equation}
where $x = \lambda/\lambda_0$. The pivot wavelength $\lambda_0 = 5000$~ \AA\ is held fixed throughout, as it acts as a reference point and is degenerate with $A$ under renormalization.  $\alpha$ is the base spectral slope and $\delta$ controls the spectral curvature strength and shape. The parameter $\delta$ determines both the strength and direction of any deviation from a pure power-law that produces additional spectral curvature: positive values produce flattening at the shorter wavelength end (characteristic of low-frequency synchrotron turnover), whereas negative values produce steepening (characteristic of cooling breaks or high-energy cutoffs). Parameter constraints and initialization values are described in Section \ref{sec:chi^2}. 

At redshift $z\gtrsim 2$, neutral hydrogen along the line of sight attenuates flux blueward of Ly$\alpha$ (rest-frame 912--1216~\AA), producing the characteristic Lyman forest and Lyman limit features. We account for this by applying a wavelength-dependent IGM medium  transmission function $T_{\rm IGM}(\lambda_{\rm obs}, z)$. The IGM transmission function  is computed using \citet{madau1995radiative} prescription implemented through the $\texttt{etau\_madau}$ package in the \textsc{synphot} library \citep{2018ascl.soft11001S}. The optical depth follows:
\begin{equation}
\tau_{\rm forest} \approx 0.0025 ~ (1 + z)^{3.7} \left(\frac{1216~\text{\AA}}{\lambda_{\rm rest}}\right)^{3.46}
\end{equation}
where the resulting transmission $T_\mathrm{IGM} = \exp(-\tau_\mathrm{forest})$. We apply this multiplicatively to all model components at each trial redshift, ensuring the Lyman break is correctly captured in high-redshift fits. Without this correction, the $\chi^2$ minimisation (described in Section~\ref{sec:chi^2}) could systematically underestimate redshifts for $z > 2.5$ sources by misinterpreting IGM absorption as intrinsic spectral curvature.

The three multi-component model families combine the power-law with a thermal template component. A PL + galaxy combined model which accounts for both the non-thermal jet emission and an underlying host galaxy contribution: 
\begin{equation}
F(\lambda_{\rm obs}) = \left[s_{\rm gal} \cdot T_{\rm gal}\!\left(\frac{\lambda_{\rm obs}}{1+z}\right) 
+ F_{\rm PL}(\lambda_{\rm obs})\right] \cdot T_{\rm IGM}(\lambda_{\rm obs}, z)
\end{equation}
The PL+QSO and PL+Line-only models takes the same form, with the galaxy template replaced by a QSO template or an emission line.
\begin{equation}
F(\lambda_{\rm obs}) = \left[s_{\rm QSO} \cdot T_{\rm QSO}\!\left(\frac{\lambda_{\rm obs}}{1+z}\right) 
+ F_{\rm PL}(\lambda_{\rm obs})\right] \cdot T_{\rm IGM}(\lambda_{\rm obs}, z)
\end{equation}

\begin{equation}
F(\lambda_{\rm obs}) = \left[s_{\rm line} \cdot T_{\rm line}\!\left(\frac{\lambda_{\rm obs}}{1+z}\right) 
+ F_{\rm PL}(\lambda_{\rm obs})\right] \cdot T_{\rm IGM}(\lambda_{\rm obs}, z)
\end{equation}

These two models represent typical FSRQ cases where sources exhibit both non-thermal jet emission (PL continuum) and significant thermal contributions from the accretion disk or BLR (QSO template or emission lines only). The PL+line-only configuration serves two purposes. First, it enhances the identification of the so-called masquerading BL Lacs --- FSRQs in which strong jet dilution suppresses the thermal disk continuum below detection, leaving only emission lines -- both broad features from the BLR and narrow forbidden lines from the NLR as evidence of the FSRQ  nature \citep{padovani2019txs}. Second, it captures sources where the accretion-disk continuum is intrinsically weak or absent, but emission remains detectable above the power-law continuum. All amplitude parameters ($s_{\rm gal}$, $s_{\rm QSO}$, $A_{\rm PL}$) are constrained to the range $[0, 100 F_{\rm med}]$ to ensure physical solutions while providing sufficient dynamic range, where $F_\mathrm{med}$ is the median observed flux of the fitted pixels, providing a data-driven scaling reference. 

This model hierarchy guards against misclassification: when a single-component galaxy or QSO model fits adequately, the observed optical emission is dominated by thermal radiation with no statistically required jet contribution. When PL-only or multi-component models are required, PL+Galaxy characterizes BL Lacs (jet-diluting host galaxy features), while PL+QSO and PL+line-only capture FSRQs (jet coexisting with disk/BLR emission). This framework captures the jet-to-disk ratios expected in the blazar sequence \citep{fossati1998unifying, ghisellini2011transition}. We note, however, that in the PL+Line-only model, the power-law 
component may absorb a mixture of jet synchrotron and smooth accretion-disk continuum emission, particularly in sources where jet dominance exceeds $\sim$90\%. In such cases the derived jet fraction $f_\mathrm{jet}$ should be treated as an upper limit on the true non-thermal contribution (Section~\ref{sec:decomp}).

\subsection{$\chi^{2}(z)$ Minimization} 
\label{sec:chi^2}
In this section, we describe the procedure to minimise $\chi^{2}$ by fitting all models on a logarithmic redshift grid. 
We first define the fitting pixels for each spectrum before performing the grid search. For each SDSS spectrum, we restrict the fitting region to $3800$--$10400$\, \AA\  corresponding to a reliable wavelength coverage of the BOSS spectrograph \citep{bolton2012spectral}. We apply standard SDSS-V spectroscopic quality masking to exclude bad pixels flagged by the pipeline, including BADSKYCHI (sky residuals significantly worse than Poisson noise), BRIGHTSKY (sky brightness dominates object flux), and pixels with negative flux exceeding $10\sigma$ significance.

We define the goodness-of-fit statistic as: 

\begin{equation}
\chi^2 = \sum_i \frac{[F_{\rm data}(\lambda_i) - F_{\rm model}(\lambda_i)]^2}{\sigma_i^2}
\label{eq:chi^2}
\end{equation}
where $[F_{\rm data}(\lambda_i)$ is the observed flux, $\sigma_i$ the uncertainty, and $F_{\rm model}(\lambda_i)$ the model flux at pixel $i$. For each point on the redshift grid, we seek the model parameters that minimize $\chi^2$. The nature of this optimisation depends on the model complexity. 

For single-component models (galaxy-only, QSO-only), the redshift is fixed at each grid point $z_{\rm grid}$, and the sole free parameter is the template normalisation $s$, of the template, which is solved analytically using weighted least squares:
\begin{equation}
s = \frac{\sum_i y_i \cdot T_i(\lambda_i, z_{\rm grid}) \cdot T_{{\rm IGM},i} / \sigma_i^2}{\sum_i [T_i(\lambda_i, z_{\rm grid}) \cdot T_{{\rm IGM},i}]^2 / \sigma_i^2}
\end{equation}
where $T_i$ is the template flux at pixel $i$ shifted to the observed frame at $z_{\rm grid}$, and $y_i$, $\sigma$ are as defined in Equation ~\ref{eq:chi^2}. Redshift is subsequently treated as a free parameter during the local refinement described in Section ~\ref{sec:refinement}

For multi-component models (PL+Galaxy, PL+QSO, PL+Line-only), we use \textsc{lmfit}  \citep{newville2016lmfit}, a nonlinear least-squares minimisation curve fitting tool, to jointly optimize the free parameters at each redshift grid point. 
For the PL+Galaxy model, the free parameters are: template scaling $s_{\rm gal}$, power-law normalisation $A$, spectral slope $\alpha$, and curvature $\delta$ (4 parameters). The PL+QSO and PL+Line-only models have identical parameter sets with $s_{\rm QSO}$ and $s_{\rm line}$ respectively. For the PL-only model, the free parameters are $A$, $\alpha$ and $\delta$ (3 parameters).
We initialize $\alpha = -1.5$ (typical synchrotron value; \citealt{tramacere2011stochastic}) and $\delta = 0$ (no curvature), with constraints $-2.5 < \alpha < 1.2$ and $-1.2 < \delta < 1.2$. The lower bound accommodates steep synchrotron spectra and cooling breaks, while the upper bound allows for inverted spectra characteristics of self-absorbed synchrotron emission \citep{urry1995unified}.  The PL normalization $A$ and template scalings are initialized to fractions of the median observed flux and allowed to vary freely. 

Performing $\chi^{2}$ minimization at each redshift grid point yields $\chi^{2}(z)$ curves for each model family. A sharp minimum indicates strong spectral features (e.g., 4000 \AA\ break, emission lines) that uniquely constrain redshift, while a flat $\chi^{2}(z)$ curve suggests a featureless jet-dominated spectrum where redshift information is weak or absent.

\subsection{Model Marginalisation and ${\rm z_{MAP}}$ Selection}
\label{sec:marginalization}
To obtain a robust redshift estimate that marginalises over template uncertainty within each model family, we construct a family-level likelihood by combining the $\chi^{2}(z)$ curves of all member templates.  We first determine a global normalisation constant $C$ as the minimum $\chi^{2}$ value across all templates in all six families. This stabilises the exponential likelihood and ensures all families remain on a numerically consistent scale for direct comparison. Each template's $\chi^{2}(z)$ curve is then converted to a likelihood across all redshifts points: 

\begin{equation}
\mathcal{L}_i(z) = \exp\!\left(-\frac{\chi^2_i(z) - C}{2}\right)
\end{equation}

Within each family, the  family-level likelihood is computed as a weighted sum over templates: 

\begin{equation}
    \mathcal{L}_{\rm family}(z) = \sum^{N_{\rm templates}}_{i = 1} w_{i} \mathcal{L}_{i}(z)
\end{equation}
with uniform weights $w_i = 1/N_{\rm templates}$. The template counts are: $N_{\rm gal} = 3$ (SWIRE ellipticals), $N_{\rm QSO} = 8$ (synthetic QSO spectra), $N_{\rm PL} = 1$, $N_{\rm PL+ gal} = 3$, $N_{\rm PL+QSO} = 8$, $N_{\rm PL+line} = 3$. Each family likelihood is normalised to proper probability density:

\begin{equation}
p_{\rm family}(z) = \frac{\mathcal{L}_{\rm family}(z)}{\int \mathcal{L}_{\rm family}(z') ~ dz'}
\end{equation}
We combine all six families using equal priors $W_f = 1/6$, reflecting the absence of prior information favouring any particular physical model-- the relative weight of each family in the global posterior is determined entirely by how well it fits the observed spectrum:

\begin{equation}
p_\mathrm{total}(z) = \frac{\sum_f W_f~\mathcal{L}_f(z)}
{\int \sum_f W_f~\mathcal{L}_f(z')~\mathrm{d}z'}
\end{equation}
Note that we combine the raw family likelihoods $\mathcal{L}_\text{family}(z)$ rather than the normalized family density $p_{\rm family}(z)$, so that families with poor overall fits contribute less to the global posterior in proportion to their absolute likelihood. The maximum a posteriori redshift $z_{\rm MAP}$ is extracted from this global posterior. At $z_{\rm MAP}$, the fractional responsibility of each template $i$ within its family is computed as:
\begin{equation}
    r_i(z_{\rm MAP}) = \frac{w_i ~ \mathcal{L}_i(z_{\rm MAP})}{\sum_j w_j ~ \mathcal{L}_j(z_{\rm MAP})}
\end{equation}
where $w_{i} = 1/N_{\rm templates}$ are the intra-family template weights, distinct from the inter-family priors $W_f$. The template with the highest responsibility is selected as the representative for local refinement. This marginalisation approach naturally downweights poor-fitting templates and is particularly important for blazars, where the correct physical model is often ambiguous due to featureless or weakly featured continua.

\subsection{Local Refinement and Model Selection via AICc} 
\label{sec:refinement}

Using $z_{\rm MAP}$ as the initial guess, we perform a refined local optimisation with  \textsc{lmfit} for the representative template of each family--- the template with the highest responsibility $r_i(z_{\rm MAP})$ identified in Section ~\ref{sec:marginalization}. In this refinement step, redshift is treated as a free parameter constrained within $\pm 0.15$ of $z_{\rm MAP}$. All other parameters (template scalings, power-law normalisation, spectral index $\alpha$, and curvature $\delta$) remain free within their original bounds. This local optimization refines the redshift estimate while avoiding spurious local minima far from the globally preferred solution.

To select the overall best-performing model, we compare the six refined fits using the corrected Akaike Information Criterion \citep[AIC;][]{sugiura1978further, hurvich1989regression} :
\begin{equation}
    {\rm AICc} = \chi^2 + 2k + \frac{2k(k+1)}{N_{\rm data} - k - 1}
    \label{eq:aicc} 
\end{equation}
where $k$ is the number of free parameters and $N_{\rm data}$ is the number of fitted pixels. The parameter counts are: $k = 2$ for single-component models (Galaxy, QSO: template scale and redshift); $k = 4$ for power-law only (normalisation $A$, spectral index $\alpha$, curvature $\delta$, and redshift); and $k = 5$ for combined models (PL+Galaxy, PL+QSO, PL+Line: template scale, $A$, $\alpha$, $\delta$, and redshift), with the pivot wavelength $\lambda_0 = 5000$~ \AA\ held fixed throughout as before. 

The AICc penalises model complexity, balancing goodness of fit against overfitting. The model with the lowest AICc is selected as the best-fit classification. The redshift of the best-fitting model after local refinement, $z_{\rm best}$, is taken as the value of $z$ from the \textsc{lmfit} parameter set of the AICc-selected model. When a combined model is selected, the best-fit label directly informs the physical classification: PL+Galaxy indicates a BL Lac candidate with a detectable host galaxy contribution, while PL+QSO or PL+Line-only indicates an FSRQ candidate with thermal disk or BLR emission coexisting with the jet. When a single-component galaxy model is selected, the observed optical emission is dominated by starlight from the host galaxy, with no statistically required jet contribution. When a single-component QSO model is selected, the spectrum is dominated by accretion disc 
continuum and broad emission lines, again with no evidence for a significant power-law jet component above the thermal and line emission. Conversely, when a pure power-law model provides the best fit, the featureless 
continuum prevents a robust redshift determination, and the redshift is poorly constrained or fixed at an initial estimate. 

\subsection{Component Flux Decomposition}
\label{sec:decomp}
If the best-fit model is one of the three combined families (PL+Galaxy, PL+QSO, PL+Line-only), we quantify the fractional contribution of the jet to the total observed optical flux. Here $F_{\rm jet}(\lambda)$ refers to the power-law component of the best-fit model, representing the non-thermal synchrotron emission from the relativistic jet, and $F_{\rm total}(\lambda)$ is the sum of all model components. The jet fraction is computed by integrating both components over the fitted wavelength range

\begin{equation}
f_{\rm jet} = \frac{\int_{3600}^{10400} F_{\rm jet}
(\lambda)~{\rm d}\lambda}
{\int_{3600}^{10400} F_{\rm total}(\lambda)~{\rm d}\lambda}
\label{eq:fjet}
\end{equation}
The jet fraction $f_\mathrm{jet}$ quantifies the fractional contribution of non-thermal synchrotron emission to the total observed optical flux. Unlike the discrete equivalent width  criterion used to separate BL Lacs from FSRQs 
\citep{stickel1991complete}, $f_{\rm jet}$ provides a continuous measure of jet dominance that captures the full range of jet-to-thermal flux ratios observed across the blazar sequence.   

\subsection{Equivalent Width Measurements}
\label{sec:ew_methods}
We measure the rest-frame equivalent widths (EWs) for absorption and emission spectral features listed in Table~\ref{tab:spectral_lines_ew}, adopted from the \citet{kasai2023optical}, which are commonly detected in blazar optical spectra. These measurements serve to validate our model-based classification against the traditional EW boundary of $5$~\AA, which separates BL Lacs (EW $< 5$~ \AA)  from FSRQs (EW$\geq 5$~ \AA) \citep{stickel1991complete, marcha1996optical}. Since the optimal integration window varies between sources depending on spectral line width and local continuum shape, we adopt an adaptive window selection approach described below. 
For each spectral feature, we define continuum sidebands spanning $50$ -- $100$~ \AA\ from the line centre for absorption features and  $50$ - $250$~ \AA\  for emission features, chosen to be sufficiently distant from the line profile while remaining within the local spectral region. All line centres are computed from the rest-frame wavelengths shifted to the best-fit redshift $z_{\rm best}$ from Section~\ref{sec:refinement}.

We fit a linear continuum to these sideband regions via weighted least squares ($\chi^{2}$ minimisation) with inverse variance weighting, then normalise the entire spectrum by this local continuum model. Dividing by the local linear continuum removes any residual power-law slope within the sideband region and provides a natural reference level of unity for the EW integration \citep{burstein1984old, shaw2012spectroscopy}, which is particularly important for jet-dominated sources where the underlying continuum is steep and variable.

\begin{figure*}
    \centering
    \includegraphics[width=\textwidth]{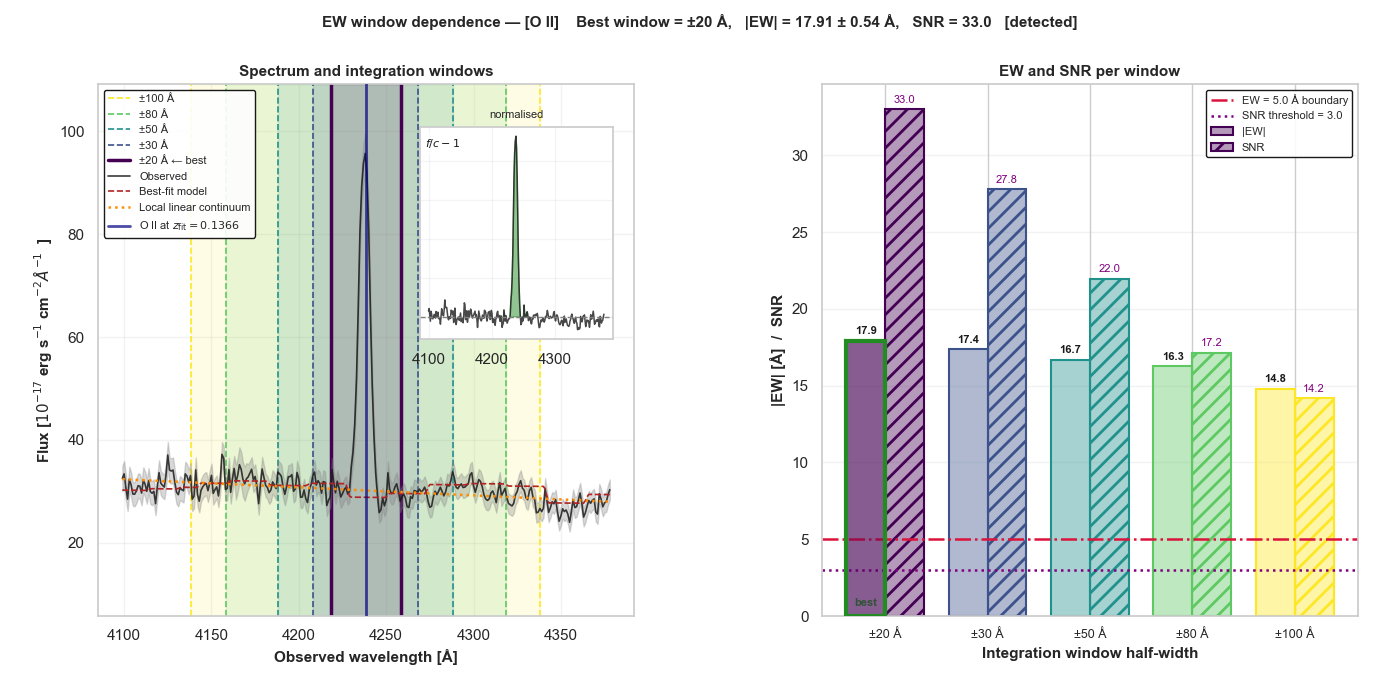}
    \caption{Illustration of the adaptive integration window selection for the [O~II] $\lambda$3729 emission line in a BL Lac candidate SDSS ID 79336239 ($z_{\rm fit} = 0.1366$). \textit{Left panel:} Observed spectrum (black) with $\pm 1\sigma$ uncertainty (grey shading) and local linear continuum fit (red dashed). The five integration windows tested are shown as  overlapping shaded regions with corresponding boundary lines, colour-coded from narrow (purple, $\pm 20$~\AA) to wide (yellow, 
    $\pm 100$~\AA). The blue vertical solid line marks the observed line centre at $\lambda_{\rm obs} = 4238.3$~\AA. The inset shows the continuum-normalised integrand $f/c - 1$, with the shaded green area corresponding to the EW integrand and directly representing the quantity integrated in Equation~\ref{eq:ew}. \textit{Right panel:} $|{\rm EW}|$ (solid bars) and S/N (hatched bars) for each integration window. The best window ($\pm 20$~\AA, highlighted with a green border)  is selected as the half-width as it yields the highest S/N. The measured $|{\rm EW}| = 17.91 \pm 0.54$~\AA\ lies well above  the $5$~\AA\ classification boundary (red dash-dot line), confirming this source as an FSRQ candidate despite its SDSS classification. The S/N threshold of 3 is shown as the purple  dotted line.}
    \label{fig:ew_demo}
\end{figure*}

We measure EWs on the normalised spectrum by integrating over wavelength windows centred on the observed line position. To optimise the S/N ratio of each measurement, we trial multiple integration windows: for absorption lines, we test integration windows of $5$, $10$, $15$, $20$, and $30$~ \AA\  and for emission lines, we test $20$, $30$, $50$, $80$, and $100$ (see Figure \ref{fig:ew_demo} for a demonstration)~ \AA. We adopt the standard signal-to-noise estimator ${\rm S/N} =|\rm EW|/\sigma_{\rm EW}$ and select the integration window that maximises this quantity for each line. A line is considered detected if ${\rm S/N} \ge 3$. The EW is computed as 
\begin{equation}
{\rm EW} = \int \left(1 - F_{\rm norm}\right) {\rm d}\lambda
\label{eq:ew}
\end{equation}
where $F_{\rm norm}$ is the continuum-normalised flux. This convention yields positive EW for absorption ($F_{\rm norm} < 1$) and negative EW for emission ($F_{\rm norm} > 1$). 

Uncertainties are propagated from the spectral flux errors as:
\begin{equation}
\sigma_{\rm EW} = \sqrt{[\Sigma (\sigma_{i} /C_{i})^{2}]} \Delta \lambda  
\end{equation}
where $\sigma_{i}$ are pixel uncertainties, $C_i$ are continuum values used for the normalisation, $\Delta \lambda$ is wavelength sampling, and the summation is carried out over each pixel (index $i$) within the windows described above.

\begin{table}
\centering
\begin{threeparttable}
\caption{Absorption and emission features measured for equivalent width analysis.}
\label{tab:spectral_lines_ew}
\begin{tabular}{lcc}
\hline
Feature name & Wavelength (\AA) & Type \\
(1) & (2) & (3) \\ 
\hline
Ly$\alpha$        & 1215   & Absorption/Emission\\
C~IV            & 1549   & Emission \\
C~III]          & 1909   & Emission \\
Fe II           & 2600   & Intervening\\
Mg~II           & 2796   & Intervening/Emission \\ 
                & 2803   & Intervening/Emission\\
{[}O~II{]}      & 3727   & Emission \\ 
                & 3729   & Emission \\
Ca~II~K         & 3933.7 & Absorption \\
Ca~II~H         & 3968.5 & Absorption \\
Ca~I~G          & 4304.4 & Absorption \\
H$\beta$        & 4861.3 & Absorption/Emission \\
{[}O~III{]}     & 4959   & Emission \\
                & 5007   & Emission \\
Mg~b            & 5184   & Absorption \\
Ca~Fe           & 5269   & Absorption \\  
Na~I~D          & 5892.5 & Absorption \\
{[} N~II{]}     & 6548.1 & Emission\\
H$\alpha$       & 6562.8 & Absorption/Emission \\
{[}N~II{]}      & 6583.6 & Emission \\
\hline
\end{tabular}
\begin{tablenotes}
    \small
    \item \textit{Notes.} The columns are: (1) Name of spectral line, (2) Rest-frame wavelength in \AA, (3) Spectral feature type. Emission lines trace either the BLR (e.g., Mg II, H$\alpha$, C IV) or the NLR (e.g., [O II], [O III], [N II). Absorption lines trace the host galaxy stellar population. Lines marked "Absorption/Emission" can appear in either form depending on the dominant spectral component. For closely-spaced doublets and multiplets (e.g., Mg II), we only list the strongest component in the automated EW estimation pipeline. 
\end{tablenotes}
\end{threeparttable}
\end{table}

\section{Results} 
\label{sec:results}
In this section, we present the results of our multi-component fitting approach described in Section~\ref{sec: 3}, applied to the sample of \Fermi\ counterparts with SDSS-V spectroscopy described in Section~\ref{sec:samp_defini}. In Section~\ref{sec:fit_ex}, we present examples demonstrating how our approach isolates host galaxy and thermal emission features that also serve as robust anchors for redshift estimation and present representative cases of BL Lac and FSRQ candidates and their SDSS automated pipeline misclassifications. 
In Section~\ref{sec:chi}, we compare the goodness of fit between the SDSS automated pipeline and the multi-component spectral fitting using the reduced chi-squared statistic $\chi^{2}_{\rm r}$. Section~\ref{sec:class_compare} presents the model distribution of the final 707 well-fitted sources and compares the jet-aware classifications against the original SDSS spectroscopic classes. Of the 707 well-fitted sources, 597 (84.4\%) have a median spectral ${\rm S/N} \geq 3$, as reported in the SDSS-V catalogue. The remaining 110 (15.6\%) sources  that fall below this threshold are retained in the sample, however, all population median calculations are restricted to sources with ${\rm S/N} \geq 3$ unless otherwise stated. 
Appendix~\ref{append:tables} provides tables of our multi-component fitting results for all 707 well-fitted sources and tabulates those where we retain the SDSS or 4FGL classifications.
In Section~\ref{jet_z_dis}, we present the redshift and optical jet fraction distributions for BL Lac candidates (PL+Galaxy models) and FSRQ candidates (PL+QSO and PL+Lines models), demonstrating that the multi-component classification naturally recovers the known redshift separation between the two blazar populations \citep{fossati1998unifying, ajello2014cosmic}. 

In Section~\ref{lumino}, we estimate the $\gamma$-ray luminosity of both populations using sources with ${\rm S/N} \geq 3$ and show approximately one order of magnitude separation in $\gamma$-ray luminosity between BL Lac and FSRQ candidates
combined with consistent optical jet fractions across both classes, confirming that the model selection captures a genuine physical distinction rather than an artefact of the fitting procedure.

\subsection{Exemplary Spectral Fitting Results}
\label{sec:fit_ex}
\begin{figure*}
     \centering
     \includegraphics[width=1\linewidth]{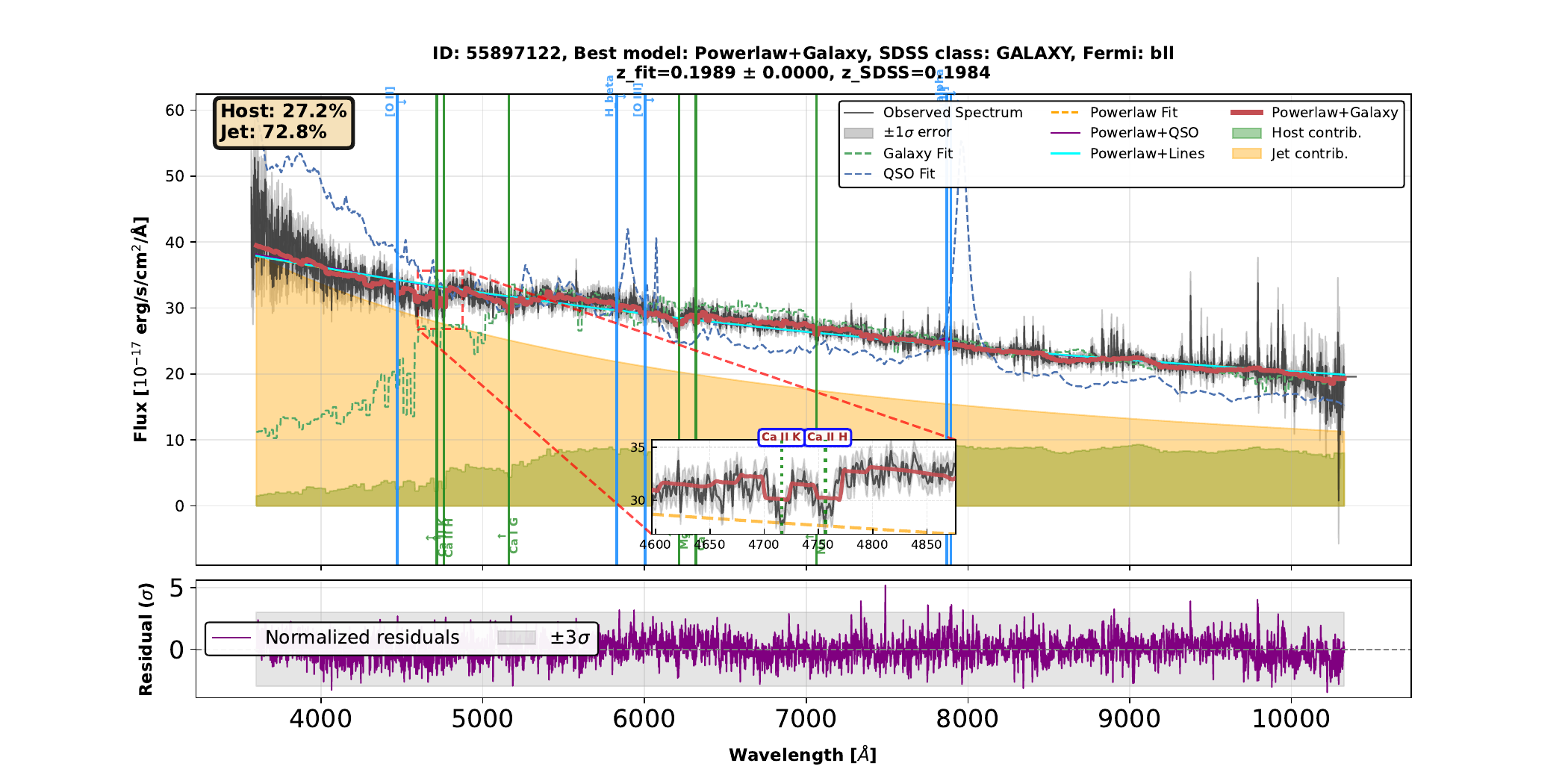}
     \caption{Multi-component spectral decomposition of SDSS J084712.93+113350.2, a \textit{Fermi}-detected BL Lac object (4FGL J0847.2+1134) misclassified as a GALAXY by the SDSS automated pipeline. \textit{Top panel}:  Observed SDSS spectrum (black) with multi-component model fits overlaid: pure galaxy (green dashed), pure QSO (blue dashed) power law (orange dashed), PL + galaxy (thick red), PL + QSO (purple), and PL + lines (cyan). The best-fit model is  PL+Galaxy ($\chi^{\rm 2} _{\rm r} = 0.8532$), with the host galaxy contribution (green shaded region, $27.2\%$) and non-thermal jet continuum (orange shaded region, $72.8\%$) shown explicitly. The inset shows the zoomed Ca~II H\&K doublet ($\lambda_{\rm rest} \approx  3933.7 $ and $ 3967.5 $~ \AA), whose absorption feature contributes to the anchoring of the redshift at  $z_{\rm fit} = 0.1989 \pm 0000$, consistent with the SDSS pipeline value of $z = 0.1984$} 
     \label{fig:ex_fit}
 \end{figure*}

Accurate classification and redshift estimation for blazars requires physically motivated spectral decomposition that quantifies the relative contributions of jet emission and accretion disk/host galaxy components. Figure ~\ref{fig:ex_fit} presents the spectral decomposition for SDSS J084712.93+113350.2 following our fitting procedure described in Section \ref{sec: 3}. 

The SDSS spectroscopic pipeline classifies this object as a GALAXY with $z = 0.1984$, based on the presence of stellar absorption features. Our multi-component analysis reveals a more complex origin for the observed emission: the PL+Galaxy model provides the best fit with $\chi^{2}_{\rm r} = 0.8532$, decomposing the optical spectrum into a non-thermal power-law continuum characteristic of relativistic jet emission (orange shaded region, $72.8\%$ of the observed flux) and a host galaxy contribution (green shaded region, $27.2\%$). This jet-dominated spectrum, combined with the absence of broad emission lines, identifies the source as a BL Lac object.  
 
 The inset panel highlights the Ca~II H\&K absorption feature, which contributes to the redshift determination at $z_{\rm fit} = 0.1989\pm 0.0000$, in close agreement with the SDSS pipeline value at $z_{\rm SDSS} = 0.1984$.  The systematic agreement between the two redshift estimates confirms the reliability of the spectral decomposition of this source.   This example also illustrates the fundamental limitation of the SDSS pipeline: while it correctly identifies the host galaxy contribution features and recovers an accurate redshift, it compares the observed spectrum against pure galaxy and QSO templates without accounting for the nonthermal jet continuum \citep{bolton2012spectral}, leading to a systematic misclassification of the source as a passive galaxy rather than a jet-dominated BL Lac object. The \textit{SIMBAD} astronomical database classifies this object (2MASS J08471293+1133501) as a BL Lac object at $z = 0.19817$, consistent with both our estimate and the SDSS redshift.

\begin{figure*}
     \centering
\includegraphics[width=\textwidth]{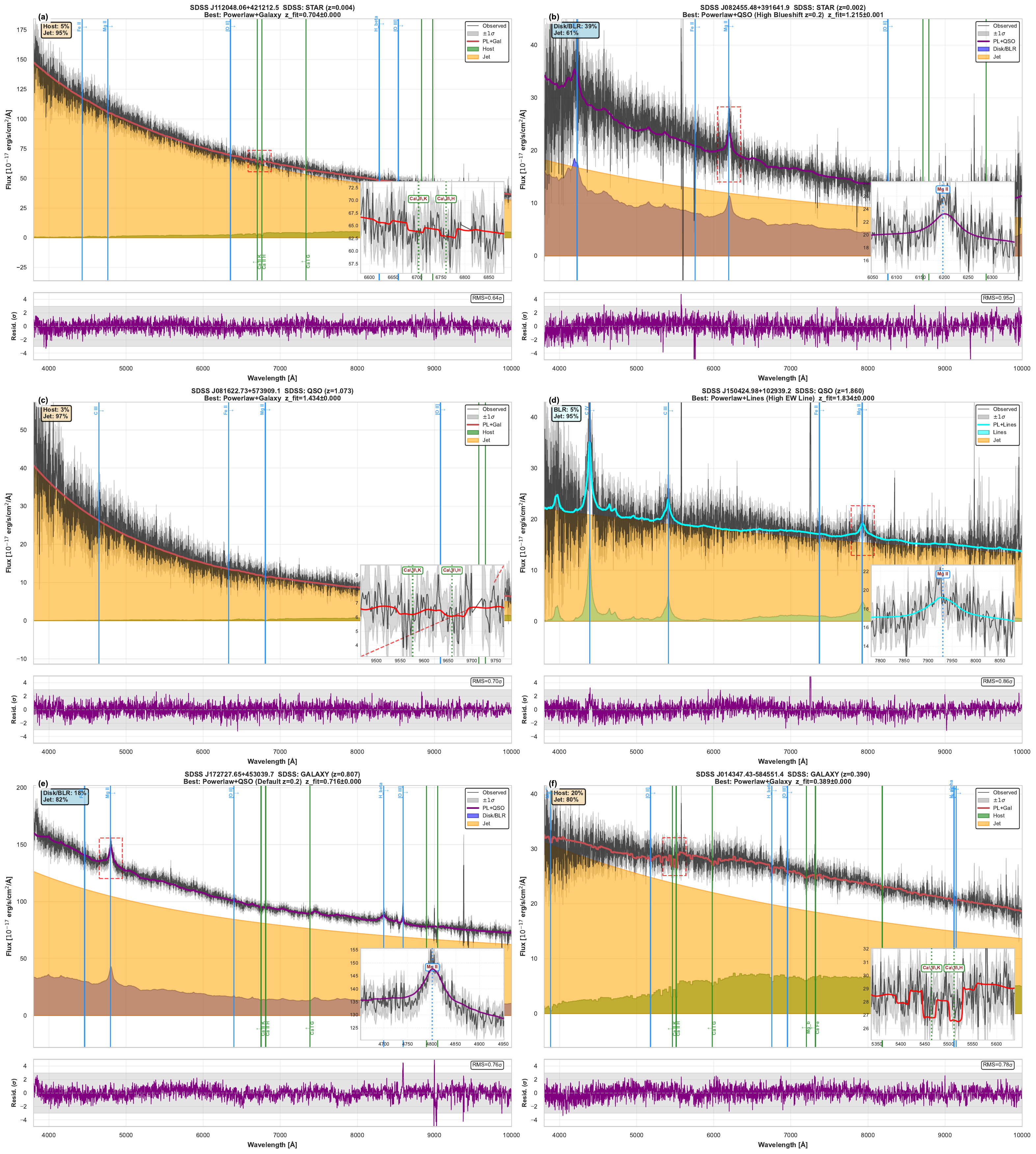}
     \caption{Multi-component spectral decomposition revealing systematic SDSS misclassifications of \textit{Fermi}-detected blazars. Each panel shows the observed SDSS spectrum (black) with our best-fit multi-component model (coloured line) decomposed into jet continuum (orange shaded region) and thermal emission components (green/blue/cyan shaded regions for galaxy/QSO/line templates, respectively). Panel labels indicate: SDSS spectroscopic classification and redshift (top), our revised classification and redshift (NEW CLASS), and the fractional flux contributions from jet and thermal components, represented by the shaded region. \textbf{(a)} BL Lac misclassified as STAR due to extreme jet dominance ($96.5\%$), producing a featureless stellar-like continuum. \textbf{(b)} FSRQ with blueshifted emission misclassified as STAR. \textbf{(c)} BL Lac misclassified as a QSO \textbf{(d)} FSRQ with strong emission lines misclassified as a generic QSO, thereby missing its blazar nature. \textbf{(e)} FSRQ misclassified as GALAXY from diluted emission lines. \textbf{(f)} BL Lac misclassified as GALAXY. Despite a correct redshift, the jet component is unrecognized by the SDSS automated pipeline, thereby assigning a false classification.}
     \label{fig:mult_examp}
 \end{figure*}

 In Appendix~\ref{app:examplespec} we provide examples of optical spectra that are best fit by each of the three main model families. Figure \ref{fig:mult_examp} showcases a diverse collection of further cases of misclassifications with the SDSS automated pipeline. These six representative examples illustrate the systematic nature of template-matching failures while demonstrating that our multi-component fitting method consistently recovers accurate classifications and redshifts.  We discuss each of these cases in turn below.

\setcounter{secnumdepth}{3} 
\subsubsection{BL Lac Object Misclassified as a STAR}
\label{sec:bl_star}
Panel (a) presents  SDSS J112048.06+421212.5, a clear example of a catastrophic pipeline failure where both classification and redshift are severely incorrect. The SDSS automated pipeline misclassified this object as a Galactic star with $z_{\rm SDSS}=0.004$, whereas our multi-component fitting recovers a redshift of $z_{\rm fit}=0.704\pm0.000$. The best-fit model is PL+Galaxy with an extreme flux decomposition of $96.5\%$ jet contribution and only $3.5\%$ host galaxy. Despite the absence of detectable emission or absorption feature above the the S/N threshold, the spectral shape provides 
a weak constraint on the redshift through the relative contribution of the host galaxy template to the overall continuum shape.
The complete absence of broad emission lines, combined with the featureless power-law continuum, places this source firmly in the BL Lac class, consistent with the Fermi association as a BL Lac. \citet{plotkin2008large} independently classify this source as a BL Lac object. The \textit{SIMBAD} astronomical database identifies this object as a BL Lac type blazar and records a spectroscopic redshift of $0.56823\pm0.00017$ from the SDSS-IV DR16 catalogue \citep{ahumada202016th}. While differing from our estimate of $z_{\rm fit} = 0.704$, it is consistent with the extragalactic nature of the source. 

\subsubsection{FSRQ Misclassified as STAR}
\label{sec:fsrq_star}  
Panel (b) presents  SDSS J082455.48+391641.9, a case where the SDSS automated pipeline misclassifies an FSRQ as a Galactic star at $z_{\rm SDSS}=0.002$, whereas our pipeline returns a redshift of $z_{\rm fit} = 1.215\pm0.001$. The best-fitting model (PL+QSO) reveals $61\%$ jet dominance with $39\%$ contribution from the accretion disk/BLR. We detect Mg II emission ($\mathrm{EW} = -12.34 \pm 0.87$~\AA, S/N $= 14.1$) and the C III] emission complex ($\mathrm{EW} = -16.43 \pm 1.56$~\AA, S/N $= 10.5$). The presence of these broad emission lines with $|{\rm EW}| > 5$~\AA\ firmly places this source in the FSRQ class, consistent with the \textit{Fermi} association. The SDSS pipeline, lacking jet-aware templates, interprets the broad-line emission as stellar absorption features at a near-zero redshift -- a direct consequence of the absence of blazar-specific templates in the automated classification scheme \citep{bolton2012spectral}. \citet{massaro2009roma} independently identify this object in the Roma-BZCAT catalogue as a confirmed blazar based on multiwavelength properties.    

\subsubsection{BL Lac Object Misclassified as QSO}
\label{sec:bllac_qso}
Panel (c) presents SDSS J081622.73+573909.1, illustrating a BL Lac incorrectly classified as a QSO by the SDSS automated pipeline at $z = 1.073$. Our multi-component fitting recovers a redshift of $z_{\rm fit} = 1.434 \pm 0.000$ with a PL+Galaxy model dominated by jet emission ($97\%$ jet, $3\%$ host). No emission or absorption features are detected above the signal-to-noise threshold (${\rm S/N} < 3$) in this spectrum, placing this source in the BL Lac class by the equivalent width criterion of \citet{stickel1991complete}. The complete absence of broad emission lines is consistent with the \textit{Fermi} association. The redshift estimate is driven by the host galaxy template contribution rather than spectral line features and therefore carries larger systematic uncertainty than for sources with detected lines 
\citep{shaw2013spectroscopy}. 
This source has been identified as a BL Lac by a number of prior studies \citep[e.g.][]{Meyer_2011ApJ...740...98M,Massaro_2015Ap&SS.357...75M,Pena-Herazo_2021AJ....161..196P}, consistent with our classification, but without a redshift estimate.\footnote{\citet{Flesch_2015PASA...32...10F} and \citet{ajello20173fhl} report a redshift of $z=0.054$ that is widely reproduced, attributed to \citet{Beckmann_2003A&A...401..927B} but absent in this reference 
\citep[as noted by][]{Cerruti_2025A&A...698A.101C}. Optical spectra of the same source were also obtained through prior SDSS programmes \citep{Abazajian_2009ApJS..182..543A,Abdurrouf_2022ApJS..259...35A} that are consistent with our BL Lac classification but were also incorrectly assigned QSO classifications (at different redshifts) by the SDSS pipeline.}
\subsubsection{FSRQ Misclassified as QSO} 
\label{sec: fsrq_qso}
Panel (d) presents the object SDSS J150424.98+102939.2 and a case where the SDSS automated pipeline classifies the source as a QSO at $z_{\rm SDSS} = 1.8602$, while our fits recover a consistent redshift of $z_{\rm fit} = 1.834\pm0.000$. The PL+lines model provides the lowest AICc among all models, with a flux decomposition showing $95\%$ jet contribution and only $5\%$ from the broad-line region. Despite the extreme jet dominance, this source exhibits strong, high-ionization broad emission lines: C~ IV (${\rm EW} =-53.18\pm2.03$~ \AA, S/N $=26.2$), C~ III (${\rm EW} =-11.27\pm 1.33$ \AA, S/N $= 8.5$), and Mg~ II (${\rm EW} = -14.27 \pm 1.19$~\AA, S/N $= 12.0$), all exceeding the $|{\rm EW}| > 5$~\AA\ FSRQ threshold \citep{stickel1991complete, marcha1996optical}. However, the jet fraction of $95\%$ should be interpreted as an upper limit on the true non-thermal contribution, as the power-law component may absorb a mixture of jet synchrotron and smooth 
accretion-disk continuum emission in the absence of a dedicated QSO template. The presence of these well-detected broad emission lines confirms active accretion onto the central supermassive black hole, firmly placing this source in the FSRQ class. This case demonstrates that multi-component modelling is essential even when the pipeline redshift is approximately correct, as the SDSS classification misses the blazar nature of the source. This source is independently identified as a blazar in the Roma-BZCAT catalogue \citep{massaro2009roma} and has a well-established spectroscopic redshift of $z=1.839$ \citep{Akiyama_2003ApJS..148..275A} including a measurement of $z = 1.83795 \pm 0.00020$ from prior SDSS observations \citep{Alam_2015ApJS..219...12A}, consistent with our $z_{\rm fit}$.

\subsubsection{FSRQ Misclassified as GALAXY}
\label{sec: fsrq_gal}
Panel (e) presents SDSS J172727.65+453039.7, a case where the SDSS automated pipeline misclassifies a jet-dominated FSRQ candidate as a GALAXY at $z = 0.807$. Our PL+QSO model recovers a redshift of $z_{\rm fit} = 0.716 \pm 
0.000$, anchored by the detection of multiple broad emission lines: Mg~ II (${\rm EW} = -5.69 \pm 0.28$~\AA, S/N $= 20.1$), H$\beta$ (${\rm EW} = -4.75 \pm 0.34$~\AA, S/N $= 14.0$), and [O~ III] (${\rm EW} = -2.75 \pm 0.20$~\AA, S/N $= 13.9$). The flux decomposition reveals $82\%$ jet contribution and  $18\%$ from the broad-line region, confirming the blazar nature of this source. The Mg~ II detection with $|{\rm EW}| > 5$~\AA\ places this source above the FSRQ threshold \citep{stickel1991complete, marcha1996optical}, while the redshift discrepancy of $\Delta z = 0.091$ between our estimate and the SDSS pipeline value indicates the inability of the pipeline to account for the broad emission lines diluted by the dominant jet continuum. 

\subsubsection{BL Lac Object Misclassified as GALAXY}
\label{bllac_gal}
Panel (f) presents SDSS J014347.43-584551.4, which exemplifies the challenge of distinguishing between passive galaxies and jet-dominated blazars when host galaxy absorption features are present. The SDSS automated pipeline classified this object as a GALAXY at $z_{\rm SDSS} = 0.390$, which our multi-component fitting confirms with excellent agreement ($z_{\rm fit}=0.389\pm0.000$). However, the spectral decomposition reveals that this is not purely a galaxy: the power-law+galaxy model shows that jet emission contributes $84\%$ of the optical flux, with only $16\%$ arising from the host stellar population, producing a very blue spectrum despite showing many of the features of a typical elliptical galaxy. We detect two 
absorption features: Mg~b (${\rm EW} = 1.54 \pm 0.34$~\AA, S/N $= 4.5$) and Ca~Fe (${\rm EW} = 1.10 \pm 0.31$~\AA, S/N $= 3.6$). Both features fall well below the $|{\rm EW}| = 5$~\AA\ BL Lac threshold \citep{stickel1991complete}, and the complete absence of broad emission lines firmly places this source in the BL Lac class.  Previous high signal-to-noise spectroscopy (${\rm S/N} = 113$) of this object by \cite{d2024optical} using SALT/RSS confirms the BL Lac classification, detecting Ca~ II H\&K absorption at a redshift of $z = 0.3902\pm0.0001$. This source is also identified as a blazar within the ROMA-BZCAT.

\subsection{Comparison with SDSS pipeline template fitting results}
\label{sec:chi}
\begin{figure*}
    \centering
    \includegraphics[width=0.99\linewidth]{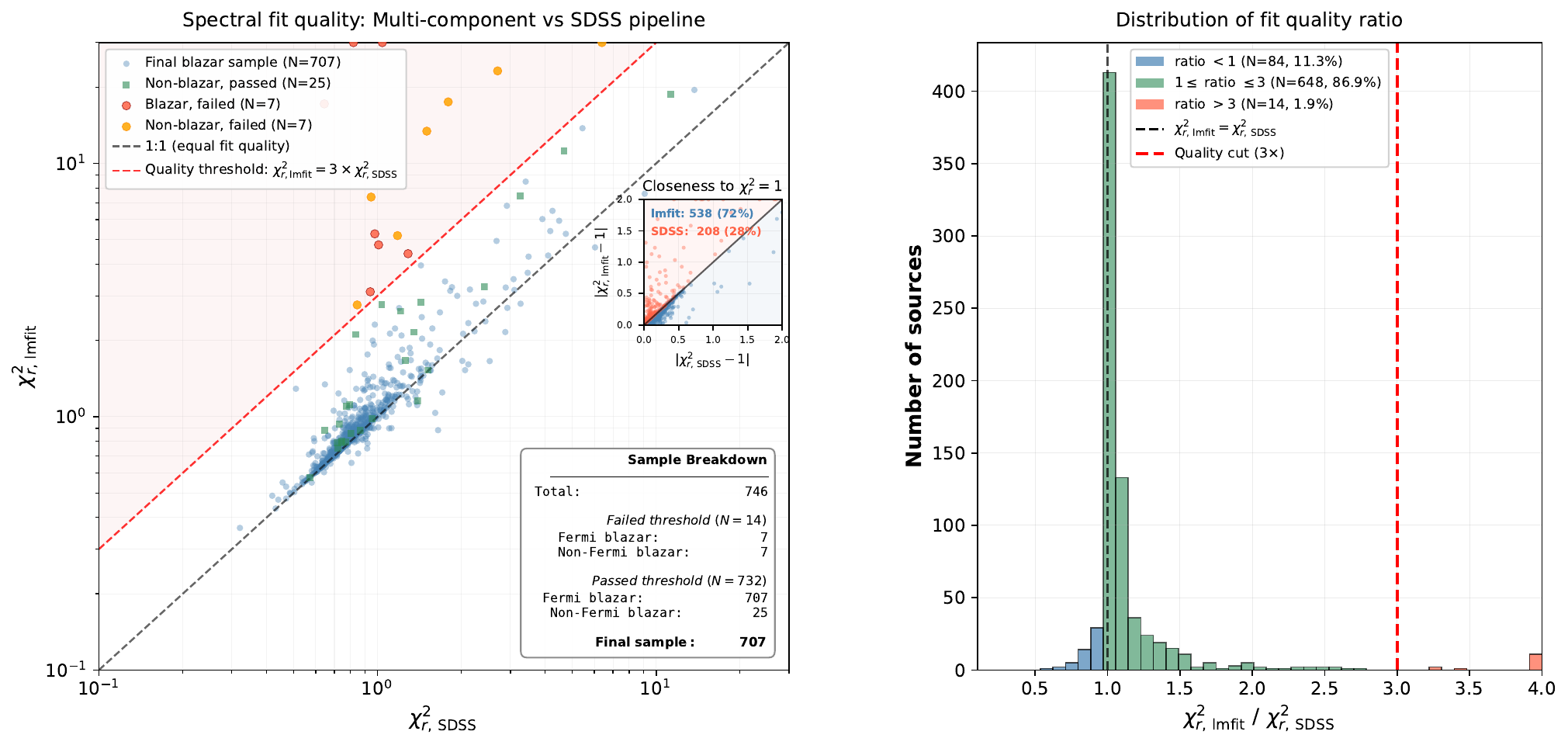}
    \caption{\textit{\textbf{Left:}} Comparison of reduced chi-squared ($\chi^{2}_{\rm r}$) values between the SDSS automated pipeline and  the multi-component spectral models developed in this work. Points are colour-coded by sample category: final blazar candidates (blue circles, $N = 707$), non-\textit{Fermi} blazar sources that passed the quality threshold (green squares, $N = 25$), \textit{Fermi} blazar sources  that failed the threshold (red circles, $N = 7$), and non-\textit{Fermi} blazar sources that failed (orange squares, $N = 7$). The red dashed line marks the 
    quality threshold $\chi^{2}_{\rm r,\,lmfit} = 3\times 
    \chi^{2}_{\rm r,\,SDSS}$, above which sources are excluded. 
    The inset shows $|\chi^2_{\rm r} - 1|$ for each pipeline, with our multi-component models achieving $\chi^2_{\rm r}$ closer to unity for 538 sources (72\%). \textit{\textbf{Right:}} Distribution of the fit quality ratio $\chi^{2}_{\rm r,\,lmfit}/\chi^{2}_{\rm r,\,SDSS}$, colour-coded by performance: blue (ratio $< 1$, $N = 134$, 11.3\%), green ($1 \leq$ ratio $\leq 3$, $N = 648$, 86.9\%), and red (ratio $> 3$, $N = 14$, 1.9\%).}
    \label{fig:rchi2}
\end{figure*}

We compare the $\chi^{\rm 2}_{\rm r}$ values obtained from our multi-template approach and the SDSS-V spectroscopic pipeline for all 746 sources (see Figure~\ref{fig:rchi2}). The left panel shows 
$\chi^{\rm 2}_{\rm r,\,lmfit}$ against $\chi^{\rm 2}_{\rm r,\,SDSS}$, where proximity to the $1:1$ line indicates that a pipeline achieves a better quality fit. Our multi-component models achieve $\chi^2_{\rm r}$ closer to unity for 538 sources (72\%), demonstrating that the physically motivated decomposition produces well-described fits for the majority of the sample. The right panel shows the distribution of the quality ratio $\chi^{\rm 2}_{\rm r,\,lmfit}/\chi^{\rm 2}_{\rm r,\,SDSS}$, which quantifies the relative fit quality between the two pipelines.

We adopt a quality threshold requiring that the $\chi^{2}_{\rm r}$ of our models does not exceed three times the SDSS pipeline 
$\chi^{\rm 2}_{\rm r}$ value, i.e.\ 
$\chi^2_{\rm r,\,lmfit} \leq 3\times \chi^2_{\rm r,\,SDSS}$, 
shown as the red dashed line in both panels. This threshold ensures that subsequent analysis of spectral components is based on reliable fits that provide a physically meaningful decomposition of the observed spectra, while allowing for the additional degrees of freedom introduced by the multi-component model families relative to the SDSS approach. 

Of the 746 sources, 732 (98.0\%) pass this quality threshold. We adopt our fits for all of these sources, as the physically motivated multi-component model families are more representative of the expected spectral signatures of blazars --- ranging from host-dominated systems with strong absorption features to jet-dominated sources with featureless continua --- compared to the SDSS pipeline's approach that is designed for different classes of sources \citep[Morrison et al. in prep]{bolton2012spectral}. For the remaining 13 sources (1.9\%; see Table~\ref{tab:sdss_better}), our models produce fits exceeding three times the SDSS $\chi^{2}_{\rm r}$ value, indicating that our model library fails to adequately describe these spectra. These failures likely arise from incomplete template coverage, particularly for unusual spectral types, transitional objects, or spectra dominated by features not captured by our model families, such as complex emission line profiles. We retain the SDSS classification for these 14 sources and exclude them from our final sample, retaining 732 sources (98.0\% of the initial sample). Of these, a further 25 lack a \textit{Fermi} blazar classification (Section~\ref{excl}) and are excluded, leaving 707 sources with reliable multi-component fits and confirmed \textit{Fermi} blazar associations. Our model library is designed to capture the dominant blazar population --- BL Lacs, FSRQs, and BCUs --- and does not include templates suitable for the diverse minority classes represented by these 25 sources (millisecond pulsars, radio galaxies, starburst galaxies, and similar; see Section~\ref{excl}). We therefore retain the \textit{Fermi} classification for these sources and exclude them from all subsequent analysis.

\subsection{Best Model Distribution compared to previous SDSS and 4FGL classifications}
\label{sec:class_compare}
\begin{figure*}
        \centering
        \includegraphics[width=0.9\linewidth]{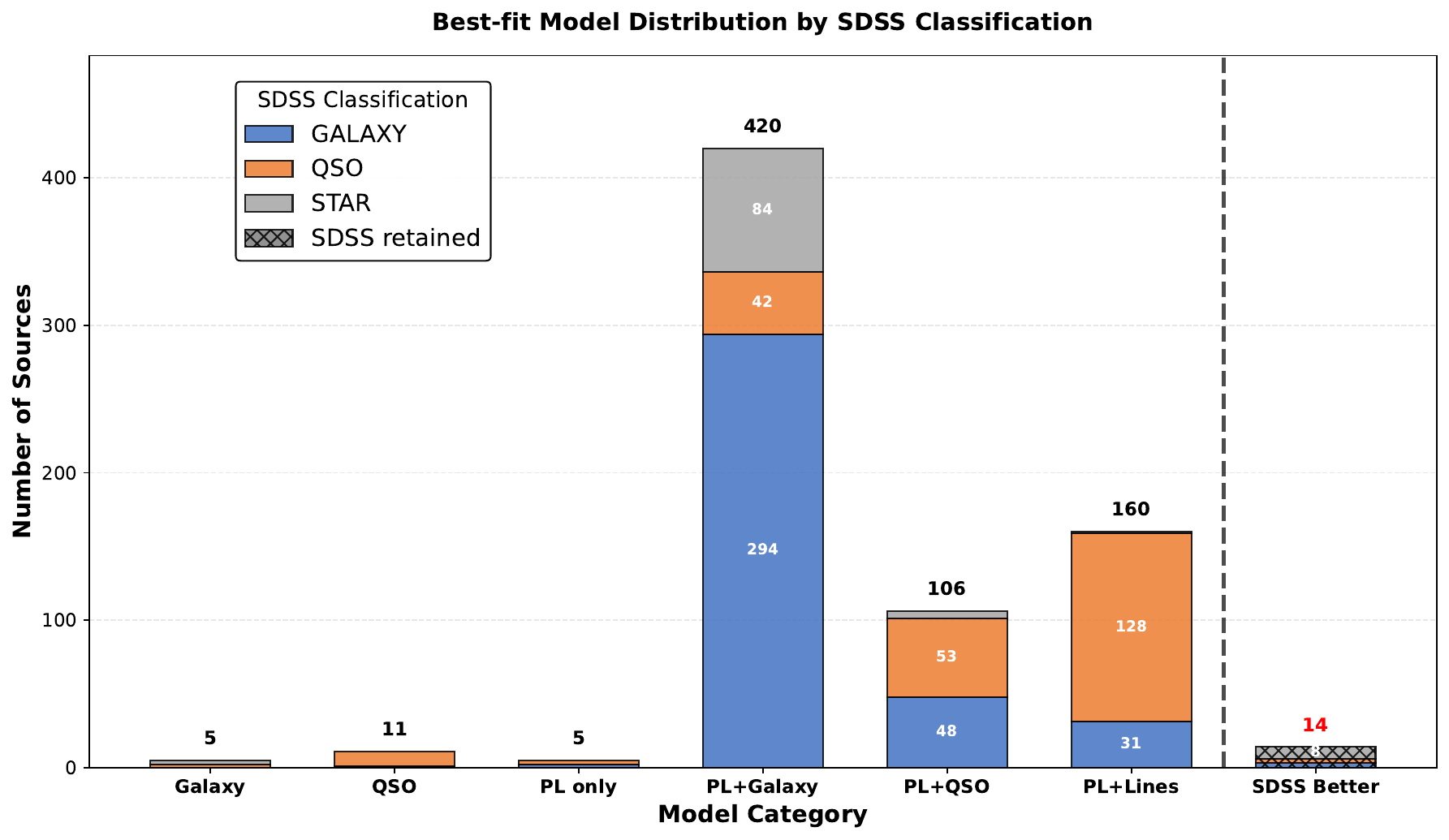}
        \caption{Best-fit spectral model distribution for 746 Fermi-SDSS sources, colour-coded by SDSS pipeline classification (GALAXY: blue, QSO: orange, STAR: grey). The final 707-source sample is dominated by PL+Galaxy (420; 59.4\%), PL+Lines (160; 22.6\%), and PL+QSO (106; 15.0\%) models. Pure single-component models account for 21 sources (3.0\%). 13 sources where the SDSS pipeline outperforms our multi-component approach are shown with cross-hatching (SDSS Better). Of the 420 PL+Galaxy sources, SDSS classifies 294 (70.0\%) as GALAXY and 84 (20.0\%) as STAR. The PL+Lines model shows the highest SDSS QSO classification rate (128; 80.0\%), though the SDSS pipeline cannot distinguish FSRQs from radio-quiet quasars. The PL+QSO model shows nearly equal SDSS classifications as QSO (50.0\%) and GALAXY (45.3\%), highlighting template-matching ambiguities when both broad emission lines and jet continuum are present.}
    \label{fig:best_mod_by_sdd}
    \end{figure*}

\begin{figure*}
        \centering
        \includegraphics[width=0.8\linewidth]{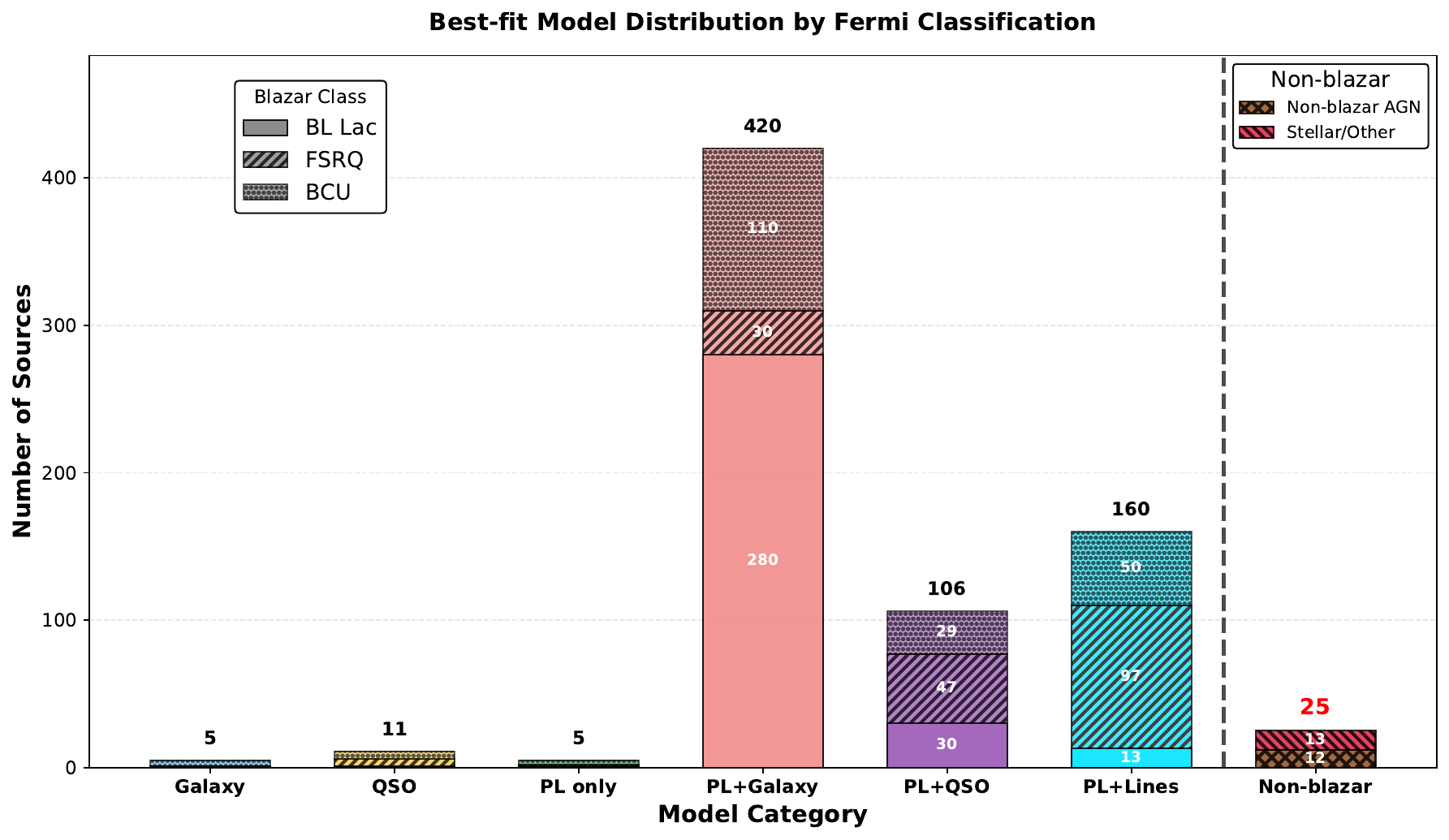}
        \caption{Best-fit spectral model distribution for 732 sources that pass the quality threshold, stratified by \textit{Fermi} blazar classification. Of the 707 confirmed \textit{Fermi} blazars: BL Lacs (pink; $N = 326$, 46.1\%), FSRQs (diagonal hatching; $N = 180$, 25.5\%), and BCUs (dots; $N = 201$, 28.4\%). The 25 non-blazar sources are shown separately to the right, split into non-blazar AGN (brown; $N = 12$) and Stellar/Other (crimson; $N = 13$). BL Lacs overwhelmingly select PL+Galaxy (280; 85.9\%), consistent with weak or absent emission lines ($|{\rm EW}| < 5$\,\AA) where host galaxy absorption anchors redshift determination. FSRQs favour PL+Lines (97; 53.9\%) and PL+QSO (47; 26.1\%), reflecting strong BLR emission ($|{\rm EW}| > 5$\,\AA) superimposed on jet continuum. BCU sources show an intermediate distribution with PL+Galaxy preferred (110; 54.7\%), suggesting most are BL Lac candidates, while 39.3\% select QSO- or line-dominated models indicating FSRQ character.}
    \label{fig:mod_by_fermi}
\end{figure*}

The distribution of best-fit models across our final sample of 707 sources reveals systematic patterns in SDSS pipeline failures and validates our multi-component approach for blazar identification. Figure~\ref{fig:best_mod_by_sdd} shows the distribution of our best-fit models stratified by SDSS pipeline classifications, while Figure~\ref{fig:mod_by_fermi} presents the same distribution across \textit{Fermi} blazar classifications assigned in the 4FGL-DR4 catalogue. These distributions demonstrate that the SDSS automated pipeline systematically misclassifies jet-dominated blazars across all three spectroscopic categories (GALAXY, QSO, and STAR) and that different blazar subclasses are assigned different spectral models in accordance with their underlying physical properties. 

\setcounter{secnumdepth}{3} 
\subsubsection{Distribution of SDSS spectroscopic Classification Compared to Best Models}
\label{sec:sdss_model_dist}
Our multi-component fitting identifies PL+Galaxy as the dominant model across the sample, selected as the best fit for 420 sources ($59.4\%$ of the final sample). This overwhelming preference reflects the prevalence of BL Lac objects in gamma-ray selected samples \citep{ajello2020fourth}, where host galaxy stellar absorption features remain detectable despite strong jet dilution. The SDSS classifications for these 420 sources were GALAXY ($294; 70.0\%$), STAR ($84; 20.0\%$), and QSO ($42; 10.0\%$), demonstrating that the pipeline correctly identified the presence of absorption features in the majority of cases but failed to recognize the underlying non-thermal continuum. The $70.0\%$ GALAXY classification rate for PL+galaxy sources initially appears adequate - these objects do exhibit strong absorption lines characteristic of galaxies. However, the physical interpretation is fundamentally incorrect: these are not  galaxies but jet-dominated BL Lac objects where the host galaxy provides redshift anchors through Mg b, Ca~ II H\&K, and Fe I absorption features, while the optical continuum is dominated by non-thermal emission from the relativistic jet. The $20.0\%$ STAR misclassification rate for PL+Galaxy sources represents the most severe pipeline failure.  This failure mode occurs when jet emission is so dominant (typically $>85\%$ of the optical flux) that the stellar absorption features are diluted below the threshold where the pipeline can distinguish between weak lines in stars and heavily jet-diluted galaxy absorption (see Figure~\ref{fig:mult_examp}, panel (a)). We confirm the extragalactic nature of these misclassified sources based on optical/IR colour diagnostics (Section \ref{bl_strip})

The PL+QSO model was selected for 106 sources ($15.0\%$), with SDSS classifications of QSO (53; $50.0\%$), GALAXY (48; $45.3\%$), and STAR (5; $4.7\%$). These sources exhibit significant emission lines from the BLR or NLR and a strong power-law continuum from the jet. The $45.3\%$ GALAXY misclassification rate is particularly striking -- these sources have detectable broad Mg II or C IV emission with equivalent widths $|{\rm EW} > 5|$\,\AA, yet the pipeline classified them as passive galaxies, likely because the jet continuum distorts the expected BLR contribution. 

The PL+Lines model -- representing FSRQs where strong broad emission lines remain detectable against an otherwise  jet-dominated continuum -- was selected for 160 sources ($22.2\%$). The SDSS pipeline predominantly assign QSO classification (128; $80.0\%$), GALAXY (31; $19.4\%$) and STAR (1; $0.6\%$) making up the remainder. The 80.0\% QSO classification rate represents the pipeline's best performance in our sample, as these sources genuinely exhibit QSO-like spectra with prominent broad lines. However, the pipeline is unable to capture the presence of a non-thermal jet component.

Pure QSO, pure Galaxy, and pure power-law models were selected for 21 sources combined ($3.0\%$ of the sample), confirming that the vast majority of \textit{Fermi}-detected blazars require a composite model incorporating both jet and thermal components. The 11 pure QSO sources were predominantly classified as QSOs by SDSS ($10; 90.9\%$), with one source classified as GALAXY ($9.1\%$), representing cases where our fitting found no significant improvement by adding a power-law component --- these may be genuine QSOs within the \Fermi\ error margin rather than the $\gamma$-ray source itself, or FSRQs with very weak jet components. Of the 5 sources assigned pure Galaxy classifications, 2 were classified as QSO by SDSS and 3 as STAR. The 5 sources best-fitted by a pure power-law model represent cases of complete jet-dominated optical emission, with their featureless continua preventing reliable redshift determination. Four of these exhibit genuinely featureless spectra or signal-to-noise ratios below $\mathrm{S/N} < 3$, for which the power-law is the only physically motivated model the pipeline can apply. The remaining source (SDSS ID\,95364441) is an exception: despite exhibiting a visible broad emission feature, the multi-component models failed to converge on a physically consistent solution, representing a catastrophic failure of our pipeline for this source. All 5 pure power-law classifications are therefore flagged as unreliable and excluded from the subsequent spectroscopic analysis.

\subsubsection{\textit{Fermi} Classification Across Best Models}
The breakdown of best-fit models by \textit{Fermi} blazar class (Figure ~\ref{fig:mod_by_fermi}) reveals a strong consistency with pre-existing blazar classification schemes. BL Lac objects (326 sources; 46.1\% of the sample) are overwhelmingly assigned the PL+Galaxy model (280; $85.9\%$), with PL+QSO (30; $9.2\%$) and PL+Lines (13; $4.0\%$) making up the remainder. This distribution directly reflects the defining characteristic of BL Lacs: weak or absent broad emission lines ($|{\rm EW}| < 5$\,\AA) due to either an intrinsically weak accretion disk luminosity or geometric effects that obscure the accretion disk/BLR. The PL+Galaxy model is the optimal best fit for these sources, as it explicitly models the two dominant spectral components --- the jet and the host galaxy --- without including BLR emission. The $9.2\%$ and $4.0\%$ of BL Lacs selecting PL+QSO and PL+Lines, respectively, likely represent sources near the $|{\rm EW}| = 5$\,\AA\ boundary between BL Lacs and FSRQs, where weak BLR emission is marginally detected \citep{padovani2019txs}.

FSRQs (180 sources; $25.5\%$) show a complementary distribution: PL+Lines dominates (97; $53.9\%$), followed by PL+QSO (47; $26.1\%$) and PL+Galaxy (30; $16.7\%$). The preference for PL+Lines over PL+QSO reflects our template construction: the ``Lines'' templates are optimized for sources with strong, high-ionization broad emission (C\,\textsc{iv}, C\,\textsc{iii}], Mg\,\textsc{ii}) but minimal continuum from the accretion disk, exactly matching the FSRQ spectral signature where jet continuum dominates but BLR lines remain strong. The $26.1\%$ of sources assigned the PL+QSO represent cases where both BLR emission and accretion disk continuum are significant. The $16.7\%$ of FSRQs selecting PL+Galaxy also likely lie near the $|{\rm EW}| = 5$\,\AA\ classification boundary.

Blazars of uncertain type (BCU; 201 sources; $28.4\%$) show an intermediate distribution: PL+Galaxy (110; $54.7\%$), PL+Lines (50; $24.9\%$), and PL+QSO (29; $14.4\%$). The $54.7\%$ PL+Galaxy preference suggests the majority of BCU sources are likely BL Lacs with particularly weak or ambiguous spectral features, while the $39.3\%$ selecting QSO- or line-dominated models suggests a substantial FSRQ contribution. Our model selection therefore provides an independent classification scheme using optical spectroscopy by SDSS that can help resolve BCU ambiguities: BCU sources assigned PL+Galaxy with $f_{\rm jet} > 70\%$ are strong BL Lac candidates, while those selecting PL+Lines with detected broad emission $|{\rm EW}| > 5$\,\AA\ are strong FSRQ candidates.

\subsubsection{SDSS Retained  and Non-blazar \textit{Fermi} Class}
\label{excl}

\begin{figure*}
    \centering
\includegraphics[width=0.8\linewidth]{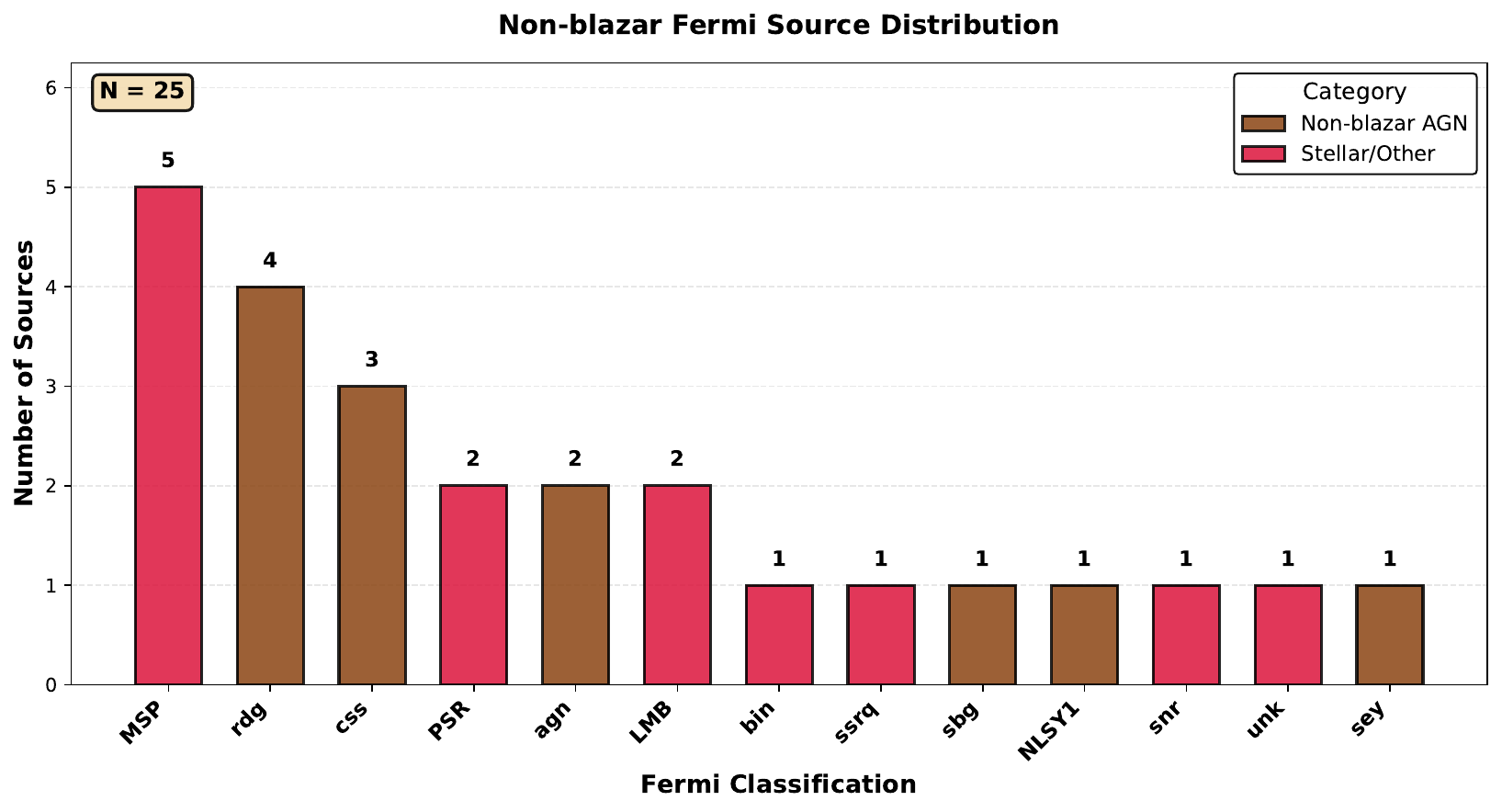}
    \caption{Distribution of the 25 non-blazar \textit{Fermi} sources. Bars are coloured by source category: brown for non-blazar AGN types (rdg: 4, css: 3, agn: 2, sbg: 1, NLSY1: 1, sey: 1; total $N = 12$) and crimson for stellar/other types (MSP: 5, PSR: 2, LMB: 2, bin: 1, ssrq: 1, snr: 1, unk: 1; total $N = 13$).}
    \label{fig:non_bl}
\end{figure*}

Of the 746 sources in the Fermi-SDSS-V crossmatch, 39 ($5.2\%$) were excluded from the final blazar sample: 14 ($1.9\%$) due to poor model performance ($\chi^{2}_{\rm lmfit} > 3 \times \chi^{2}_{\rm SDSS}$; Figure~\ref{fig:best_mod_by_sdd}) and 25 ($3.4\%$) due to non-blazar \textit{Fermi} classifications (Figure~\ref{fig:mod_by_fermi}). The 14 catastrophic failures represent cases where the SDSS template-matching pipeline, despite its limitations, outperformed our multi-component approach, predominantly sources classified as STAR (8; 57.1\%), GALAXY (3; 21.4\%), and QSO (3; 21.4\%). The 25 non-blazar \textit{Fermi} sources span a diverse range of classes (see Figure~\ref{fig:non_bl}): 12 non-blazar AGN including radio galaxies (rdg; 4), compact steep spectrum radio sources (css; 3), generic AGN (agn; 2), a starburst galaxy (sbg), a narrow-line Seyfert 1 (NLSY1), and a Seyfert galaxy (sey); and 13 stellar or unrelated sources including millisecond pulsars (MSP; 5), pulsars (PSR; 2), low-mass X-ray binaries (LMB; 2), a binary system (bin), a steep-spectrum radio-quiet quasar (ssrq), a supernova remnant (snr), and an unidentified source (unk). Our model library is designed to capture the dominant blazar population --- BL Lacs, FSRQs, and BCUs --- and does not include templates suitable for these diverse minority classes. We therefore retain the \textit{Fermi} classification for these sources and exclude them from all subsequent analysis.

\subsection{Redshift (z) and Jet Fraction Distribution}
\label{jet_z_dis}
\begin{figure*}
    \centering
    \includegraphics[width=0.90\linewidth]{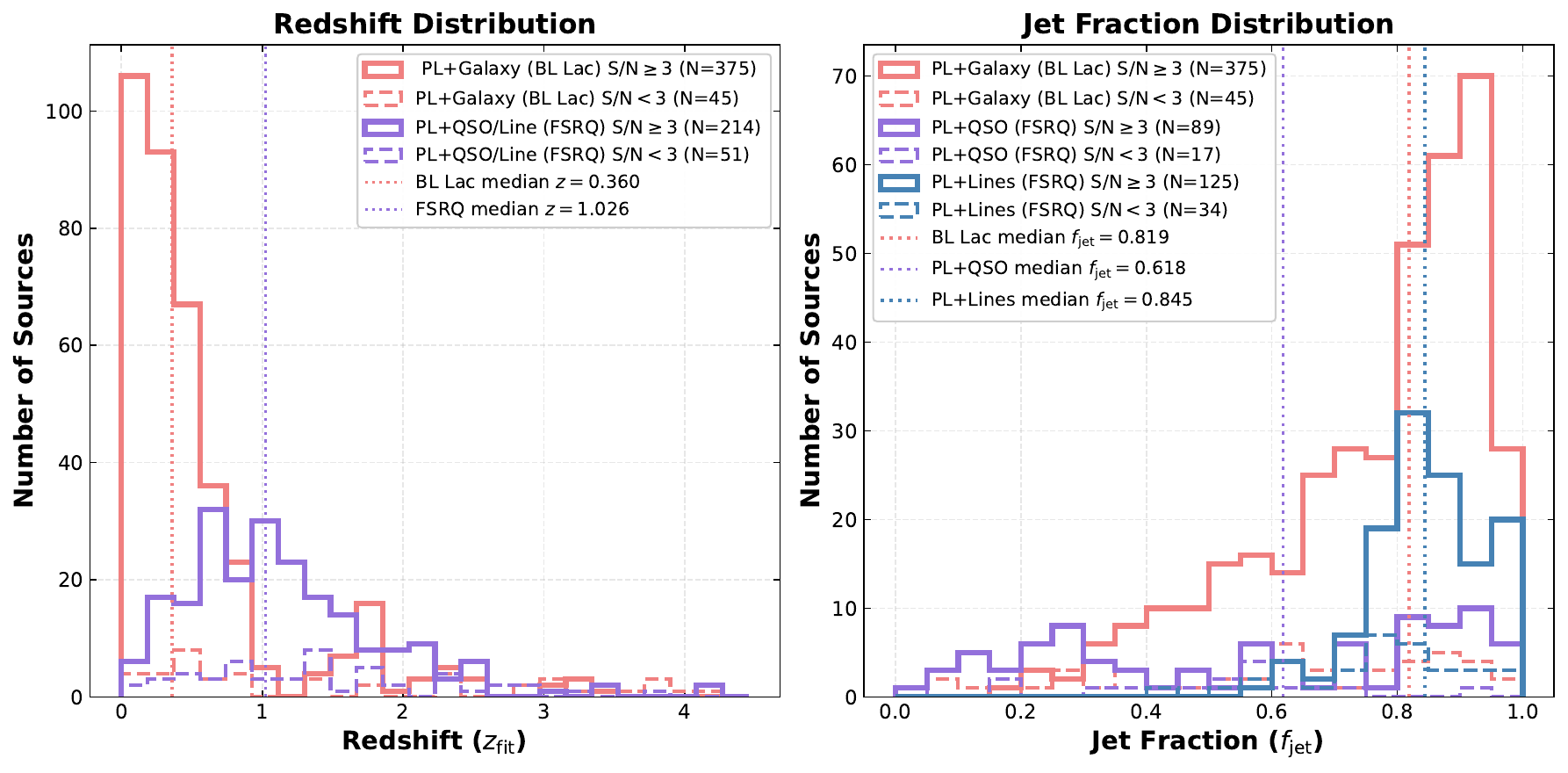}
    \caption{Redshift ($z_{\rm fit}$; left) and jet fraction ($f_{\rm jet}$; right) distributions for BL Lac candidates (PL+Galaxy; $N = 420$; red) and FSRQ candidates (PL+QSO, $N = 106$, purple; PL+Lines, $N = 159$, blue). Solid histograms show sources with spectral ${\rm S/N} \geq 3$; dashed histograms show sources with ${\rm S/N} < 3$, which are excluded from all median calculations. Dashed vertical lines demarcate the median estimate of the source population. BL Lac candidates are concentrated at low redshifts (median $z = 0.360$) while FSRQ candidates extend to higher redshifts (median $z = 1.026$), a separation confirmed by a KS test ($D = 0.548$, $p = 3.13 \times 10^{-38}$; \citealt{feller1948kolmogorov, virtanen2020scipy}). In the jet fraction panel, BL Lac candidates display a sharply peaked distribution at $f_{\rm jet} \sim 0.9$--$1.0$ (median $f_{\rm jet} = 0.819$), indicating uniform jet dominance. PL+QSO sources show a lower median $f_{\rm jet} = 0.618$, while PL+Lines sources show a median $f_{\rm jet} = 0.845$, interpreted as an upper limit in the absence of an explicit thermal template.}
    \label{fig:z_jt}
\end{figure*}

Following the redshift ($z_{\rm fit}$) and jet fraction ($f_{\rm jet}$) estimation approach described in Section~\ref{sec: 3}, we investigate the distributions of these quantities for our final sample of 707 \Fermi-detected sources with updated spectral fits. Figure~\ref{fig:z_jt} presents the $z_{\rm fit}$) and ($f_{\rm jet}$ distributions for BL Lac candidates (PL+Galaxy; $N = 420$) and FSRQ candidates (PL+QSO, $N = 106$, PL+Line; $N = 160$; total $N = 265$), after the exclusion of one FSRQ candidate (PL+Line) with unreliable redshift or jet fraction estimates. 

The median $z_{\rm fit}$ and $f_{\rm jet}$ are calculated from sources with spectral S/N $\gtrsim$ 3 only (BL Lac: $N = 375$; FSRQ: $N = 214$); low-S / N sources are shown as dashed histograms in Figure~\ref{fig:z_jt} and excluded from all population statistics.   

The 420 BL Lac candidates (PL+Galaxy) exhibit a strong concentration at low redshifts, with a sharp peak at $ z \sim 0.1-0.5$ and a median of $z = 0.360$ (S/N $\geq 3$ sources). The distribution falls off rapidly beyond $z \sim 1.0$ with very few sources detected at $z > 1.5$. PL+Galaxy sources beyond $z \gtrsim 1.8$ are retained in the sample but their redshift estimates are considered low-confidence, as the SDSS wavelength coverage provides no strong rest-frame anchors for the galaxy template at these redshifts; the majority of these sources also have spectral ${\rm S/N} < 3$ and are indicated by the shaded region in Figure~\ref{fig:lum}. FSRQ candidates on the other hand, show a broader redshift distribution extending to significantly higher redshifts, with a median of ${\rm z =1.026}$ and individual sources reaching $ z > 4$. This separation is consistent with the known cosmological evolution of blazar samples, where FSRQs are systematically found at higher redshifts than BL Lacs \citep{ajello2014cosmic}, reflecting the transition from radiatively efficient accretion at high redshift to radiatively inefficient accretion in the local universe. We assess whether the two samples are drawn from the same underlying redshift distribution using the Kolmogorov–Smirnov (KS) test \citep{feller1948kolmogorov, virtanen2020scipy}, which confirms that the two redshift distributions are statistically inconsistent with originating from the same parent population ($D = 0.548$, $p = 3.13\times 10^{-38}$).  

The preferential detection of BL Lacs at low redshifts likely reflects various selection effects. First, the weak-lined optical spectra that define BL Lacs become particularly difficult to identify at high redshifts, where spectroscopic redshift determination requires detection of faint host galaxy absorption features or weak emission lines \citep{shaw2013spectroscopy}. The FSRQ distribution extending to higher redshifts is consistent with their strong broad emission lines, which enhance redshift measurements even for distant sources.
In Figure~\ref{fig:z_jt}, we separate the FSRQ candidates by model subtype, which reveals a physically distinct jet fraction distribution. PL+QSO sources show a median $f_{\rm jet} = 0.618$ (S/N $\geq 3$ sources), reflecting genuine thermal QSO template contributions to the optical continuum, while PL+Lines sources show a median $f_{\rm jet} = 0.845$ (S/N $\geq 3$; $N = 125$), which should be interpreted as an upper limit since the power-law component absorbs all continuum emission in the absence of an explicit thermal template. BL Lac (PL+Galaxy) candidates show a median $f_{\rm jet} = 0.819$ (S/N $\geq 3$; $N = 375$), consistent with strong jet dilution of host galaxy starlight. The KS tests confirm that all three pairwise jet fraction comparisons are statistically significant with being drawn from different underlying distributions, with the strongest separation between PL+QSO and PL+Lines ($D = 0.522$, $p = 1.71 \times 10^{-13}$).

BL Lac candidates exhibit a sharply peaked, narrow jet fraction distribution concentrated at $f_{\rm jet} \sim 0.9$--$1.0$, with $\sim 70\%$ of sources having $f_{\rm jet} > 0.8$ and very few below $f_{\rm jet} = 0.6$. FSRQ candidates show a broader distribution: while the modal value peaks at $f_{\rm jet} \sim 0.8$--$1.0$, a significant tail extends to $f_{\rm jet} \sim 0.3$--$0.7$ where thermal disk emission contributes comparably to the jet continuum. The similar BL Lac and PL+Lines jet fractions but markedly lower PL+QSO jet fraction demonstrates that when an explicit thermal template is available, the power-law component is no longer required to account for the full continuum, revealing the true jet contribution. This is consistent with unified blazar schemes where BL Lacs represent systems with intrinsically weak thermal components or more extreme beaming, while FSRQs represent systems with significant disk emission superimposed on the jet continuum \citep{fossati1998unifying, ghisellini2017fermi}.

\subsection{$\gamma$-ray Luminosity} 
\label{lumino}
We calculate the $\gamma$-ray luminosities in the observed 100\, MeV--100\, GeV band using the energy fluxes reported in the 4FGL-DR4 catalogue \citep{ballet2023fermi}, applying K-corrections that account for spectral shape and cosmological redshift. We adopt a flat ${\rm \Lambda CDM}$ cosmology with ${\rm H_o = 70~kms^{-1}~Mpc^{-1}}$ and ${\rm \Omega_{m}} = 0.3$ \citep{aghanim2020planck}. The luminosity is computed as ${\rm L_{\gamma} = 4 \pi~D_{L}^{2}~S_{obs} ~K(z, \Gamma)}$, where ${\rm D_{L}}$ is the luminosity distance, ${\rm S_{obs}}$ is the observed energy flux, and ${\rm K(z, \Gamma)}$ is the K-correction that converts the observed energy flux to the equivalent rest-frame luminosity, accounting for the source redshift and spectral shape \citep{ghisellini2009unifying, ajello2012luminosity}. For sources where the best-fit  model by \Fermi is a power-law as identified in the 4FGL-DR4 catalogue, we use ${\rm K(z, \Gamma) = (1+z)^{(\Gamma -2)}}$ where $\Gamma$ is the photon index from the catalogue parameter  PL\_Index \citep{abdo2010spectral, nolan2012fermi}. For LogParabola models, we use the photon index at the pivot energy (${\rm \alpha = LP\_{Index}}$) for the K-correction, providing an approximation for sources with curved spectra. For PowerLaw with Exponential Cutoff models, we use ${\rm K(z,\Gamma) = (1+z)^{(\Gamma -2)}}$ with $\Gamma$ from PLEC\_IndexS. Sources lacking spectral model information default to the power-law K-correction using PL\_Index. 

\begin{figure*}
    \centering
    \includegraphics[width=0.99\linewidth]{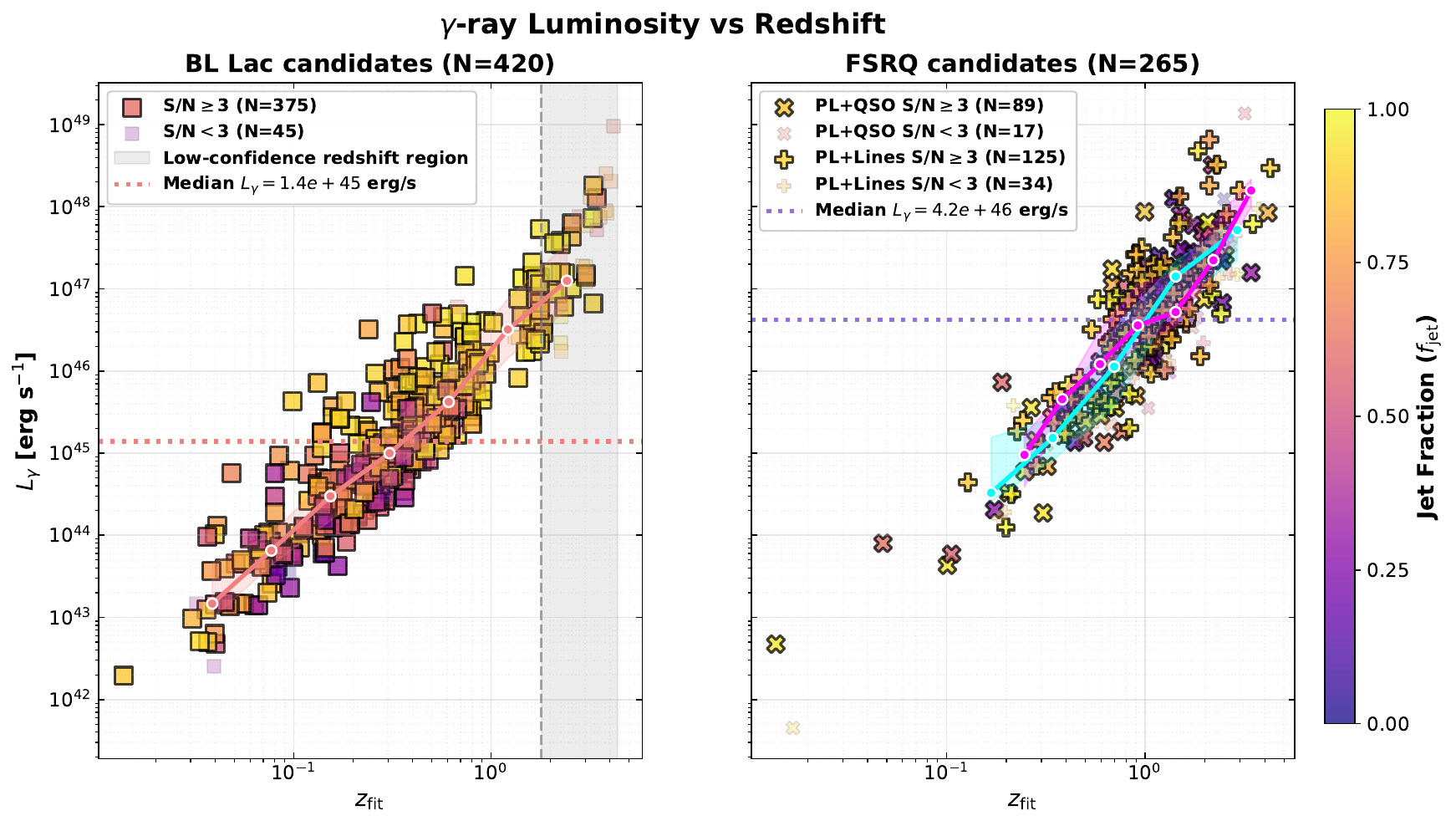}
        \caption{$\gamma$-ray luminosity versus spectroscopic redshift for BL Lac candidates (PL+Galaxy, $N = 420$, left panel) and FSRQ candidates (PL+QSO/Lines, $N = 265$, right panel), colour-coded by optical jet fraction from the multi-component spectral decomposition. Sources with spectral ${\rm S/N} < 3$ are shown with reduced opacity and smaller markers. \textit{Left}: The shaded grey region ($z \gtrsim 1.8$) indicates the low-confidence redshift region, where PL+Galaxy spectral fits become poorly constrained due to the absence of strong rest-frame features in the SDSS wavelength coverage. The solid line with shaded band shows the running median with interquartile range in redshift bins, tracing how the typical $\gamma$-ray luminosity evolves with redshift across the populations. The dotted horizontal line marks the global median $L_\gamma = 1.39 \times 10^{45}$\,erg\,s$^{-1}$ at median $z = 0.360$ (S/N $\geq 3$ sources). \textit{Right}: FSRQ candidates show a global median $L_\gamma = 4.23 \times 10^{46}$\,erg\,s$^{-1}$ at median $z = 1.026$ (S/N $\geq 3$ sources), a separation of approximately one order of magnitude ($\sim$ 1.5 dex) above the BL Lac median, consistent with established $\gamma$-ray luminosity functions \citep{ajello2014cosmic, ajello2012luminosity}. Running medians are shown separately for PL+QSO (magenta) and PL+Lines (cyan), revealing that PL+Lines sources sit systematically above PL+QSO sources at similar redshifts. Within the FSRQ population, PL+QSO sources ($N = 89$, median $z = 1.023$, median $L_\gamma = 6.78 \times 10^{46}$\,erg\,s$^{-1}$) show systematically higher luminosities than PL+Lines sources ($N = 125$, median $z = 1.034$, median $L_\gamma = 3.82 \times 10^{46}$\,erg\,s$^{-1}$), despite occupying similar redshift ranges. PL+Lines sources also show a higher median jet fraction ($f_{\rm jet} = 0.845$) compared to PL+QSO sources ($f_{\rm jet} = 0.618$).}
    \label{fig:lum}
\end{figure*}

 Figure~\ref{fig:lum} presents the $\gamma$-ray luminosity versus redshift for the BL Lac and FSRQ candidates with finite luminosity estimates, colour-coded by optical jet fraction. Medians are computed from sources with spectral ${\rm S/N} \geq 3$ only. BL Lac candidates (PL+Galaxy, $N = 420$) show a median $L_{\gamma} = 1.39 \times 10^{45}$\,erg\,s$^{-1}$ at median $z = 0.360$, while FSRQ candidates (PL+QSO and PL+Lines, $N = 265$) show a median $L_{\gamma} = 4.23 \times 10^{46}$\,erg\,s$^{-1}$ at median $z = 1.026$. These values are consistent with $\gamma$-ray luminosity functions derived independently for BL Lacs \citep{ajello2014cosmic} and FSRQs \citep{ajello2012luminosity}, confirming that the model-based classification recovers the established blazar populations. The running median reveals a clear positive $L_\gamma$--$z$ correlation in both populations, consistent with the Malmquist bias expected in a flux-limited $\gamma$-ray survey --- more luminous blazars are detectable at higher redshifts. The separation of $\sim$ 1.5 dex in median $\gamma$-ray luminosity between BL Lac and FSRQ candidates likely reflects the different $\gamma$-ray emission mechanisms operating in the two subclasses: BL Lacs are predominantly powered by synchrotron self-Compton processes \citep{fossati1998unifying}, whereas the $\gamma$-ray emission from FSRQs is boosted by inverse Compton scattering of external photon fields from the broad-line region and accretion disc \citep{ghisellini2009canonical}. Within the FSRQ population, PL+QSO sources sit systematically above PL+Lines sources in the running median at comparable redshifts, indicating higher $\gamma$-ray luminosities despite similar redshift ranges. PL+QSO sources ($N = 89$, median $z = 1.023$, median $L_\gamma = 6.78 \times 10^{46}$\,erg\,s$^{-1}$) show higher absolute luminosities than PL+Lines sources ($N = 125$, median $z = 1.034$, median $L_\gamma = 3.82 \times 10^{46}$\,erg\,s$^{-1}$), and PL+Lines sources show a higher median jet fraction ($f_{\rm jet} = 0.845$) compared to PL+QSO sources ($f_{\rm jet} = 0.618$), suggesting that the latter are more jet-dominated with weaker accretion disc contributions relative to the non-thermal continuum.

Despite the luminosity differences, both BL Lac and FSRQ populations exhibit similar jet fractions (BL Lacs: median $f_{\rm jet} = 0.819$;\, FSRQs: median $f_{\rm jet} = 0.828$). This suggests that jet dominance in the optical continuum is not the discriminator between the two subclasses. Rather, the presence of thermal disk and broad line region emission, captured by the PL+QSO and PL+Lines models--provides the fundamental distinction \citep{padovani2017active, ghisellini2011transition, ghisellini2017fermi}. BL Lacs with $f_{\rm jet} > 0.9$ represent extreme dilution cases where host galaxy absorption features become difficult to detect, and the optical spectrum becomes essentially featureless --- precisely the regime where the SDSS pipeline most frequently misclassifies these sources as stars. 

\section{Validation of Multi-Component Fitting
Approach}
\label{sec:discuss}
In this section, we investigate the physical validity of the multi-component spectral classifications and redshift estimates derived in Section~\ref{sec: 3} and presented in Section~\ref{sec:results}, by testing them against three independent diagnostics. In Section~\ref{bl_strip}, we compare the spectral reclassifications against IR photometry from the \textit{Wide-field Infrared Survey Explorer} \citep[\textit{WISE};][]{wright2010wide}, assessing whether sources classified as BL Lac or FSRQ candidates occupy the expected regions of the \textit{WISE} colour-colour diagram. In Section~\ref{z_stats}, we validate the redshift estimates for the 111 sources in common with the LAT high-energy catalogue \citep[FHL;][]{ajello20173fhl}, quantifying the improvement in redshift accuracy relative to the SDSS automated pipeline. In Section~\ref{ew_comparison}, we examine whether the equivalent width distributions of the classified sources are consistent with the traditional $|{\rm EW}| = 5$\,\AA\ BL Lac/FSRQ boundary \citep{stickel1991complete}, and discuss the substantial population of hybrid sources showing simultaneous emission and absorption features that challenge a strict binary classification scheme.  

\subsection{Confirmation of new classifications using the WISE-Colour diagram} 
\label{bl_strip}
\begin{figure*}
    \centering
    \includegraphics[width=0.7\linewidth]{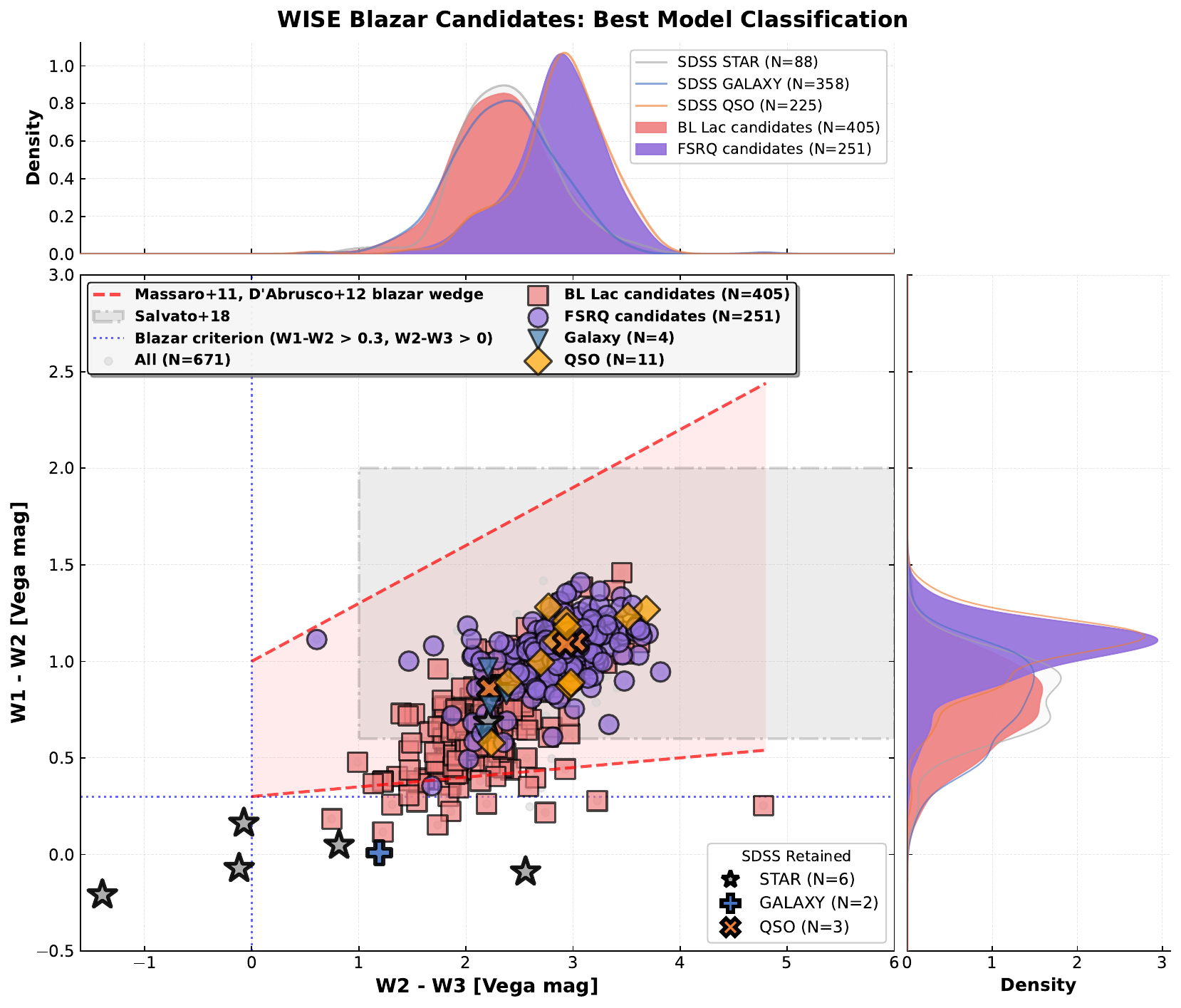}
    \caption{\textit{WISE} colour-colour diagram for 671 sources with valid $W1, W2, and W3$ magnitudes, coloured by best-fit model classification: BL Lac candidates (PL only and PL+Galaxy; lightcoral squares, $N = 405$ of 420 with valid \textit{WISE} photometry), FSRQ candidates (PL+QSO and PL+Lines; mediumpurple circles, $N = 251$ of 266), pure QSO (orange diamonds, $N = 11$), and Galaxy (blue triangles, $N = 4$). Large markers with black edges indicate the 11 sources where SDSS classifications are retained (STAR: $N = 6$; GALAXY: $N = 2$; QSO: $N = 3$). Canonical blazar regions from \citet{massaro2011identification} and \citet{d2012infrared} (red dashed wedge) and \citet{salvato2018finding} (grey box) are overlaid. Marginal KDE distributions show the SDSS pipeline classifications (STAR: $N = 88$; GALAXY: $N = 358$; QSO: $N = 225$) as thin coloured outlines, with deep filled regions indicating the BL Lac (red) and FSRQ (purple) model compositions.}
    \label{fig:blazar_stripe}
\end{figure*}

Figure~\ref{fig:blazar_stripe} presents the \textit{WISE} colour-colour diagram for the 671 sources with valid \textit{WISE} magnitudes in the W1, W2 and W3 bands where we adopt our new classification ($\chi^2_{\rm r,\,lmfit} < 3\times\chi^2_{\rm r,\,SDSS}$), coloured by our best-fit model. This can be compared with Figure~\ref{fig:sdss_wise}, which shows the same sources coloured by their original SDSS classifications. The overwhelming majority of sources that we classify as BL Lac candidates (PL+Galaxy; $N = 405$ of 420 with valid \textit{WISE} photometry) and all of our FSRQ candidates (PL+QSO/Lines; $N = 251$) occupy the canonical blazar region of the \textit{WISE} colour-colour diagram, defined by the blazar strip of \citet{massaro2011identification} and \citet{d2012infrared} and the AGN region of \citet{salvato2018finding}. This independent multi-wavelength validation confirms that our reclassifications are physically well-motivated: sources that the SDSS pipeline assigned to the STAR, GALAXY, and QSO categories but that our spectral decomposition identifies as blazar candidates show \textit{WISE} infrared colours fully consistent with AGN activity rather than stellar photospheres or passive galaxies.

The most striking result is the reclassification of SDSS STAR sources. Of the 88 sources originally classified as Galactic stars, 85 (96.6\%) fall within the blazar strip of \citet{massaro2011identification} and \citet{d2012infrared}, while 78 (88.6\%) lie within the AGN region of \citet{salvato2018finding}. These sources are predominantly BL Lac candidates (90.9\%), and their \textit{WISE} colours 
are entirely inconsistent with Galactic stellar photospheres and fully consistent with non-thermal AGN activity, providing unambiguous independent confirmation that these are genuine blazars misidentified by the SDSS automated pipeline. In contrast, none of the 6 SDSS STAR-classified sources where we retain the original classification fall within any of the canonical blazar selection regions, validating that these are genuinely stellar sources.

An inspection of the \textit{WISE} colour distribution reveals a systematic difference between BL Lac and FSRQ candidates: the BL Lac candidates tend to occupy slightly bluer $W1-W2$ and $W2-W3$ colours compared to FSRQ candidates, with a subset falling outside or at the blue edge of the canonical blazar selection regions. This colour offset reflects the stronger host galaxy contribution in BL Lac candidates, which dilutes the non-thermal continuum and shifts the mid-IR colours blueward relative to a pure power-law spectrum \citep{plotkin2012lack, caccianiga2015wise}. More so, the bluer mid-IR colours of BL Lac candidates are also consistent with a lack of significant dusty torus emission \citep{plotkin2012lack}. The absence of torus emission in BL Lac objects is physically consistent with their classification as radiatively inefficient accretors: without a luminous ultraviolet accretion disc to illuminate and heat the surrounding dusty material, no torus emission is produced \citep{gardner2018powers}. In this case, our BL Lac candidates -- despite being detected as \textit{Fermi} gamma-ray sources with powerful relativistic jets pointed along our line-of-sight -- likely harbour central engines in which accretion proceeds through a radiatively inefficient, advection-dominated accretion flow (ADAF) rather than a geometrically thin, optically thick disc \citep{narayan1994advection, heckman2014coevolution}.

\subsection{Assessing the quality of updated redshift estimates}
\label{z_stats}
\begin{figure*}
    \centering
    \includegraphics[width=0.99\linewidth]{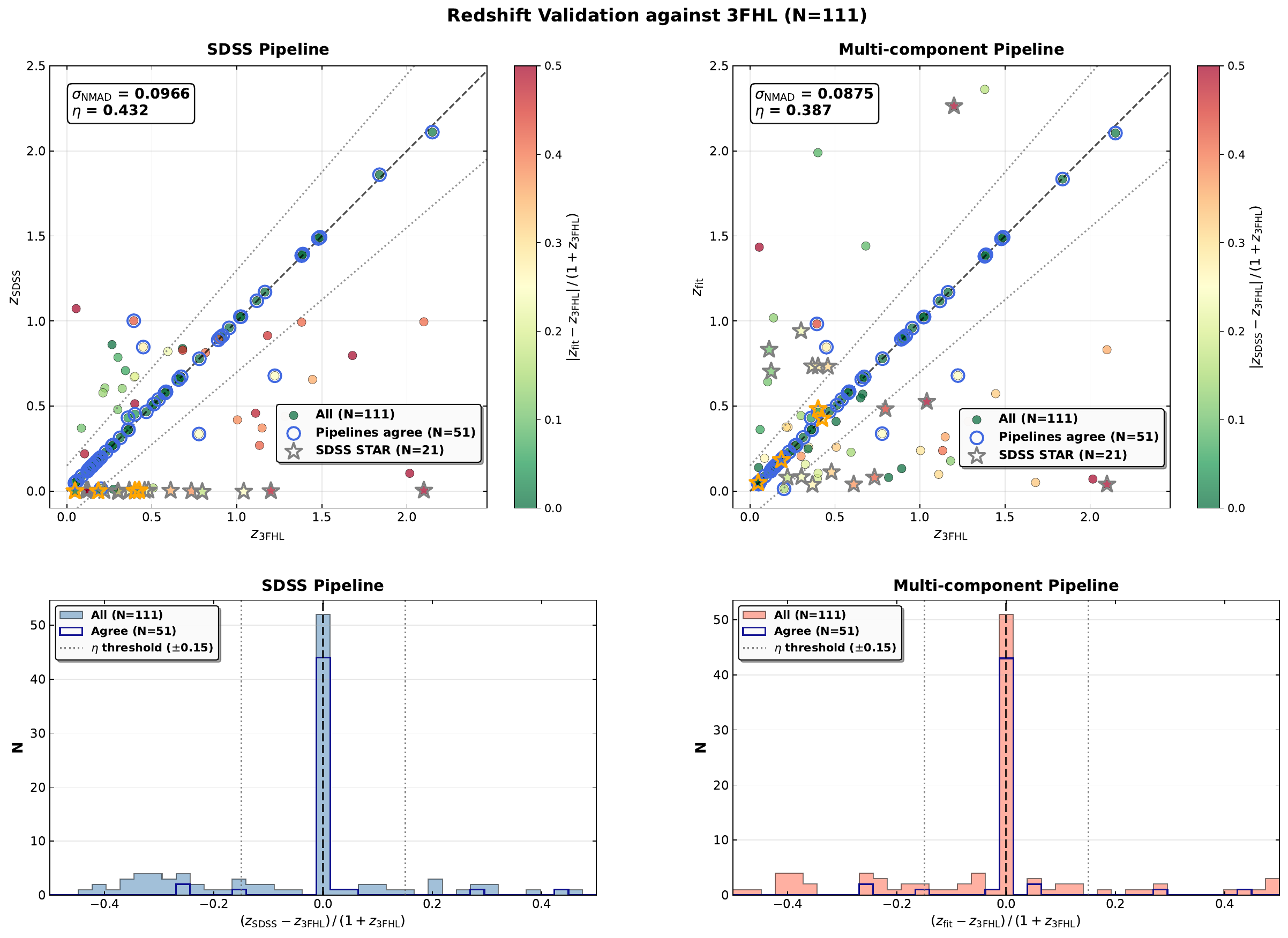}
    \caption{Redshift comparison between the multi-component fit estimates ($z_{\rm fit}$) and 3FHL spectroscopic redshifts for the 111 sources passing both the quality threshold and the ${\rm S/N} \geq 3$ criterion. \textit{Top panels}: scatter plots of $z_{\rm SDSS}$ (left) and $z_{\rm fit}$ (right) against $z_{\rm 3FHL}$, colour-coded by the normalised deviation of the opposing pipeline (green: small deviation; red: large deviation). Dotted lines mark the $|\Delta z|/(1+z) = 0.15$ threshold. Blue open circles identify the 51 sources (45.9\%) where both pipelines independently converge on consistent redshifts; grey stars mark the 21 sources classified as \texttt{STAR} by the SDSS pipeline. In the right panel, the grey stars separate into two groups relative to the dotted threshold lines: 5 sources recovered by the multi-component pipeline (within the threshold) and 16 where the multi-component redshift disagrees with the 3FHL value. \textit{Bottom panels}: distributions of the normalised redshift residual $(z_{\rm model} - z_{\rm 3FHL})\,/\,(1 + z_{\rm 3FHL})$ for the SDSS pipeline (left) and multi-component pipeline (right). Dashed vertical line marks $\Delta z = 0$; dotted vertical lines mark the $\eta$ threshold at $\pm 0.15$.}
    \label{fig:z_test}
\end{figure*}

An essential component of this work is the redshift estimation for the blazar candidates derived from the multi-component spectral fitting approach described in Section~\ref{sec: 3}. To assess the reliability of our new estimates, we compare them against prior spectroscopic redshifts from the Third Catalog of Hard \textit{Fermi}-LAT Sources \citep[3FHL;][]{ajello20173fhl}. The 3FHL contains 1556 sources detected above 10\,GeV in the first seven years of \textit{Fermi}-LAT data, characterised in the 10\,GeV to 2\,TeV energy range, and provides spectroscopic redshifts for a well-studied subset of confirmed blazars. Of the 707 sources in our final sample, 111 pass our quality filter ($\chi^{2}_{\rm r,\,lmfit} \leq 3 \times \chi^{2}_{\rm r,\,SDSS}$ and ${\rm S/N} \geq 3$) and have 3FHL redshifts. Figure~\ref{fig:z_test} compares the 3FHL redshifts to the SDSS pipeline estimates (left panel) and the estimates with our new multi-component pipeline (right panel) for these sources. 

We assess the performance of the redshift estimation for for both pipelines using two standard metrics: i) the normalised median absolute deviation \citep[NMAD;][]{ilbert2006accurate} defined as 
\begin{equation}
\sigma_{\rm NMAD}  = 1.48 \times {\rm median} (|z_{\rm model} - z_{\rm 3FHL}|)/ (1 + z_{\rm 3FHL})
\end{equation} 
where $z_{\rm model}$ corresponds to either $z_{\rm SDSS}$ or our new estimate, $z_{\rm fit}$, and $z_{\rm 3FHL}$ serves as the reference spectroscopic redshift; ii) a physically motivated catastrophic-failure fraction ($\eta$). For the multi-component pipeline, we adopt  the standard, which is defined as 
\begin{equation}
\eta_{\rm fit} = {N\!\left(|z_{\rm fit} - z_{\rm 3FHL}|\,/\,
(1 + z_{\rm 3FHL}) > 0.15\right)}/{N_{\rm total}}
\end{equation}
while for the SDSS pipeline we extend this also to include sources classified as \texttt{STAR}, since a stellar classification by SDSS implies $z \approx 0$ regardless of the normalised deviation, representing a fundamental classification failure independent of any redshift metric. Thus, 
\begin{equation}
\eta_{\rm SDSS} = \frac{N\!\left(|z_{\rm SDSS} - z_{\rm 3FHL}|\,/\,
(1 + z_{\rm 3FHL}) > 0.15 \;\text{OR}\; 
\texttt{CLASS} = \texttt{STAR}\right)}{N_{\rm total}}.
\end{equation}
 We note that the 3FHL redshifts are drawn from a broad range of catalogues and studies and, while they represent the best available external benchmark, some may themselves carry systematic uncertainties or require further confirmation; they should therefore be treated as reliable reference values rather than absolute ground truth.

On the full validation sample ($N = 111$), the multi-component pipeline achieves $\eta = 0.387$ compared to $\eta = 0.432$ for the SDSS pipeline under this definition, a 10.4\% reduction in the catastrophic outlier fraction. The corresponding $\sigma_{\rm NMAD}$ values are $0.0875$ and $0.0966$ with our pipeline and the SDSS pipeline, respectively, indicating our multi-component approach achieves improved accuracy. When both pipelines independently converge on consistent redshifts ($|z_{\rm SDSS} - z_{\rm fit}| \,/\, (1 + z_{\rm fit}) < 0.01$; $N = 51$, 45.9\%), both methods achieve excellent accuracy ($\sigma_{\rm NMAD} = 0.0010$, $\eta = 0.098$), confirming that the multi-component fitting introduces no systematic redshift biases for well-constrained sources. 

Of the 111 sources in our sample with pre-existing redshifts in the 3FHL compilation, 21 (18.8\%) are classified as \texttt{STAR} by the SDSS pipeline, representing cases where SDSS assigns $z \approx 0$ to genuine extragalactic blazar candidates.
All of these are reclassified as BL Lac candidates by our multi-component pipeline. Five of these are recovered with redshifts consistent with the 3FHL reference 
($\Delta z/(1+z) < 0.1$), while the remaining 16 show disagreement with the 3FHL value. All 16 sources are best fitted by the PL+Galaxy model with high-quality spectral decompositions; hence, the disagreement with the 3FHL reference value does not necessarily reflect a failure of our multi-component pipeline. Given that the 3FHL redshifts for these sources are drawn from various external catalogues and that SDSS itself catastrophically misclassified them as stars, the reliability of the 3FHL reference redshifts for this specific subset cannot be taken for granted. 

In addition to the sources that SDSS classified as stars, we highlight a number of other populations in Figure~\ref{fig:z_test}.
Firstly, there are 8 sources where $z_{\rm SDSS}$ is in poor agreement with $z_{\rm 3FHL}$ but our updated $z_{\rm fit}$ provides a good agreement: these correspond to the green circles in the left panel of Figure~\ref{fig:z_test} that lie off the 1:1 relation (i.e. outside the region delimited by the dotted lines indicating a relative redshift error $>0.15$). 
Notably, in all 8 cases, SDSS \emph{over}estimates the redshift and the sources thus lie above the upper dotted line. 
These sources represent additional cases where the SDSS pipeline appears to fail catastrophically but our bespoke blazar pipeline recovers the (likely) correct redshift.
In contrast, green circles in the right-hand panel of Figure~\ref{fig:z_test} that lie away from the 1:1 relation represent cases where the SDSS pipeline was in agreement with the 3FHL value but our pipeline appears to fail.
Excluding sources that SDSS classified as STAR (representing a failure of SDSS even if the 3FHL redshift is nominally consistent with $z\approx0$), there are 12 sources that fall in this category with the multi-component pipeline producing both over- and under-estimates compared to $z_{\rm 3FHL}$.
A minority of cases (2/12) appear to represent genuine catastrophic failures where our pipeline does not identify the correct line features, but the vast majority (10/12) correspond to sources with strong jet contributions and relatively weak spectral features. 
 Our new redshift measurement---using our physically motivated multi-component pipeline designed for blazars---may in fact be more reliable for these sources.\footnote{In many of these cases the 3FHL redshift may be based on SDSS I--IV observations and was estimated using an earlier version of the SDSS pipeline, resulting in inevitable agreement with the SDSS-V pipeline estimate.}
Finally, redder circles that lie away from the 1:1 relation (above or below the dotted lines) in \emph{both} panels represent cases where \emph{neither} $z_{\rm SDSS}$ nor $z_{\rm fit}$ are consistent with $z_{\rm 3FHL}$. In four of these cases (circled in blue) $z_{\rm SDSS}$ and $z_{\rm fit}$ are consistent with each other; while both pipelines were run on the same spectroscopic data, these likely correspond to incorrect values within the 3FHL compilation and we recommend adopting the consistent estimate provided by the SDSS or multi-component pipeline in future studies.
The true redshift of the remaining outliers with both pipelines verus 3FHL remains unclear -- careful investigation of the origin of the 3FHL redshift or independent spectroscopic follow-up would be needed to verify or reject $z_{\rm fit}$.

\subsection{Comparison of template-based approach to equivalent-width-based classifications}
\label{ew_comparison} 
The traditional blazar classification separates BL Lacs (${\rm EW} < 5$~\AA\ for emission line features) from FSRQs (${\rm EW}> 5$\,\AA\ for broad emission lines) based on the rest-frame equivalent width measurements \citep{stickel1991complete, urry1995unified}. We investigate the population-level EW distribution of BL Lac candidates with host features (PL+Galaxy; $N=420$) and FSRQ candidates (PL+QSO/Lines; $N = 265$). In all sources, we attempt to measure emission features (C~IV, C~III], Mg~ II, [O~II], H$\beta$, [O~III]) tracing the ionised broad and narrow-line region and  and absorption features (Mg~b, Ca~II K/H, Ca~ I G, Na~ I D) tracing the host galaxy stellar population, using the detection threshold of ${\rm S/N} \geq 3$ as described in Section~\ref{sec:ew_methods}. Spectral features satisfying this threshold are identified in 350 BL Lac sources ($83.3\%$) and 251 FSRQ sources ($94.7\%$). The higher detection fraction in FSRQs reflects the presence of strong broad emission lines, which are more readily detected above the noise threshold than the weak absorption features that dominate BL Lac spectra. 

To mitigate the influence of spurious detections on the reported mean values, an iterative $3\sigma$ clipping is applied to the distribution of equivalent width measurements for each spectral line independently. In each iteration, a measurement ${\rm EW}_i$ is rejected if $|{\rm EW}_i - \overline{\rm EW}| > 3\sigma_{\rm EW}$, where $\overline{\rm EW} = N^{-1}\sum_{i} {\rm EW}_i$ is the sample mean and $\sigma_{\rm EW} = \sqrt{(N-1)^{-1}\sum_{i}({\rm EW}_i - \overline{\rm EW})^2}$ is the sample standard deviation. This is repeated on the retained sample until no further measurements are rejected in a given iteration, or until a maximum of five iterations is reached. Mean values are then reported as $\bar{\rm EW} \pm {\rm SEM}(\text{standard error of the mean})$, where ${\rm SEM} = \sigma_{\rm EW}/\sqrt{N_{\rm ret}}$ and $N_{\rm ret}$ is the number of detections retained after clipping. Table~\ref{tab:ew_comparison} presents the full results. 

\begin{figure*}
    \centering
    \includegraphics[width=0.8\textwidth]{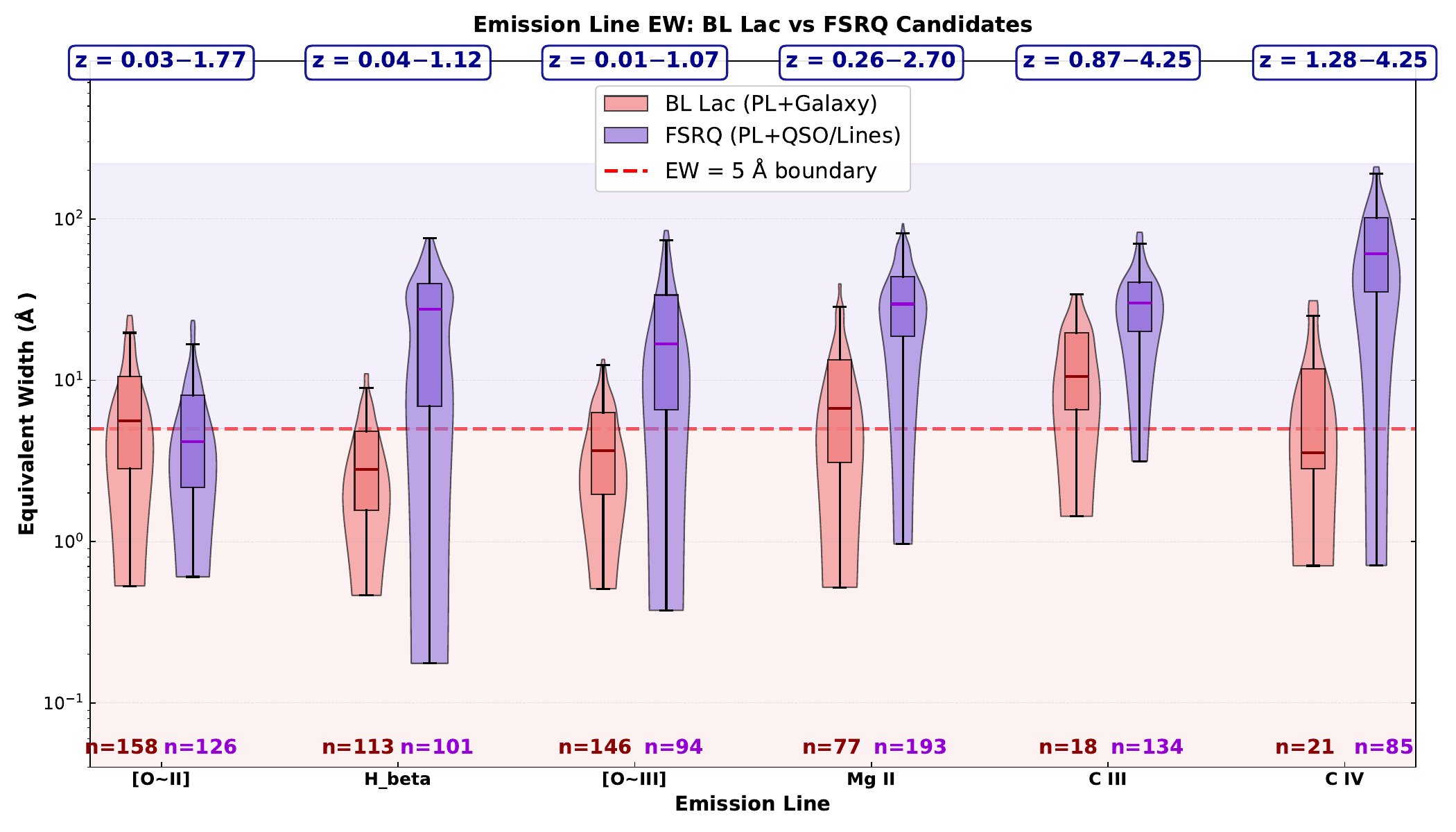}
    \caption{Emission line EW distributions for BL Lac (PL+Galaxy; ${\rm N}=18$ to $168$ per line) and FSRQ (PL+QSO/Lines; ${\rm N= 85-193}$ per line) candidates after iterative $3\sigma$ clipping. The violin plots show kernel density estimates of the EW distribution, with overlaid box plots, which shows median (thick line), interquartile range (box), and ${\rm 1.5\times IQR ~whiskers}$. Sample sizes (n) are shown below each distribution. The horizontal red dashed line marks the traditional ${\rm EW} = 5$\,\AA\ classification boundary. Redshift ranges (blue boxes) indicate the observable redshift range for each spectral feature: C IV, C III are only detectable at high-z ($z > 0.86$) while lines [O II], H-beta, [O III] are detected at low-z ($z < 1.8$). The strongest contrast appears in H$\beta$ (BL Lac median $\approx 2.81$~\AA\ vs FSRQ median $\approx 27.48$~\AA) and [O~III] (BL Lac median $\approx 4.10$~\AA\ vs FSRQ median $\approx 17.77$~\AA).}
    \label{fig:em_comp}
\end{figure*}

Figure~\ref{fig:em_comp} shows the distribution of the EW measurements for any detected emission lines within both the FSRQ and BL Lac candidate samples. The distributions reflect observational selection effects: ultraviolet lines (C~IV, C~III]) are only accessible at high redshift ($z > 0.87$) when redshifted into the optical window, while optical lines ([O~II], H$\beta$, [O~III]) are detectable at lower redshifts ($z < 1.8$). Despite these selection effects, a clear separation is observed between BL Lac and FSRQ populations across all emission lines. FSRQ candidates exhibit stronger emission lines with an overall mean EW of $22.86 \pm 0.86$~\AA\ (median $26.44$\,\AA) across all lines, with the strongest contrast in C~IV (BL Lac median $\approx 3.57$~\AA\ vs FSRQ median $\approx 60.62$~\AA) and H$\beta$ (BL Lac median $\approx 2.81$~\AA\ vs FSRQ median $\approx 27.48$~\AA). When an emission line is detected in an FSRQ candidate, it tends to exceed the canonical 5\,\AA\ threshold in the majority of cases, confirming that the PL+QSO/line template fits tend to correspond to traditional FSRQ classifications (although there is often a tail to lower EWs, most notably in the [O~II] line that is detectable at lower redshifts).

\begin{figure*}
    \centering
    \includegraphics[width=0.9\textwidth]{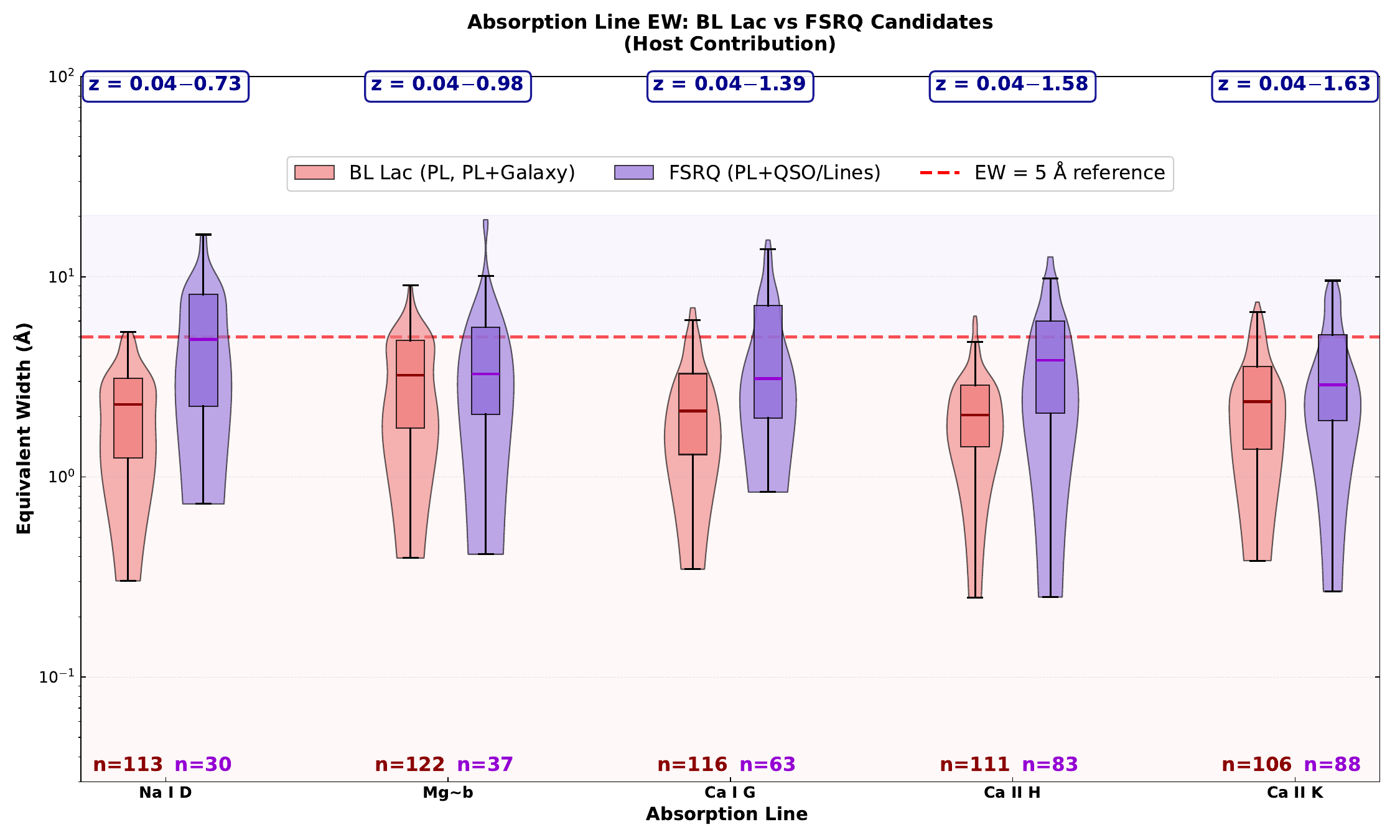}
    \caption{Absorption line equivalent width distributions for the same BL Lac candidate (PL+Galaxy) and FSRQ candidate (PL+QSO/Lines) samples after iterative $3\sigma$ clipping. Violin and box plot formatting follows Figure~\ref{fig:em_comp}. Unlike emission lines, absorption features (Ca~II K/H, Ca~I G, Mg~b, Na~I D) show substantial overlap between BL Lac and FSRQ populations, with both classes exhibiting weak absorption (medians $\sim 2$--$4$~\AA) characteristic of host galaxy stellar populations in massive elliptical galaxies. The redshift ranges indicate that Na~I D and Mg~b are only observable at $z < 1.0$, while Ca~I G and the Ca~II H+K doublet are detectable up to $z \approx 1.39$ and $z \approx 1.59$ respectively. The similarity in absorption feature strengths between the two populations confirms that host galaxy stellar contributions are present in both blazar classes, and that the BL Lac/FSRQ classification is driven by the broad-line region emission properties rather than the host galaxy absorption \citep{stickel1991complete}.}
    \label{fig:ab_comp}
\end{figure*}

Figure~\ref{fig:ab_comp} shows the absorption line distributions. Both populations show broadly similar absorption feature strengths, with mean EW values below $|5|$\,\AA\ for all lines in both classes. BL Lac candidates show absorption mean EW ranging from $2.25 \pm 0.11$\,\AA\ (median $\approx 2.04$; Ca~II H) to $3.38 \pm 0.18$\,\AA\ (median $\approx 3.22$; Mg~b), with $> 79\%$ of detections below 5\,\AA\ for all lines. FSRQ candidates show slightly higher absorption EWs --- particularly Na~I D with mean $5.50 \pm 0.70$\,\AA\ (median $\approx 4.86$; $50.0\%$ above 5\,\AA) --- consistent with more luminous host galaxies at higher redshifts contributing detectable stellar features despite the jet continuum. The overlap in absorption properties between the two populations confirms that host-galaxy presence alone does not distinguish BL Lacs from FSRQs; classification is driven by the BLR emission-line properties.

\begin{figure*}
    \centering
    \includegraphics[width=0.85\textwidth]{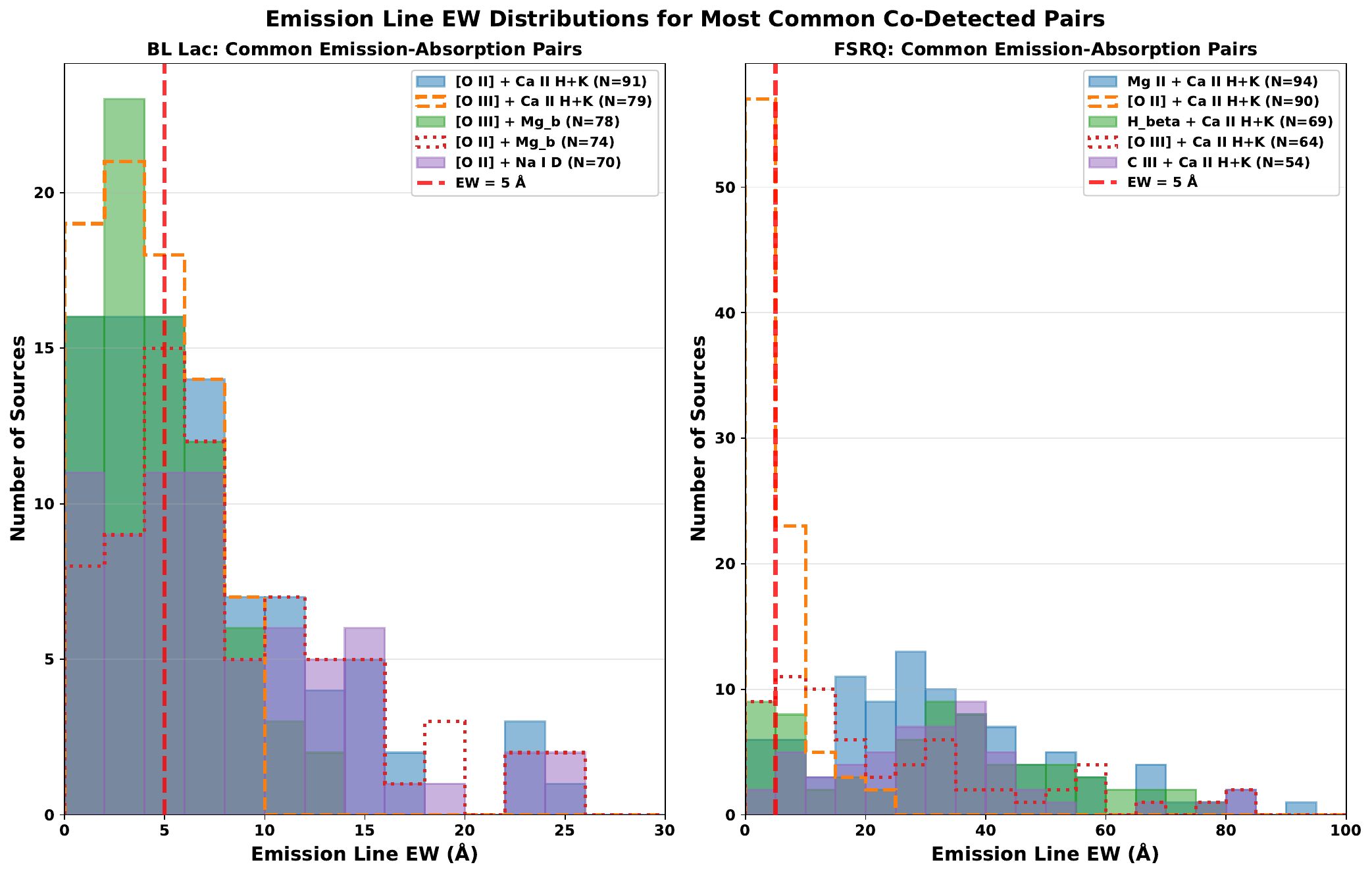}
    \caption{Emission line EW distributions for the five most common co-detected emission-absorption pairs in BL Lac candidates (left panel) and FSRQ candidates (right panel), using $3\sigma$ clipped measurements for consistency with Table~\ref{tab:ew_comparison}. Each histogram shows the distribution of emission line EW values for sources where both the labelled emission and absorption lines are simultaneously detected. The red dashed vertical line marks the $|{\rm EW}| = 5$\,\AA\ classification boundary \citep{stickel1991complete}. For BL Lac candidates, the dominant pairs involve low-ionisation NLs ([O~III], [O~II]) co-detected with host galaxy absorption (Mg~b, Ca~II H+K), with emission EW distributions broadly peaking below 5 \AA\ . For FSRQ candidates, Mg~II and [O~II] dominate the co-detected pairs, with distributions extending well above 5\,\AA\ confirming active broad-line region emission alongside detectable host galaxy absorption.}
    \label{fig:both_detection} 
\end{figure*}

Figure~\ref{fig:both_detection} shows the emission line EW distribution for the most common co-detected emission-absorption pairs for both BL Lac and FSRQ candidates. The dominant co-detected pairs in BL Lac candidates involve low-ionisation NLs ([O~III], [O~II]) with host galaxy absorption (Mg~b, Ca~II H+K), with emission EW distributions broadly peaking below 5\,\AA. For FSRQ candidates, Mg~II and [O~II] dominate the co-detected pairs with distributions extending well above 5\,\AA. The simultaneous presence of both emission and absorption features in $51.4\%$ of BL Lac candidates and $52.6\%$ of FSRQ candidates demonstrates that both populations retain measurable host galaxy contributions despite jet continuum dilution. Among PL+Galaxy sources, emission lines are detected in only a fraction of sources despite being within the observable redshift window for a given line. For example, Mg~II is detected in only 77 of the 256 PL+Galaxy sources within its observable redshift range ($z = 0.26$--$2.70$; $30.1\%$), and when detected, $\sim 58.4\%$ show $|{\rm EW}| > 5$\,\AA\ (mean $9.63 \pm 1.01$\,\AA), suggesting these sources sit near or above the traditional BL Lac/FSRQ classification boundary.

\begin{figure*}
    \centering
    \includegraphics[width=0.85\textwidth]{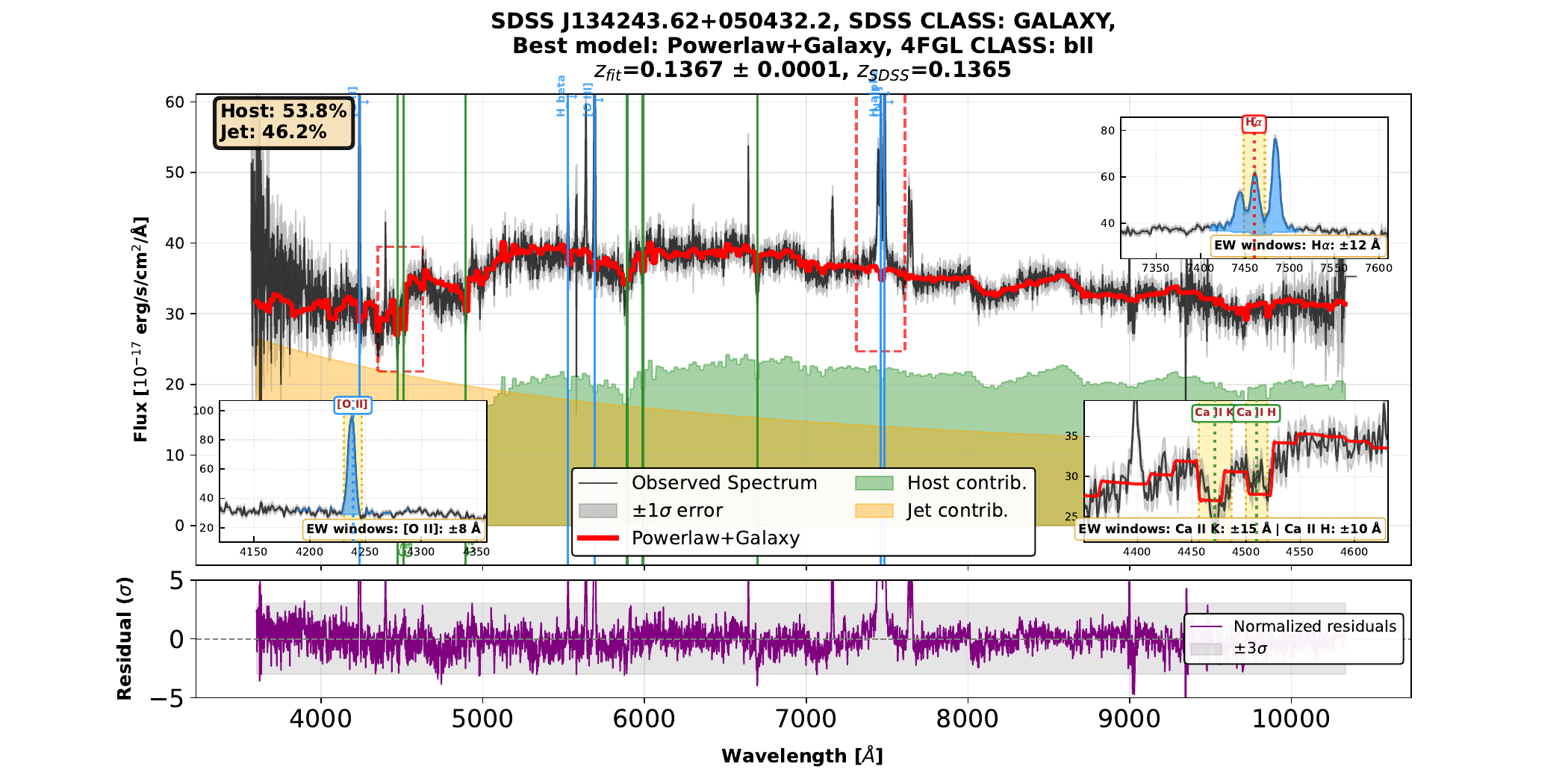}
    \caption{A representative \texttt{Powerlaw+Galaxy} source at $z = 0.1366$ with simultaneous emission and absorption line detections. Strong [O\,II] ($|{\rm EW}| = 16.76 \pm 0.53$\,\AA, S/N $= 31.5$) and a strong 
    H$\alpha$ emission line ($|{\rm EW}| = 8.93 \pm 0.18$\,\AA, S/N $= 48.3$) within the H$\alpha$-[NII] complex doublet,
        both well above the 5\,\AA\ classification boundary, coexist with host galaxy Ca\,II\,K/H absorption, highlighting the complexity of applying the traditional EW criterion to blazars with strong jets.}
    \label{fig:em_abs}
\end{figure*}

It is important to note an aspect of the BL Lac emission line detection that reveals a genuine complexity in the traditional classification criterion: in a BL Lac candidate that is fitted with the PL+Galaxy template (adopting elliptical galaxy templates without emission lines), we are also able to detect a prominent emission line. Crucially, when an emission line is detected in a PL+Galaxy source, it frequently exceeds the $|{\rm EW}| = 5$\,\AA\ threshold (see Figure~\ref{fig:em_abs}). This pattern -- infrequent detection but high EW when detected -- is consistent across other lines: [O~III] is detected in 161 PL+Galaxy sources with $42.9\%$ exceeding 5\,\AA, while [O~II] is detected in 168 sources with $61.3\%$ exceeding 5\,\AA. These cases likely represent a combination of masquerading BL Lacs \citep{padovani2019txs} and sources near the classification boundary where the EW criterion is inherently ambiguous, consistent with \citet{ghisellini2011transition} who argue that a small EW does not necessarily imply intrinsically weak emission lines, being instead the result of a particularly beamed non-thermal continuum at the epoch of observation. Figure~\ref{fig:em_abs} presents an exemplary PL+Galaxy source at $z = 0.1366$ where strong [O~II] ($|{\rm EW}| = 16.76 \pm 0.53$\,\AA, S/N $= 31.5$) and H$\alpha$ ($|{\rm EW}| = 8.93 \pm 0.18$\,\AA, S/N $= 48.3$, within the H$\alpha$-[NII] complex doublet) emission lines exceeding the $5$\,\AA\ threshold are simultaneously detected alongside host galaxy absorption features, illustrating the complexity of applying the traditional EW criterion to blazars with strong jets contributing to their optical spectra.


\begin{table}
\centering
\caption{Equivalent width measurements for commonly detected spectral features in BL~Lac candidates (PL+galaxy models) and FSRQ candidates (PL+QSO/Lines models) after $3\sigma$ clipping. For all detected lines (absorption and emission), BL~Lac candidates show systematically weak features (${\rm EW_{mean} = 4.47 \pm 0.15}$~\AA) with $66.7\%$ below 5~\AA, while FSRQ candidates exhibit strong broad emission lines (${\rm EW_{mean} = 22.86\pm0.86}$~\AA) with $83.9\%$ exceeding 5~\AA. Values are reported as mean $\pm$ SEM (Standard Error of the Mean.)}
\label{tab:ew_comparison}
\begin{tabular}{lcccc}
\hline
Line & Line Type & $N_{\rm ret}$ & Mean EW (\AA) & Range (\AA) \\
\hline \\[3pt]
\multicolumn{5}{c}{\textbf{BL Lac Candidates (PL+Galaxy)}} \\
\hline
Ca I G    & Abs & 116 & $2.52 \pm 0.14$ & $0.35$--$7.00$ \\
Ca II H   & Abs & 111 & $2.25 \pm 0.11$ & $0.25$--$6.36$ \\
Ca II K   & Abs & 106 & $2.70 \pm 0.16$ & $0.38$--$7.47$ \\
Mg~b      & Abs & 122 & $3.38 \pm 0.18$ & $0.39$--$9.08$ \\
Na I D    & Abs & 113 & $2.30 \pm 0.12$ & $0.30$--$5.30$ \\
C III]    & Em  &  18 & $13.27 \pm 2.16$ & $1.43$--$33.98$ \\
C IV      & Em  &  21 & $8.93 \pm 2.03$ & $0.71$--$31.13$ \\
H$\beta$  & Em  & 113 & $3.48 \pm 0.23$ & $0.46$--$10.97$ \\
Mg II     & Em  &  77 & $9.63 \pm 1.01$ & $0.52$--$39.51$ \\
{[O III]} & Em  & 146 & $4.48 \pm 0.25$ & $0.51$--$13.46$ \\
{[O II]}  & Em  & 158 & $7.64 \pm 0.49$ & $0.53$--$25.21$ \\
\hline \\ [2pt]
\multicolumn{5}{c}{\textbf{FSRQ Candidates (PL+QSO/Line)}} \\
\hline
C III]    & Em  & 134 & $32.04 \pm 1.55$ & $3.15$--$82.66$ \\
C IV      & Em  &  85 & $72.09 \pm 5.49$ & $0.71$--$210.15$ \\
H$\beta$  & Em  & 101 & $27.53 \pm 1.98$ & $0.18$--$76.21$ \\
Mg II     & Em  & 193 & $32.06 \pm 1.41$ & $0.97$--$93.42$ \\
{[O III]} & Em  &  94 & $23.07 \pm 2.17$ & $0.37$--$84.72$ \\
{[O II]}  & Em  & 126 & $5.99  \pm 0.47$ & $0.60$--$23.46$ \\
Ca I G    & Abs &  63 & $4.71 \pm 0.45$ & $0.84$--$15.27$ \\
Ca II H   & Abs &  83 & $4.44 \pm 0.32$ & $0.25$--$12.56$ \\
Ca II K   & Abs &  88 & $3.76 \pm 0.26$ & $0.27$--$9.55$ \\
Mg~b      & Abs &  37 & $4.53 \pm 0.67$ & $0.41$--$19.26$ \\
Na I D    & Abs &  30 & $5.50 \pm 0.70$ & $0.73$--$16.23$ \\
\hline
\end{tabular}
\end{table}

\section{Conclusions}
\label{sec:conclusions}
In this work, we have developed a physically motivated multi-component spectral pipeline to accurately classify and characterise blazar candidates within SDSS-V DR20 that have been systematically misclassified by the SDSS automated pipeline. Cross-matching the \textit{Fermi}/4FGL-DR4 catalogue \citep{ballet2023fermi} with the SDSS-V DR20 yields 746 optical counterparts with available spectroscopy obtained through SDSS-V. After excluding 14 sources with bad fits ($\chi^2_{\rm r,\,lmfit} > 3\times\chi^2_{\rm r,\,SDSS}$) and 25 non-blazar \textit{Fermi} classes, our final sample comprises 707 well-fitted spectra. Our pipeline employs six model families combining power-law, elliptical Galaxy, and synthetic QSO templates \citep{polletta2007spectral, temple2021modelling} with optional emission line components, yielding 420 sources (59.4\%) best fitted by PL+Galaxy models, 160 (22.6\%) by PL+Lines, 106 (15.0\%) by PL+QSO, and 21 by single-component models. This distribution directly shows the fundamental limitation of the SDSS automated pipeline \citep{bolton2012spectral}, which fits only stellar, galaxy, or QSO templates without explicitly introducing a non-thermal jet continuum component, leading to the systematic misclassification of jet-dominated sources as galactic stars, galaxies, or typical quasars.

Independent validation of the reclassifications using the \textit{WISE} colour-colour diagram (see Section~\ref{bl_strip}, Figure~\ref{fig:blazar_stripe}) confirms that the overwhelming majority of our BL Lac candidates (PL only, PL+Galaxy) and FSRQ candidates (PL+QSO, PL+Lines) occupy the canonical blazar region defined by \citet{massaro2011identification} and \citet{d2012infrared}. Of the 671 sources with valid \textit{WISE} magnitudes, the most striking result is among the 88 SDSS STAR-classified sources: 96.6\% (85/88) fall within the blazar strip, with 90.9\% reclassified as BL Lac candidates --- sources misidentified as Galactic stars by the SDSS pipeline likely due to their featureless power-law continua. The BL Lac candidates tend to occupy slightly bluer \textit{WISE} colours than FSRQ candidates, consistent with stronger host galaxy contributions and a lack of significant dusty torus emission \citep{plotkin2012lack}, pointing towards radiatively inefficient accretion in these jet-dominated sources \citep{narayan1994advection, heckman2014coevolution}.
 
We evaluate the redshift accuracy of our new blazar-specific pipeline using the 111 sources with prior spectroscopic redshifts in the Third Catalog of Hard \textit{Fermi}-LAT Sources \citep[3FHL;][]{ajello20173fhl} that satisfy both the quality threshold ($\chi^2_{\rm r,\,lmfit} \leq 3\times\chi^2_{\rm r,\,SDSS}$) and the spectral ${\rm S/N} \geq 3$ criterion (see Section~\ref{z_stats}, Figure~\ref{fig:z_test}). On the full validation sample, our multi-component pipeline achieves an accuracy of $\sigma_{\rm NMAD} = 0.0875$ and a catastrophic failure rate of $\eta = 0.387$. Our accuracy is comparable to adopting the SDSS pipeline redshifts for the same validation sample, but the rate of catastrophic failures is around 10\% lower (counting both discrepant redshifts and stellar classifications as failures in SDSS).

For the 51 sources (45.9\%) where both methods independently converge on consistent redshifts, both achieve excellent accuracy ($\sigma_{\rm NMAD} = 0.0010$, $\eta = 0.098$), confirming that the blazar-specific decomposition introduces no systematic redshift biases. A more physically motivated assessment is provided by the stellar misclassification failure rate: 21 of the 111 validation sources (18.9\%) are classified as \texttt{STAR} by the SDSS pipeline, representing cases where SDSS assigns $z \approx 0$ to genuine extragalactic blazar candidates. Our multi-component pipeline provides a physically motivated classification for all 21 of these sources, reclassifying them as BL Lac candidates with well-constrained PL+Galaxy fits. Of these 21 sources, the multi-component pipeline recovers a redshift consistent with the 3FHL reference ($\Delta z/(1+z) < 0.1$) for 5 sources. For the remaining 16, all best fitted by the PL+Galaxy model with high-quality spectral decompositions, the disagreement with the 3FHL reference value does not necessarily reflect a failure of our pipeline given that the 3FHL redshifts for this specific subset cannot be taken for granted.

The redshift and $\gamma$-ray luminosity distributions confirm that the model-based classification recovers the established separation between blazar populations (see Section~\ref{lumino}, Figure~\ref{fig:lum}). BL Lac candidates show a median $z = 0.360$ and median $L_\gamma = 1.39\times10^{45}$\,erg\,s$^{-1}$, while FSRQ candidates exhibit a median $z = 1.026$ and median $L_\gamma = 4.23\times10^{46}$\,erg\,s$^{-1}$, a separation of ${\sim}1.5$\,dex in median $\gamma$-ray luminosity, consistent with established $\gamma$-ray luminosity functions \citep{ajello2014cosmic, ajello2012luminosity}. Separating FSRQ candidates by model subtype reveals physically distinct jet fraction distributions: PL+QSO sources show a median $f_{\rm jet} = 0.618$, reflecting genuine thermal continuum contributions from the QSO template, while PL+Lines sources show a median $f_{\rm jet} = 0.845$, interpreted as an upper limit in the absence of an explicit thermal template. BL Lac candidates show a median $f_{\rm jet} = 0.819$. The presence or absence of thermal accretion disc and broad-line region emission --- captured by the PL+QSO and PL+Lines models --- provides the fundamental physical distinction between the two subclasses \citep{ghisellini2011transition, ghisellini2017fermi}.

Analysis of the traditional equivalent width classification criterion (Section~\ref{ew_comparison}) shows that our multi-component spectral classification scheme robustly validates the traditional $|{\rm EW}| = 5$\,\AA\ boundary at the population level: BL Lac candidates show a mean EW of $4.47\pm 0.15$\,\AA\ across all detected lines compared to $22.86\pm0.86$\,\AA\ for FSRQ candidates, a factor of ${\sim}5.1$ separation, with 66.7\% of BL Lac detections below 5\,\AA\ and 83.9\% of FSRQ detections above it \citep{stickel1991complete}. However, the boundary is not a sharp physical threshold: 49.5\% of BL Lac candidates (208/420) have at least one detected emission line with $|{\rm EW}| > 5$\,\AA, predominantly [O II] or [O III], while only 3.4\% of FSRQ candidates (9/266) have all detected emission lines below the threshold, suggesting that these may represent obscured or low-accretion systems where the broad-line region contribution is temporarily suppressed \citep{ghisellini2011transition}.

A particularly significant finding emerges from the simultaneous detection of both emission and absorption features in a substantial fraction of both populations. Of the 350 BL Lac candidates with any detected spectral features ($83.3\%$ of the full BL Lac sample), 216 ($51.4\%$) show simultaneous emission and absorption line detections and of the 251 FSRQ candidates with detected features ($94.7\%$), 140 ($55.8\%$) show both. These hybrid sources --- exhibiting host galaxy stellar absorption alongside emission lines characteristic of an active broad-line region --- cannot be adequately described by a purely binary BL Lac/FSRQ classification. They likely represent a physically diverse population including masquerading BL Lacs \citep{padovani2019txs}, transitional sources where accretion state changes drive spectral evolution \citep{ghisellini2011transition}, and objects near the classification boundary where the traditional criterion is inherently ambiguous \citep{ghisellini2011transition}. The existence of this hybrid population motivates the need to move beyond a strict binary classification and towards a more physically motivated, continuous description of blazar properties.

This work shows that bespoke, physically motivated spectral templates are essential for the reliable classification of blazars in large scale optical surveys such as SDSS-V. The automated multi-component fitting approach developed in this work successfully recovers non-thermal jet contributions, host galaxy properties, and accretion disc signatures across a diverse sample of 707 \textit{Fermi}-detected sources, yielding classifications that broadly follow the traditional BL Lac/FSRQ division while revealing a more continuous range of physical properties within the blazar population. The identification of a substantial population of hybrid sources -- showing signatures of both host galaxy stellar populations and radiatively efficient accretion -- points towards the need for future work incorporating multi-epoch spectroscopy and broadband SED modelling to fully characterise the accretion and jet properties of these transitional objects. The pipeline developed here is directly scalable to the full SDSS-V footprint and future large spectroscopic surveys, offering a robust foundation for blazar population studies in the era of next-generation multi-wavelength facilities.

\section*{Acknowledgements}
MIN acknowledges support from the Development in Africa with Radio Astronomy (DARA) project through an MSc by Research studentship.
JA acknowledges support from a UKRI Future Leaders Fellowship (grant code: MR/Y019539/1).
EK thanks DARA for travel support to engage in this work.
ALR acknowledges support from a Leverhulme Early Career Fellowship.
LHG acknowledges financial support from ANID program FONDECYT Iniciaci\'on 11241477.

Funding for the Sloan Digital Sky Survey V has been provided by the Alfred P. Sloan Foundation, the Heising-Simons Foundation, the National Science Foundation, and the Participating Institutions. SDSS acknowledges support and resources from the Center for High-Performance Computing at the University of Utah. SDSS telescopes are located at Apache Point Observatory, funded by the Astrophysical Research Consortium and operated by New Mexico State University, and at Las Campanas Observatory, operated by the Carnegie Institution for Science. The SDSS website is \url{www.sdss.org}.

SDSS is managed by the Astrophysical Research Consortium for the Participating Institutions of the SDSS Collaboration, including the Carnegie Institution for Science, Chilean National Time Allocation Committee (CNTAC) ratified researchers, Caltech, the Gotham Participation Group, Harvard University, Heidelberg University, The Flatiron Institute, The Johns Hopkins University, L'Ecole polytechnique f\'{e}d\'{e}rale de Lausanne (EPFL), Leibniz-Institut f\"{u}r Astrophysik Potsdam (AIP), Max-Planck-Institut f\"{u}r Astronomie (MPIA Heidelberg), Max-Planck-Institut f\"{u}r Extraterrestrische Physik (MPE), Nanjing University, National Astronomical Observatories of China (NAOC), New Mexico State University, The Ohio State University, Pennsylvania State University, Smithsonian Astrophysical Observatory, Space Telescope Science Institute (STScI), the Stellar Astrophysics Participation Group, Universidad Nacional Aut\'{o}noma de M\'{e}xico, University of Arizona, University of Colorado Boulder, University of Illinois at Urbana-Champaign, University of Toronto, University of Utah, University of Virginia, Yale University, and Yunnan University. 

\section*{Data Availability}

All of the data used for this paper are available through the SDSS Data Release 20, {\bf link to be added},
which includes a Value-added Catalogue (VAC) that provides the bespoke blazar spectral fitting results for the 707 sources examined here. In addition, an example python notebook showing how to access the results along with further metadata to present the spectral results is available here {\bf link to be added}.

\clearpage
\appendix
\onecolumn
\section{Tables providing final source classifications}
\label{append:tables}
 
Appendix~\ref{append:tables} presents the full spectral classification and redshift results for all sources in the sample. 
Table~\ref{app:good_table} lists the first 20 of the 707 sources 
passing the quality threshold 
($\chi^2_{\rm r,\,lmfit} \leq 3 \times \chi^2_{\rm r,\,SDSS}$) 
with confirmed \textit{Fermi} blazar classifications (the full table of 707 sources is available online). 
Table~\ref{tab:sdss_better} presents the 14 sources where the 
SDSS classification is retained due to poor multi-component model 
performance. Table~\ref{tab:fermi_sources} lists the 25 sources 
with non-blazar \textit{Fermi} classifications excluded from the 
final blazar sample. All tables include the following columns:
 
\begin{itemize}
 
    \item \textbf{SDSS ID} --- the unique SDSS source identifier.
 
    \item \textbf{SDSS Name} --- the full SDSS coordinate-based 
    source name.
 
    \item \textbf{4FGL Name} --- the \textit{Fermi}/4FGL-DR4 
    catalogue designation \citep{ballet2023fermi}.
 
    \item \textbf{4FGL Class} --- the \textit{Fermi} classification 
    from the 4FGL-DR4 catalogue.
 
    \item \textbf{Best Model} --- the best-fit multi-component model 
    family selected by the Akaike Information Criterion with 
    correction for finite sample size (AICc; 
    Section~\ref{sec: 3}). Model labels are abbreviated as: 
    PL+Gal (Powerlaw+Galaxy), PL+QSO (Powerlaw+QSO), PL+Lines 
    (Powerlaw+Lines), and PL (Powerlaw only). Where applicable, 
    the specific template variant is given in parentheses.
 
    \item \textbf{SDSS Class} --- the original SDSS spectroscopic 
    classification (STAR, GALAXY, or QSO).
 
    \item \textbf{$z_{\rm SDSS}$} --- the redshift reported by the 
    SDSS automated pipeline.
 
    \item \textbf{$z_{\rm fit}$} --- the multi-component redshift 
    estimate with its associated $1\sigma$ uncertainty. PL+Galaxy 
    sources beyond $z \gtrsim 1.8$ have low-confidence redshift 
    estimates: at these redshifts the observed spectra are largely 
    featureless and the SDSS wavelength coverage provides no strong 
    rest-frame anchors for the elliptical galaxy template, making 
    the redshift determination unreliable (Section~\ref{jet_z_dis}). 
    These sources are indicated by $^{\dagger}$ in the tables.
 
    \item \textbf{$\chi^2_{\rm r,\,lmfit}$} --- the reduced 
    chi-squared of the best multi-component model fit.
 
    \item \textbf{$\chi^2_{\rm r,\,SDSS}$} --- the reduced 
    chi-squared of the SDSS pipeline fit.
 
    \item \textbf{Jet Fraction} --- the optical jet fraction 
    ($f_{\rm jet}$) derived from the multi-component spectral 
    decomposition, representing the fractional contribution of the 
    power-law jet continuum to the total optical flux. For sources 
    best fitted by the PL+Lines model family, this quantity is 
    presented as an upper limit (indicated by $<$) on the true 
    non-thermal contribution, as the power-law component may absorb 
    a mixture of jet synchrotron and smooth accretion-disc continuum 
    emission in the absence of an explicit thermal template 
    (Section~\ref{sec:decomp} and Section~\ref{jet_z_dis}).
 
    \item \textbf{$\alpha$} --- the power-law spectral slope of the 
    jet continuum component.
 
    \item \textbf{$\delta$} --- the power-law curvature parameter.
 
    \item \textbf{S/N} --- the median spectral signal-to-noise ratio 
    per pixel as reported by the SDSS pipeline. Sources with 
    ${\rm S/N} < 3$ are retained in the sample but their fitted 
    parameters carry greater uncertainty and are excluded from all 
    population median calculations (Section~\ref{jet_z_dis}). 
    They are indicated by $^{\ddagger}$ in the tables.
 
\end{itemize}

\begin{landscape}
{\scriptsize \begin{longtable}{llllllllllllll}
\caption{4FGL sources with SDSS-V optical spectroscopy where the blazar multi-component models provide improved \\($\chi^2_{\rm r,lmfit} < 3 \times \chi^2_{\rm r,SDSS}$).
\label{app:good_table}
}\\

\hline
SDSS ID & SDSS Name & 4FGL Name & 4FGL Class & Best Model & SDSS Class & $z_{\rm SDSS}$ & $z_{\rm fit}$ & $\chi^2_{\rm r,\,lmfit}$ & $\chi^2_{\rm r,\,SDSS}$ & Jet Fraction & $\alpha$ & $\delta$ & S/N \\
\hline
\endfirsthead

\multicolumn{14}{c}{\textit{Continued from previous page}} \\
\hline
SDSS ID & SDSS Name & 4FGL Name & 4FGL Class & Best Model & SDSS Class & $z_{\rm SDSS}$ & $z_{\rm fit}$ & $\chi^2_{\rm r,\,lmfit}$ & $\chi^2_{\rm r,\,SDSS}$ & Jet Fraction & $\alpha$ & $\delta$ & S/N \\
\hline
\endhead

\hline
\multicolumn{14}{r}{\textit{Continued on next page}} \\
\endfoot

\hline
\multicolumn{14}{l}{$^{\dagger}$ Low-confidence redshift: PL+Galaxy source at $z \gtrsim 1.8$ where the SDSS wavelength coverage provides no strong rest-frame anchors for the galaxy template.} \\
\multicolumn{14}{l}{$^{\ddagger}$ Low spectral S/N ($<3$): fitted parameters carry greater uncertainty and are excluded from population median calculations.} \\
\multicolumn{14}{l}{$<$ Jet fraction upper limit: PL+Lines sources where the power-law component may absorb a mixture of jet synchrotron and accretion-disc continuum emission.} \\
\endlastfoot

13362106 & SDSS J083251.50+330010.1 & 4FGL J0833.0+3300 & bll & Powerlaw+Galaxy & GALAXY & 0.6718 & $0.6739 \pm 0.0002$ & 0.890 & 0.864 & 0.295 & 0.519 & 1.200 & 2.3$^{\ddagger}$ \\
15606475 & SDSS J184806.09-423025.2 & 4FGL J1848.1-4230 & bcu & Galaxy & STAR & -0.0001 & $0.3927 \pm 0.0002$ & 0.771 & 0.745 & -- & -- & -- & 1.7$^{\ddagger}$ \\
20570296 & SDSS J074554.01-004417.4 & 4FGL J0746.0-0039 & fsrq & Powerlaw+Lines (High Blueshift Line) & QSO & 0.9997 & $0.9961 \pm 0.0002$ & 0.728 & 0.727 & $< 0.907$ & -1.297 & -1.200 & 10.5 \\
32622057 & SDSS J134105.15+395945.6 & 4FGL J1341.2+3958 & bll & Powerlaw+Galaxy & GALAXY & 0.1716 & $0.1714 \pm 0.0000$ & 1.030 & 0.763 & 0.610 & -0.117 & -1.200 & 29.0 \\
54748037 & SDSS J022324.13+541333.3 & 4FGL J0223.5+5415 & bcu & Powerlaw+Galaxy & GALAXY & 0.5943 & $0.2285 \pm 0.0002$ & 0.685 & 0.666 & 0.878 & -0.270 & 1.200 & 11.6 \\
54984385 & SDSS J042525.35+632001.7 & 4FGL J0425.3+6319 & bcu & Powerlaw+Galaxy & QSO & 0.8152 & $1.4310 \pm 0.0004$ & 0.781 & 0.754 & 0.552 & 1.200 & 1.199 & 1.6$^{\ddagger}$ \\
55028445 & SDSS J050756.17+673724.2 & 4FGL J0507.9+6737 & bll & Powerlaw+Galaxy & GALAXY & 0.0895 & $0.1274 \pm 0.0001$ & 1.927 & 1.545 & 0.668 & -0.966 & 1.200 & 20.9 \\
55088509 & SDSS J033356.74+653656.2 & 4FGL J0333.9+6537 & bll & QSO (Red z=0.5 (ebv = 0.2)) & QSO & 0.8879 & $0.9820 \pm 0.0019$ & 0.743 & 0.735 & -- & -- & -- & 1.1$^{\ddagger}$ \\
55122816 & SDSS J042650.07+682552.9 & 4FGL J0426.7+6826 & bcu & Powerlaw+Galaxy & STAR & 0.0001 & $2.2653 \pm 0.0008$$^{\dagger}$ & 1.088 & 1.059 & 0.910 & 1.022 & -1.053 & 4.1 \\
55156843 & SDSS J051631.23+735108.7 & 4FGL J0516.4+7350 & bll & Powerlaw+Galaxy & GALAXY & 0.2512 & $0.2382 \pm 0.0002$ & 0.820 & 0.813 & 0.481 & -0.528 & 1.200 & 4.4 \\
55501941 & SDSS J001131.90+704531.6 & 4FGL J0012.0+7043 & bcu & Powerlaw+Galaxy & GALAXY & 0.1939 & $3.4242 \pm 0.0004$$^{\dagger}$ & 0.993 & 0.869 & 0.662 & 1.200 & 1.200 & 3.9 \\
55606068 & SDSS J024330.89+712017.9 & 4FGL J0243.4+7119 & bll & Powerlaw+Galaxy & GALAXY & 0.9582 & $0.1337 \pm 0.0002$ & 0.790 & 0.772 & 0.724 & 1.111 & -0.047 & 7.2 \\
55754756 & SDSS J014935.26+860115.5 & 4FGL J0151.3+8601 & bll & Powerlaw+Galaxy & GALAXY & 0.1547 & $3.2778 \pm 0.0708$$^{\dagger}$ & 1.840 & 1.227 & 1.000 & 1.034 & 1.200 & 9.5 \\
55759307 & SDSS J085920.56+004712.2 & 4FGL J0859.2+0047 & bll & Powerlaw+Galaxy & GALAXY & 0.9026 & $0.9040 \pm 0.0004$ & 0.853 & 0.845 & 0.952 & -1.539 & 0.107 & 13.7 \\
55761530 & SDSS J085749.80+013530.3 & 4FGL J0857.7+0137 & bll & Powerlaw+Galaxy & GALAXY & 0.2813 & $0.2810 \pm 0.0001$ & 1.386 & 1.149 & 0.833 & -0.079 & -1.200 & 26.3 \\
55840060 & SDSS J093239.36+104235.2 & 4FGL J0932.7+1041 & bll & Powerlaw+Galaxy & GALAXY & 0.3615 & $0.3622 \pm 0.0001$ & 0.807 & 0.764 & 0.657 & -0.563 & -1.197 & 12.5 \\
55847533 & SDSS J090700.09+085751.4 & 4FGL J0907.1+0856 & bll & Powerlaw+Galaxy & GALAXY & 0.4452 & $0.4458 \pm 0.0006$ & 0.757 & 0.747 & 0.813 & -1.308 & 1.200 & 3.2 \\
55897122 & SDSS J084712.93+113350.2 & 4FGL J0847.2+1134 & bll & Powerlaw+Galaxy & GALAXY & 0.1984 & $0.1990 \pm 0.0000$ & 0.994 & 0.883 & 0.719 & -1.082 & -0.195 & 30.9 \\
55902603 & SDSS J082930.32+085821.3 & 4FGL J0829.4+0857 & fsrq & Powerlaw+Lines (High Blueshift Line) & QSO & 0.8455 & $0.8455 \pm 0.0003$ & 1.126 & 1.065 & $< 0.782$ & 0.490 & 1.200 & 2.6$^{\ddagger}$ \\
55909650 & SDSS J082054.81+100609.4 & 4FGL J0821.1+1007 & bcu & Powerlaw+Lines (Median Line) & QSO & 0.9518 & $0.9523 \pm 0.0004$ & 0.985 & 0.940 & $< 0.420$ & 1.182 & 0.061 & 1.2$^{\ddagger}$ \\
\thispagestyle{empty}
\end{longtable}

} 

{\scriptsize \begin{longtable}{llllllllllllll}
\caption{The 14 4FGL sources with SDSS-V optical spectroscopy where we retain the SDSS pipeline classification as it produces a substantially better fit than our blazar multi-component analysis ($\chi^2_{\rm r,SDSS} < 2 \times \chi^2_{\rm r,lmfit}$)} \label{tab:sdss_better} \\
\hline
SDSS ID & SDSS Name & 4FGL Name & 4FGL Class & Best Model & SDSS Class & $z_{\rm SDSS}$ & $z_{\rm fit}$ & $\chi^2_{\rm r,\,lmfit}$ & $\chi^2_{\rm r,\,SDSS}$ & Jet Fraction & $\alpha$ & $\delta$ & S/N \\
\hline
\endfirsthead

\multicolumn{14}{c}{\textit{Continued from previous page}} \\
\hline
SDSS ID & SDSS Name & 4FGL Name & 4FGL Class & Best Model & SDSS Class & $z_{\rm SDSS}$ & $z_{\rm fit}$ & $\chi^2_{\rm r,\,lmfit}$ & $\chi^2_{\rm r,\,SDSS}$ & Jet Fraction & $\alpha$ & $\delta$ & S/N \\
\hline
\endhead

\hline
\multicolumn{14}{r}{\textit{Continued on next page}} \\
\endfoot

\hline
\multicolumn{14}{l}{$^{\dagger}$ Low-confidence redshift: PL+Galaxy source at $z \gtrsim 1.9$ where the SDSS wavelength coverage provides no strong rest-frame anchors for the galaxy template.} \\
\multicolumn{14}{l}{$^{\ddagger}$ Low spectral S/N ($<3$): fitted parameters carry greater uncertainty and are excluded from population median calculations.} \\
\multicolumn{14}{l}{$<$ Jet fraction upper limit: PL+Lines sources where the power-law component may absorb a mixture of jet synchrotron and accretion-disc continuum emission.} \\
\endlastfoot

54742039 & SDSS J021210.47+532138.8 & 4FGL J0212.1+5321 & bin & Powerlaw+Galaxy & STAR & -0.0002 & $0.0553 \pm 0.0000$ & 23.186 & 2.698 & 0.078 & -1.779 & 1.199 & 58.3 \\
66949726 & SDSS J203213.12+412724.3 & 4FGL J2032.2+4127 & PSR & Powerlaw+Galaxy & STAR & 0.0002 & $3.4234 \pm 0.0005$$^{\dagger}$ & 5.188 & 1.179 & 0.782 & 1.200 & 1.200 & 20.7 \\
67558800 & SDSS J181635.93+451033.8 & 4FGL J1816.5+4510 & MSP & Powerlaw+QSO (Red z =0.5 (ebv = 0.1)) & STAR & 0.0008 & $2.2397 \pm 51397.4229$ & 2.763 & 0.845 & 1.000 & -2.500 & -0.853 & 10.8 \\
69702773 & SDSS J235836.84-180717.5 & 4FGL J2358.5-1808 & bll & Powerlaw+Lines (High Blueshift Line) & QSO & 1.2113 & $0.1169 \pm 0.0007$ & 17.178 & 0.644 & $< 0.978$ & -1.166 & -1.200 & 5.4 \\
70174185 & SDSS J014833.80+012901.4 & 4FGL J0148.6+0127 & bll & QSO (Default z=0.2) & GALAXY & 0.3614 & $1.3280 \pm 0.0399$ & 4646.083 & 0.819 & -- & -- & -- & 3.1 \\
74302545 & SDSS J084014.69-022711.4 & 4FGL J0840.1-0225 & bcu & Powerlaw+Galaxy & GALAXY & 0.0497 & $0.0499 \pm 0.0000$ & 5.272 & 0.977 & 0.364 & 1.196 & 1.200 & 27.3 \\
76013694 & SDSS J050650.15+032358.8 & 4FGL J0506.9+0323 & bll & Powerlaw+QSO (Default z=0.2) & STAR & -0.0010 & $1.3280 \pm 0.1446$ & 1453231.328 & 1.040 & 0.218 & -2.484 & -1.200 & 10.6 \\
82110263 & SDSS J173233.55-313124.4 & 4FGL J1732.5-3131 & PSR & Powerlaw+Galaxy & STAR & -0.0001 & $3.8693 \pm 0.0000$$^{\dagger}$ & 13.407 & 1.502 & 0.719 & 1.200 & 1.200 & 28.8 \\
82408579 & SDSS J172733.14-304807.9 & 4FGL J1727.6-3050 & glc & Powerlaw+Galaxy & STAR & 0.0005 & $3.5470 \pm 0.0002$$^{\dagger}$ & 17.494 & 1.793 & 0.055 & 1.199 & 1.198 & 23.0 \\
83271794 & SDSS J174804.74-244645.3 & 4FGL J1748.0-2446 & glc & Powerlaw & STAR & -0.0009 & $4.9305 \pm 0.0209$ & 91.666 & 6.390 & -- & 1.200 & 1.200 & 10.1 \\
91564676 & SDSS J053600.03-673507.4 & 4FGL J0535.2-6736 & HMB & Powerlaw+QSO (Red z=0.5 (ebv = 0.2)) & STAR & 0.0010 & $2.4459 \pm 625064.7554$ & 7.364 & 0.948 & 1.000 & -2.500 & -1.200 & 31.5 \\
92560104 & SDSS J054045.85-541822.1 & 4FGL J0540.8-5415 & fsrq & Powerlaw+Galaxy & QSO & 1.1911 & $0.2699 \pm 0.0001$ & 3.113 & 0.942 & 0.480 & -2.489 & 1.200 & 15.8 \\
95793687 & SDSS J063546.51-751616.8 & 4FGL J0635.6-7518 & fsrq & Powerlaw+QSO (High EW z=1.5) & QSO & 0.6574 & $2.7789 \pm 0.0007$ & 4.769 & 1.009 & 0.880 & -2.457 & -0.338 & 20.3 \\
121807820 & SDSS J084014.69-022711.5 & 4FGL J0840.1-0225 & bcu & Powerlaw+Galaxy & GALAXY & 0.0497 & $0.0500 \pm 0.0000$ & 4.402 & 1.287 & 0.233 & 1.200 & 1.200 & 30.6 \\
\thispagestyle{empty}
\end{longtable}
}

{\scriptsize \begin{longtable}{llllllllllllll}
\caption{The 25 4FGL sources with SDSS-V spectroscopy but with non-blazar classifications in the 4FGL-DR4 catalogue that we therefore exclude from analysis with our blazar model pipeline.} \label{tab:fermi_sources} \\
\hline
SDSS ID & SDSS Name & 4FGL Name & 4FGL Class & Best Model & SDSS Class & $z_{\rm SDSS}$ & $z_{\rm fit}$ & $\chi^2_{\rm r,\,lmfit}$ & $\chi^2_{\rm r,\,SDSS}$ & Jet Fraction & $\alpha$ & $\delta$ & S/N \\
\hline
\endfirsthead

\multicolumn{14}{c}{\textit{Continued from previous page}} \\
\hline
SDSS ID & SDSS Name & 4FGL Name & 4FGL Class & Best Model & SDSS Class & $z_{\rm SDSS}$ & $z_{\rm fit}$ & $\chi^2_{\rm r,\,lmfit}$ & $\chi^2_{\rm r,\,SDSS}$ & Jet Fraction & $\alpha$ & $\delta$ & S/N \\
\hline
\endhead

\hline
\multicolumn{14}{r}{\textit{Continued on next page}} \\
\endfoot

\hline
\multicolumn{14}{l}{$^{\dagger}$ Low-confidence redshift: PL+Galaxy source at $z \gtrsim 1.9$ where the SDSS wavelength coverage provides no strong rest-frame anchors for the galaxy template.} \\
\multicolumn{14}{l}{$^{\ddagger}$ Low spectral S/N ($<3$): fitted parameters carry greater uncertainty and are excluded from population median calculations.} \\
\multicolumn{14}{l}{$<$ Jet fraction upper limit: PL+Lines sources where the power-law component may absorb a mixture of jet synchrotron and accretion-disc continuum emission.} \\
\endlastfoot

55831735 & SDSS J093520.72+090035.9 & 4FGL J0935.3+0901 & bin & Powerlaw+Galaxy & STAR & 0.0004 & $0.0839 \pm 0.0003$ & 0.796 & 0.770 & 0.389 & -1.461 & 0.989 & 1.9$^{\ddagger}$ \\
55927207 & SDSS J084047.59+131223.6 & 4FGL J0840.8+1317 & ssrq & Powerlaw+Lines (Median Line) & QSO & 0.6802 & $0.6804 \pm 0.0002$ & 2.608 & 1.209 & $< 0.618$ & 0.461 & 1.200 & 4.0 \\
57183869 & SDSS J090933.50+425346.5 & 4FGL J0910.0+4257 & css & Powerlaw+QSO (High Blueshift z=0.2) & GALAXY & 0.7430 & $0.2729 \pm 0.0006$ & 0.980 & 0.957 & 0.947 & -1.204 & 1.058 & 37.7 \\
57408185 & SDSS J112833.62+583346.7 & 4FGL J1128.2+5831 & sbg & Powerlaw+Galaxy & GALAXY & 0.0105 & $2.0840 \pm 0.0005$$^{\dagger}$ & 18.792 & 11.283 & 0.866 & 1.200 & 0.640 & 21.4 \\
57904400 & SDSS J075828.11+374711.8 & 4FGL J0758.7+3746 & rdg & Powerlaw+Galaxy & GALAXY & 0.0409 & $0.0408 \pm 0.0000$ & 2.764 & 1.035 & 0.302 & 1.199 & 1.193 & 23.4 \\
58639432 & SDSS J084957.98+510829.0 & 4FGL J0850.0+5108 & NLSY1 & Powerlaw+QSO (Red z=0.5 (ebv = 0.2)) & GALAXY & 0.5840 & $0.5847 \pm 0.0002$ & 1.153 & 1.392 & 0.628 & -0.032 & 1.198 & 17.0 \\
62724568 & SDSS J145907.58+714019.9 & 4FGL J1459.0+7140 & css & Powerlaw+Lines (High EW Line) & QSO & 0.9033 & $0.9024 \pm 0.0001$ & 1.668 & 1.260 & $< 0.737$ & -0.830 & 1.200 & 8.8 \\
65673015 & SDSS J221532.69+513536.4 & 4FGL J2215.6+5135 & MSP & Powerlaw+Lines (High Blueshift Line) & QSO & 1.7286 & $3.5130 \pm 0.0069$ & 0.573 & 0.572 & $< 0.926$ & -0.021 & -1.007 & 0.9$^{\ddagger}$ \\
66235391 & SDSS J195840.10+284554.0 & 4FGL J1958.7+2846 & PSR & Powerlaw+Galaxy & GALAXY & 1.0013 & $3.4179 \pm 0.0003$$^{\dagger}$ & 0.801 & 0.745 & 0.274 & 1.200 & 1.200 & 1.8$^{\ddagger}$ \\
66834014 & SDSS J201603.36+371207.5 & 4FGL J2016.2+3712 & snr & Powerlaw+Galaxy & STAR & 0.0000 & $3.1604 \pm 0.0007$$^{\dagger}$ & 0.790 & 0.741 & 0.916 & -0.042 & 1.200 & 7.0 \\
67608140 & SDSS J182931.78+484446.2 & 4FGL J1829.5+4845 & css & Powerlaw+QSO (Default z=0.2) & QSO & 0.6923 & $2.8759 \pm 0.0007$ & 7.440 & 3.255 & 0.689 & -1.830 & 1.198 & 18.2 \\
70126366 & SDSS J023737.97+020742.5 & 4FGL J0237.7+0206 & rdg & Powerlaw+Galaxy & GALAXY & 0.3149 & $0.3149 \pm 0.0001$ & 0.861 & 0.803 & 0.640 & -0.409 & -1.200 & 10.9 \\
70267966 & SDSS J003820.53-020740.5 & 4FGL J0038.7-0204 & rdg & Powerlaw+Lines (High EW Line) & QSO & 0.2205 & $0.2207 \pm 0.0001$ & 11.168 & 4.672 & $< 0.911$ & -0.380 & 1.200 & 19.3 \\
76639662 & SDSS J060835.79+114941.0 & 4FGL J0608.6+1149 & unk & Powerlaw+Galaxy & GALAXY & 0.6685 & $0.7982 \pm 0.0002$ & 0.886 & 0.866 & 0.465 & 1.199 & 1.197 & 2.0$^{\ddagger}$ \\
79648099 & SDSS J102438.66-071919.8 & 4FGL J1024.5-0719 & MSP & Powerlaw+Galaxy & STAR & 0.0007 & $0.3186 \pm 0.0001$ & 1.116 & 0.795 & 0.192 & 1.200 & 1.193 & 6.8 \\
79706031 & SDSS J111812.46-041324.4 & 4FGL J1118.2-0415 & agn & Powerlaw+Lines (Median Line) & QSO & 2.5539 & $2.5575 \pm 0.0002$ & 3.255 & 2.417 & $< 0.789$ & -1.428 & -1.200 & 11.9 \\
79933375 & SDSS J102347.69+003840.7 & 4FGL J1023.7+0038 & LMB & Powerlaw+Lines (High Blueshift Line) & STAR & 0.0006 & $2.5749 \pm 0.0034$ & 0.884 & 0.646 & $< 0.988$ & -2.058 & 1.199 & 9.2 \\
91255187 & SDSS J042749.65-670435.1 & 4FGL J0427.8-6704 & LMB & Powerlaw+Galaxy & QSO & 1.0123 & $3.1588 \pm 0.0011$$^{\dagger}$ & 2.157 & 1.352 & 0.859 & -2.500 & -1.200 & 6.0 \\
96784938 & SDSS J112555.34-601406.9 & 4FGL J1126.4-6011 & MSP & Powerlaw+Galaxy & GALAXY & 0.3702 & $3.1340 \pm 0.0000$$^{\dagger}$ & 0.753 & 0.721 & 0.803 & 1.059 & 0.650 & 2.0$^{\ddagger}$ \\
103453487 & SDSS J145930.11-605320.6 & 4FGL J1459.5-6053 & PSR & Powerlaw+Galaxy & STAR & -0.0002 & $3.7165 \pm 0.0006$$^{\dagger}$ & 2.111 & 0.835 & 0.836 & 1.200 & 1.200 & 9.1 \\
103793482 & SDSS J154344.21-514954.9 & 4FGL J1543.6-5148 & MSP & Powerlaw+Galaxy & STAR & -0.0001 & $3.7181 \pm 0.0008$$^{\dagger}$ & 0.937 & 0.728 & 0.808 & 1.200 & 1.200 & 3.8 \\
109563674 & SDSS J144941.89-091000.8 & 4FGL J1449.7-0910 & agn & Powerlaw+Galaxy & GALAXY & 0.1835 & $0.1835 \pm 0.0001$ & 0.791 & 0.717 & 0.624 & 0.770 & -1.200 & 14.6 \\
110627802 & SDSS J203934.97-561709.3 & 4FGL J2039.5-5617 & MSP & Powerlaw+Galaxy & STAR & 0.0001 & $0.0250 \pm 0.0001$ & 1.098 & 0.773 & 0.126 & -1.722 & 1.198 & 7.4 \\
114642944 & SDSS J211852.96-073227.6 & 4FGL J2118.8-0723c & sey & Powerlaw+Lines (High EW Line) & QSO & 0.2598 & $0.2601 \pm 0.0001$ & 2.833 & 1.431 & $< 0.833$ & -0.145 & 1.200 & 7.2 \\
120505898 & SDSS J160612.69+000027.2 & 4FGL J1606.0+0011 & rdg & Powerlaw+Galaxy & GALAXY & 0.0580 & $0.0578 \pm 0.0001$ & 1.527 & 1.519 & 0.168 & 1.198 & 1.199 & 5.8 \\
\thispagestyle{empty}
\end{longtable}
}

\end{landscape}

\clearpage
\section{Example optical spectra for sources with updated blazar model fits, classifications and redshifts}
\label{app:examplespec}

The following figures present exemplary spectral fits for BL Lac and FSRQ candidates in the final blazar sample, grouped by best-fit model family. Sources are selected across a range of redshifts within each group to illustrate the diverse spectral characteristics encountered at different epochs. Each panel shows the observed SDSS  spectrum (black) with $\pm 1\sigma$ uncertainty (grey shading), the best-fit multi-component model (red/purple/cyan), and the individual jet and thermal components where applicable. Detected spectral features are indicated by vertical markers, and the zoomed inset highlights the key spectral feature that anchors the redshift determination for each source. 

Two aspects of the spectral fits deserve particular attention when inspecting these figures. First, PL+Galaxy sources beyond $z \gtrsim 1.8$ (Section~\ref{jet_z_dis}) fall within the low-confidence redshift regime. At those redshifts, the SDSS wavelength coverage provides no strong rest-frame anchors for the elliptical galaxy template, and the derived redshift should be treated with caution. These sources are flagged in Table~\ref{app:good_table}. For PL+Lines sources, the optical jet fraction $f_{\rm jet}$ reported in the figure panels is an upper limit on the true non-thermal contribution, as discussed in Section~\ref{sec:decomp} and Section~\ref{jet_z_dis}.  

\begin{figure*}
    \centering
\includegraphics[height=0.90\textheight]{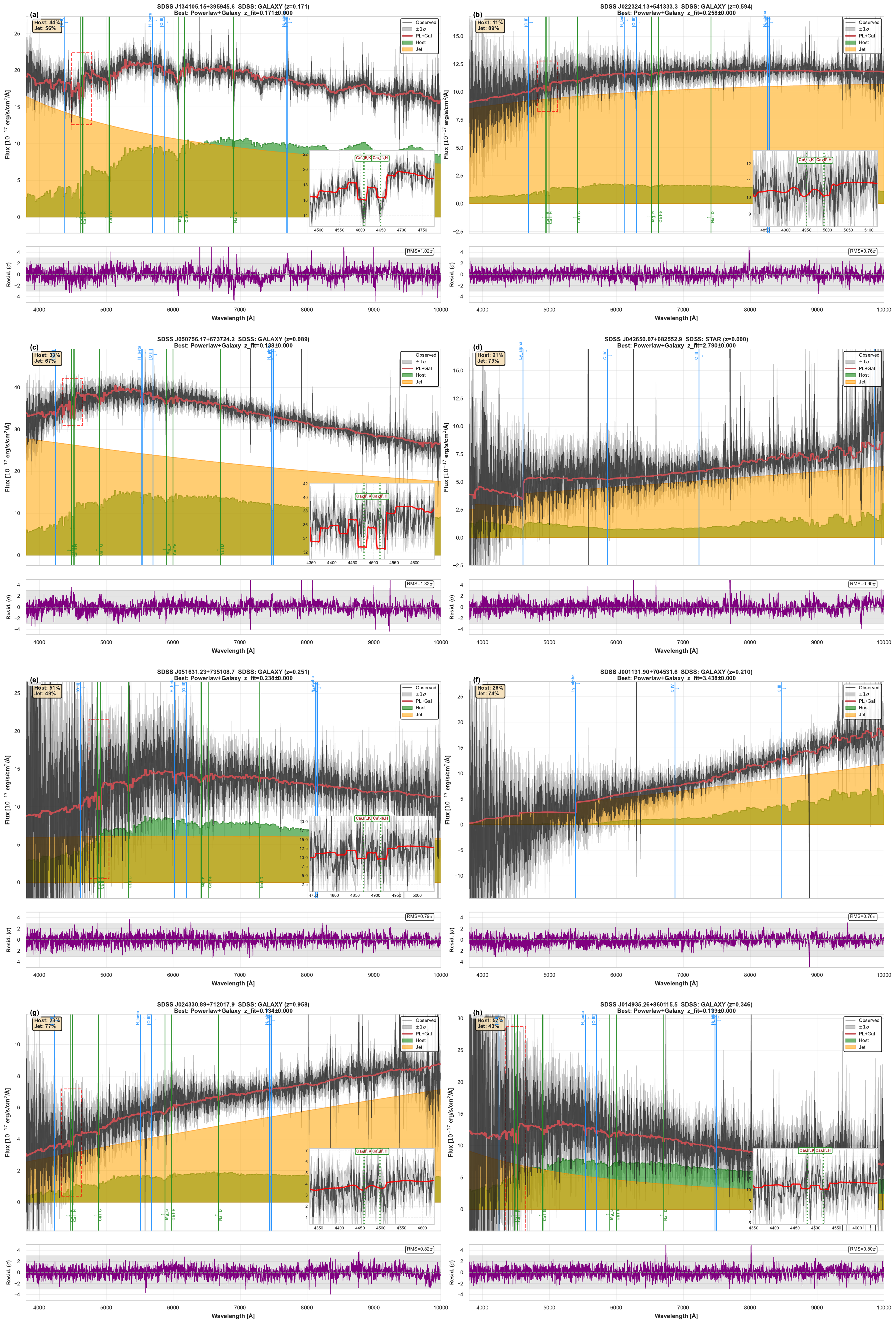}
    \caption{Representative spectral fits for power law + Galaxy sources showing the decomposition of jet (orange) and host galaxy (green) components. Sources are ordered by increasing redshift.}
    \label{append:plgal}
\end{figure*}

\begin{figure*}
    \centering
\includegraphics[height=0.90\textheight]{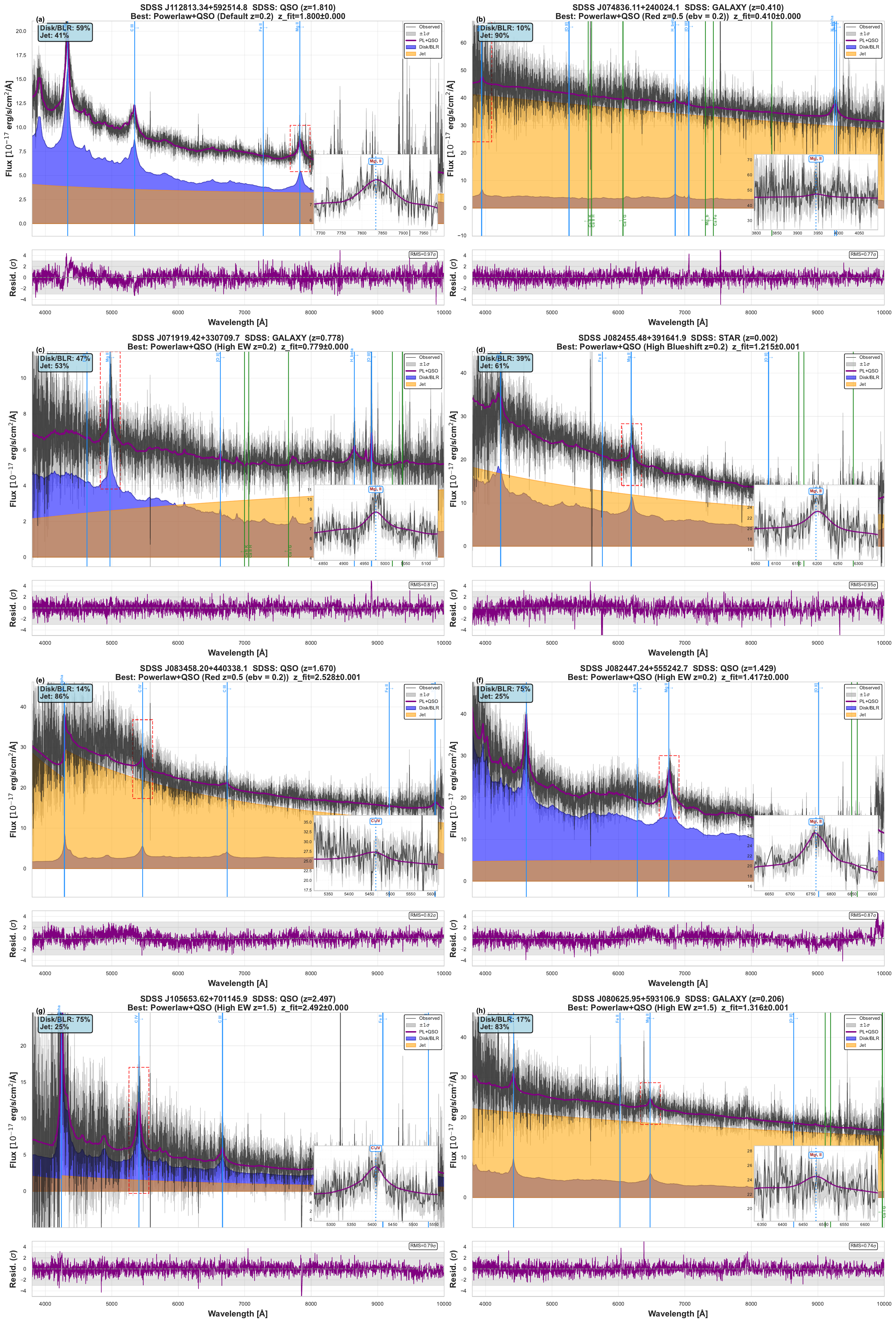}
    \caption{Representative spectral fits for power law + QSO sources showing the decomposition of jet (orange) and BLR/Disk (purple) components. Sources are ordered by increasing redshift.}
    \label{append:plqso}
\end{figure*}

\begin{figure*}
    \centering
\includegraphics[height=0.90\textheight]{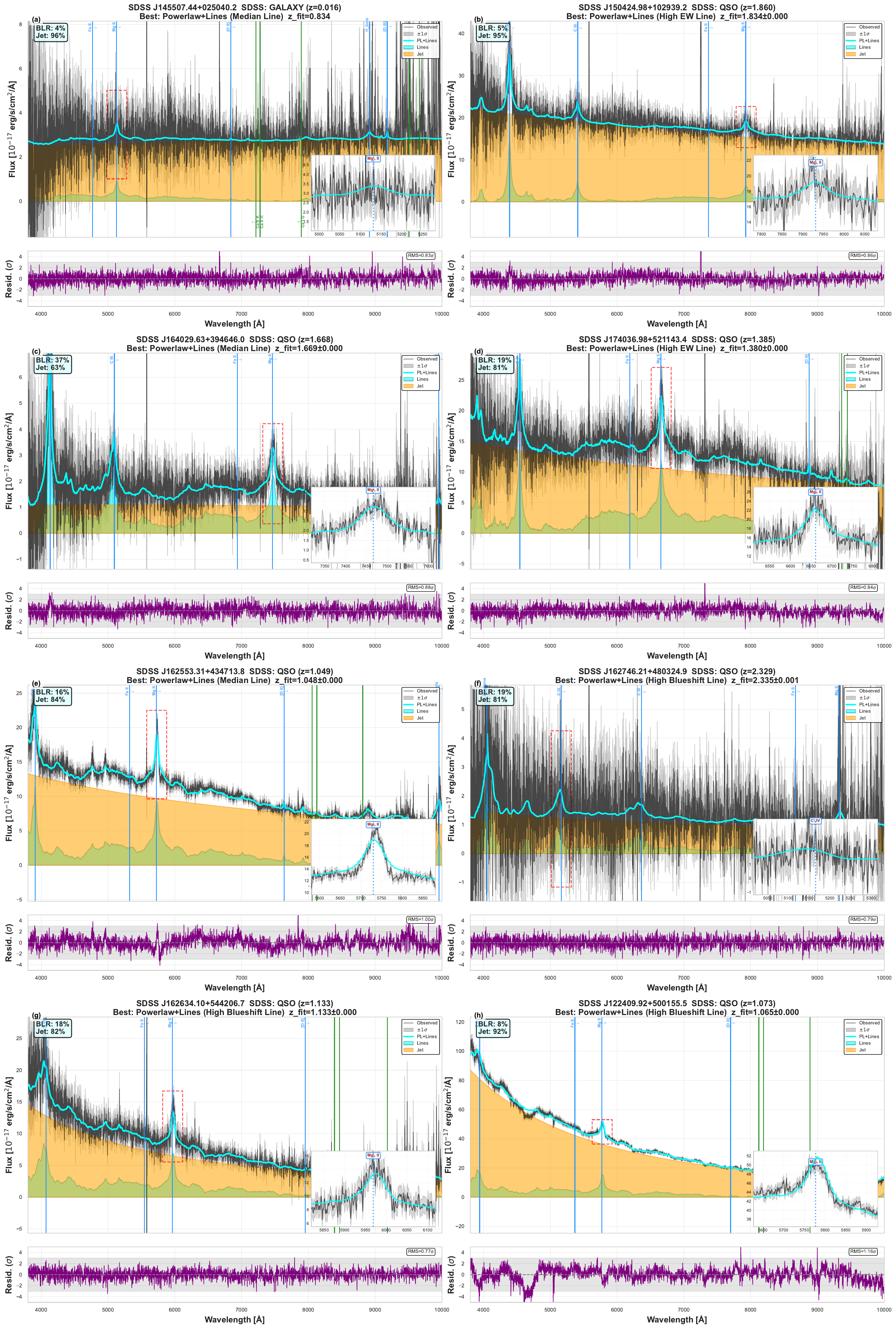}
    \caption{Representative spectral fits for power law + Lines-only sources showing the decomposition of jet (orange) and BLR (lightblue) components. Sources are ordered by increasing redshift.}
    \label{append:pllines}
\end{figure*}

The exemplary spectral fits are organised into 3 sections: BL Lac candidates fitted by the PL+Galaxy model (Figure~\ref{append:plgal}), FSRQ candidates fitted by the PL+QSO model (Figure~\ref{append:plqso}), sources fitted by the PL+Lines model family (Figure~\ref{append:pllines}), which include FSRQ candidates with suppressed accretion continuum contributions as well as potential masquerading BL Lac candidates \citep{padovani2019txs} where strong emission lines are detected against an otherwise jet-dominated continuum.

\bsp	\label{lastpage}
\end{document}